\documentclass{emulateapj}
\usepackage{apjfonts}
\usepackage{graphicx}

\def\mic              {\hbox{$\mu{\rm m}$}}

\shorttitle{Milky Way Tomography IV: Dissecting Dust}
\shortauthors{Berry et al.}

\begin{document}

\title{ The Milky Way Tomography with SDSS: IV. Dissecting Dust}

\author{
Michael Berry\altaffilmark{\ref{Washington},\ref{Rutgers}},
\v{Z}eljko Ivezi\'{c}\altaffilmark{\ref{Washington}},
Branimir Sesar\altaffilmark{\ref{Caltech}},
Mario Juri\'{c}\altaffilmark{\ref{HarvardMario}},
Edward F. Schlafly\altaffilmark{\ref{Harvard}},
Jillian Bellovary\altaffilmark{\ref{Michigan}},
Douglas Finkbeiner\altaffilmark{\ref{Harvard}},
Dijana Vrbanec\altaffilmark{\ref{PMF}},
Timothy C. Beers\altaffilmark{\ref{JINA}},
Keira J. Brooks\altaffilmark{\ref{Washington}},
Donald P. Schneider\altaffilmark{\ref{PennState}},
Robert R. Gibson\altaffilmark{\ref{Washington}},
Amy Kimball\altaffilmark{\ref{NRAO}},
Lynne Jones\altaffilmark{\ref{Washington}},
Peter Yoachim\altaffilmark{\ref{Washington}},
Simon Krughoff\altaffilmark{\ref{Washington}},
Andrew J. Connolly\altaffilmark{\ref{Washington}},
Sarah Loebman\altaffilmark{\ref{Washington}},
Nicholas A. Bond\altaffilmark{\ref{Rutgers}},
David Schlegel\altaffilmark{\ref{LBNL}},
Julianne Dalcanton\altaffilmark{\ref{Washington}},
Brian Yanny\altaffilmark{\ref{FNAL}},
Steven R. Majewski\altaffilmark{\ref{UVa}},
Gillian R. Knapp\altaffilmark{\ref{Princeton}}, 
James E. Gunn\altaffilmark{\ref{Princeton}},
J. Allyn Smith\altaffilmark{\ref{AustinPeay}},
Masataka Fukugita\altaffilmark{\ref{UT3}},
Steve Kent\altaffilmark{\ref{FNAL}},
John Barentine\altaffilmark{\ref{APO}}, 
Jurek Krzesinski\altaffilmark{\ref{APO}}, 
Dan Long\altaffilmark{\ref{APO}}
}

\altaffiltext{1}{Department of Astronomy, University of Washington, Box 351580, Seattle, WA 98195
\label{Washington}}
\altaffiltext{2}{Physics and Astronomy Department, Rutgers University Piscataway, NJ 08854-8019, U.S.A.
\label{Rutgers}}
\altaffiltext{3}{Division of Physics, Mathematics and Astronomy, Caltech, Pasadena, CA 91125
\label{Caltech}}
\altaffiltext{4}{Hubble Fellow; Harvard College Observatory, 60 Garden St., Cambridge, MA 02138
\label{HarvardMario}}
\altaffiltext{5}{Harvard-Smithsonian Center for Astrophysics, 60 Garden Street, Cambridge, MA 02138
\label{Harvard}}
\altaffiltext{6}{Departmentof Astronomy, University of Michigan, Ann Arbor, MI, USA
\label{Michigan}}
\altaffiltext{7}{Department of Physics, Faculty of Science, University of Zagreb, Bijeni\v{c}ka cesta 32, 10000 Zagreb, Croatia
\label{PMF}}
\altaffiltext{8}{National Optical Astronomy Observatories, Tucson, AZ, 85719, Department of Physics \& Astronomy 
and JINA: Joint Institute for Nuclear Astrophysics, Michigan State University, East Lansing, MI 48824, USA
\label{JINA}}
\altaffiltext{9}{Department of Astronomy and Astrophysics, The Pennsylvania State University, University Park, PA 16802
\label{PennState}}
\altaffiltext{10}{National Radio Astronomy Observatory, 520 Edgemont Road, Charlottesville, VA 22903-2475
\label{NRAO}}
\altaffiltext{11}{Lawrence Berkeley National Laboratory, One Cyclotron Road, MS 50R5032, Berkeley, CA, 94720 
\label{LBNL}}
\altaffiltext{12}{Fermi National Accelerator Laboratory, P.O. Box 500, Batavia, IL 60510
\label{FNAL}}
\altaffiltext{13}{Department of Astronomy, University of Virginia,
       P.O. Box 400325, Charlottesville, VA 22904-4325
\label{UVa}}
\altaffiltext{14}{Princeton University Observatory, Princeton, NJ 08544
\label{Princeton}}
\altaffiltext{15}{Dept. of Physics \& Astronomy, Austin Peay State University, 
Clarksville, TN 37044
\label{AustinPeay}}
\altaffiltext{16}{
Institute for Cosmic Ray Research, University of Tokyo, Kashiwa, Chiba, Japan
\label{UT3}}
\altaffiltext{17}{Apache Point Observatory, 2001 Apache Point Road, P.O. Box 59, 
Sunspot, NM 88349-0059
\label{APO}}

\begin{abstract}
We use SDSS photometry of 73 million stars to simultaneously 
obtain best-fit main-sequence stellar energy distribution (SED) and 
amount of dust extinction along the line of sight towards each 
star. Using a subsample of 23 million stars with 2MASS photometry, 
whose addition enables more robust results, we show that SDSS photometry 
alone is sufficient to break degeneracies between intrinsic stellar 
color and dust amount when the shape of extinction curve is fixed. 
When using both SDSS and 2MASS photometry, the ratio of the total 
to selective absorption, $R_V$, can be determined with an uncertainty
of about 0.1 for most stars in high-extinction regions. These fits enable detailed 
studies of the dust properties and its spatial distribution, and of 
the stellar spatial distribution at low Galactic latitudes ($|b|<30^\circ$). 
Our results are in good agreement with the extinction normalization
given by the \citet[][SFD]{SFD98} dust maps at 
high northern Galactic latitudes, but indicate that the SFD extinction 
map appears to be consistently overestimated by about 20\% in the
southern sky, in agreement with recent study by \citet{Sch2010}. 
The constraints on the shape of the dust extinction
curve across the SDSS and 2MASS bandpasses disfavor the reddening law 
of \citet{ODonnell}, but support the models by \citet{Fitz99} and 
\citet{CCM}. For the latter, we find a ratio of the total 
to selective absorption to be $R_V=3.0\pm0.1$(random)$\pm0.1$ (systematic) 
over most of the high-latitude sky. At low Galactic 
latitudes ($|b|<5^\circ$), we demonstrate that the SFD map cannot be reliably 
used to correct for extinction because most stars are embedded in dust, rather 
than behind it, as is the case at high Galactic latitudes. We analyze three-dimensional 
maps of the best-fit $R_V$ and find that $R_V=3.1$ cannot be ruled out in any
of the ten SEGUE stripes at a precision level of $\sim0.1-0.2$.  Our best estimate 
for the intrinsic scatter of $R_V$ in the regions probed by SEGUE stripes is $\sim0.2$. 
We introduce a method for efficient selection of candidate red giant stars in the disk,
dubbed ``dusty parallax relation'', which utilizes a correlation between distance and
the extinction along the line of sight. We make these best-fit parameters, as well as all 
the input SDSS and 2MASS data, publicly available
in a user-friendly format. These data can be used for studies of stellar
number density distribution, the distribution of dust properties, for
selecting sources whose SED differs from SEDs for high-latitude main sequence
stars, and for estimating distances to dust clouds and, in turn, to molecular gas clouds.  
\end{abstract}
\keywords{
methods: data analysis ---
stars: statistics ---
Galaxy: disk, stellar content, structure, interstellar medium
}

\section{                        INTRODUCTION                             }

From our vantage point inside 
the disk of the Milky Way, we have a unique opportunity to study an $\sim L^*$ spiral 
galaxy in great detail. By measuring and analyzing the properties of large numbers of 
individual stars, we can map the Milky Way in a nine-dimensional space spanned by the 
three spatial coordinates, three velocity components, and the three main stellar 
parameters -- luminosity, effective temperature, and metallicity.
In a series of related studies, we used data obtained by
the Sloan Digital Sky Survey \citep{SDSS} to study in detail the distribution of
tens of millions of stars in this multi-dimensional space. In \citet[hereafter 
J08]{Juric08} we examined the spatial distribution of stars in the Galaxy; 
in \citet[hereafter I08]{Ivezic08} we extended our analysis to include the
metallicity distribution; and in \citet[hereafter B10]{Bond10} 
we investigated the distribution of stellar velocities. In Juri\'{c} et al. (in prep) we estimate 
stellar luminosity functions for disk and halo stars, and describe an empirical Galaxy model and 
corresponding publicly available modelling code that encapsulate these SDSS-based results.

All of the above studies were based on SDSS data at high Galactic latitudes 
($|b|>30^\circ$). Meanwhile, the second phase of SDSS has delivered imaging data for 
ten $\sim2.5^{\circ}$ degree wide stripes 
\citep[in SDSS terminology, two independent observing runs produce two interleaving strips, 
which form a stripe, see][]{EDR} that cross the Galactic plane
\citep[the so-called SEGUE data, see][]{Yanny09}. 
At least in principle, these data can be used to extend 
the above analysis much closer to the mid-plane of the Galaxy, and to search for evidence
of effects such as disk warp and disk flare. 

However, at low Galactic latitudes sampled by SEGUE data, 
there are severe problems with the interstellar dust extinction corrections. 
High-latitude SDSS data are typically corrected for interstellar
extinction using maps from \citet[][hereafter SFD]{SFD98}. 
When the full SFD extinction correction is 
applied to low-latitude data, the resulting color-magnitude and color-color diagrams 
have dramatically different morphology than those observed at high
Galactic latitudes. Models developed by J08 suggest that these
problems are predominantly due to the fact that stars are embedded in
the dust layer, rather than behind it (the latter is an excellent approximation
for most stars at high Galactic latitudes), and thus the SFD extinction value
is an overestimate for most stars. This conclusion is also supported by
other Galaxy models, such as Besan\c{c}on \citep{Robin03} and TRILEGAL \citep{trilegal05}.
Therefore, in order to fully exploit SEGUE data, both the intrinsic
colors of a given star and the amount of dust extinction along the line-of-sight 
to the star have to be known. Distances to stars,
which can be derived using appropriate photometric parallax relations (see I08),  
would then enable mapping of the stellar spatial distribution. The interstellar medium (ISM)
dust distribution and dust extinction properties are interesting in their own right e.g.,
\citep[e.g.,][and references therein]{FM09,draineBook}. 
An additional strong motivation for quantifying stellar 
and dust distribution close to the Galactic plane is to inform the planning
of the Large Synoptic Survey Telescope (LSST) survey, which is considering deep 
multi-band coverage of the Galactic plane\footnote{
\hbox{See also Chapters 6 and 7 in the LSST Science Book available from}
www.lsst.org/lsst/scibook.} \citep{IvezicLSST}. 

The amount of dust can be constrained by measuring dust
extinction and/or reddening, typically at UV, optical and near-IR wavelengths, 
by measuring dust emission at far-IR wavelengths, and by employing a tracer
of interstellar medium (ISM), such as HI gas. For example, in their pioneering
studies in the late 1960s, Shane \& Wirtanen used galaxy counts, and \citet{KK74} exploited
a correlation between dust and HI column densities to infer the amount of dust extinction.
The most widely used contemporary dust map (SFD) is derived from observations of dust
emission at 100 \mic\ and 240 \mic, and has an angular resolution
of $\sim$6 arcmin (the temperature correction applied to IRAS 100 \mic\ data
is based on DIRBE 100 \mic\ and 240 \mic\ data, and has a lower angular
resolution of $\sim1^{\circ}$; see SFD for more details). It has been found that the 
SFD map sometimes overestimates the dust column by 20-30\% when the dust extinction in 
the SDSS $r$ band, $A_r\sim0.85A_V$, exceeds 0.5 mag \citep[e.g.,][]{Arce99}.
Such an error may be due to confusion of the background emission and that from point sources. 
A generic shortcoming of the far-IR emission-based methods is that they cannot 
provide constraints on the three-dimensional distribution of dust; instead, only the total 
amount of dust along the line of sight to infinity is measured. In addition, the far-IR data provide no 
constraints for the wavelength dependence of extinction at UV, optical and near-IR 
wavelengths. 

With the availability of wide-angle digital sky surveys at optical and near-IR wavelengths, 
such as SDSS and 2MASS (see \S\ref{sec:methodology} for more details), it is now possible 
to study the effects of dust extinction using many tens of millions of sources. For example, 
\citep[][ hereafter Sch2010]{Sch2010} utilized colors of blue stars,
and \citet{PG2010} utilized colors of passive red galaxies, to estimate errors in the 
SFD map at high Galactic latitudes. 
In both studies, {\it the dust reddening is assumed
constant within small sky patches,} and the color distribution for a 
large number of sources from a given patch is used to infer the mean reddening
(Peek \& Graves dub this approach ``standard crayon'' method). Traditional 
dust reddening estimation methods where the ``true'' color of a star is determined 
using spectroscopy were extended to the extensive SDSS spectroscopic dataset
by \citet{SF2010} and \citet{JWF2011}; they obtained results consistent with the above 
``standard crayon'' methods. Studies of dust extinction with SDSS data are limited
to $A_V \la 10$; 2MASS data alone can be used to trace dust up to $A_V\sim 20$
using near-infrared color excess method \citep{Lombardi2001,Lombardi2011,Majewski2011},
though estimates of stellar distances are not as reliable as with SDSS data. 

In this work, we extend these studies to low Galactic latitudes where stars are embedded
in dust, and also investigate whether optical and near-IR photometry are sufficient to 
constrain the shape of the dust extinction curve.
We estimate dust extinction along the line of sight to {\it each detected star} by simultaneously 
fitting its observed optical/IR spectral energy distribution (SED) using an empirical library 
of intrinsic reddening-free SEDs, a reddening curve described by the standard 
parameters: $R_V=A_V/E(B-V)$, and the dust extinction along the line of sight in the SDSS
$r$ band, $A_r$. We first select a dust extinction model using high Galactic latitude
data and another variation of the ``standard crayon'' method that incorporates the
eight-band SDSS-2MASS photometry. Our SED fitting method that treats each star 
separately allows an estimation of the three-dimensional 
spatial distributions of both stars and dust. The dataset and methodology, including various 
tests of the adopted algorithm, are described in \S2. 
Results are analyzed in \S3, and a preliminary investigation of the three-dimensional 
stellar count distribution and the distribution of dust properties is presented in \S4.
The main results are summarized and discussed in \S5.

\section{DATA AND METHODOLOGY}
\label{sec:methodology}

We first describe the data used in this work, and then discuss methodology, including 
various tests of the adopted algorithm. All datasets used in this study are defined using 
SDSS imaging data for unresolved sources. Objects that are positionally associated with 
2MASS sources are a subset of the full SDSS sample. Although the SDSS-2MASS dataset 
is expected to provide better performance than SDSS data alone when estimating dust 
properties and intrinsic stellar colors, we also consider the SDSS dataset alone (hereafter
referred to as ``only-SDSS'') because it is effectively deeper (unless the dust extinction
in the SDSS $r$ band is larger than several magnitudes). We start by briefly describing 
the SDSS and 2MASS surveys.

\subsection{SDSS Survey} 

The properties of the SDSS are documented in
\cite{F96, Gunn98, Hogg02, Smith02, EDR, Pier03, Ivezic04, Tucker06} and \cite{Gunn06}. 
In addition to its imaging survey data, SDSS has obtained well over half a
million stellar spectra, many as part of the Sloan Extension for Galactic
Understanding and Exploration \citep[SEGUE;][]{Yanny09}. 
Here we only reiterate that the survey photometric catalogs are 95\% complete to a 
depth of $r\sim22$, with photometry accurate to $\sim$0.02 mag (both absolute and 
rms error) for sources not limited by Poisson statistics. Sources with $r<20.5$ have 
astrometric errors less than 0.1 arcsec per coordinate \citep[rms;][]{Pier03}, and robust 
star/galaxy separation is achieved for $r\la 21.5$ \citep{Lupton01}.

The SDSS Data Release 7 \citep{DR7}.  used in this work contains photometric and astrometric data
for 357 million unique objects\footnote{For more details, see http://www.sdss.org/dr7/},
detected in 11,663 sq. deg. About half of these objects are unresolved, and are dominated
by stars (quasars contribute about 1\%, see J08). 
A full discussion of the photometric quality control for the SEGUE scans is
detailed in \citet{DR7}. Briefly, median reddening-free
colors ($Q_{gri}$ and $Q_{riz}$) were calculated for each field using
magnitudes computed by both the SDSS {\it photo} \citep{Lupton01} and Pan-STARRS {\it PS} \citep{Magnier2010}
image processing pipelines, and the position and width of the locus of points
(corresponding to the stellar main sequence) were computed. Fields
within 15$^\circ$ of the Galactic plane had a wider distribution
($\sigma_Q$(photo,PS) $\sim$ 0.035, 0.027 mag) than fields outside the
plane ($\sigma_Q$(photo,PS) $\sim$ 0.021, 0.20 mag). It can
therefore be inferred that (unsurprisingly) the photometric precision
in the plane is slightly poorer than that at higher latitudes. A more
direct comparison is provided by magnitude differencing the {\tt
photo} and {\tt PS} photometry. For stars with $14 < u,g,r,i,z <
20$, the median PSF magnitude difference was found to be 0.014 mag
within the plane versus 0.010 mag outside the plane.

\subsection{2MASS Survey} 

The Two Micron All Sky Survey used two 1.3 m telescopes to survey the entire sky 
in near-infrared light \citep{Skrutskie97}. Each telescope
had a camera with three 256$\times$ 256 arrays of HgCdTe detectors, and 
observed simultaneously in the $J$ (1.25 $\mu$m), $H$ (1.65 $\mu$m), 
and $K_s$ (2.17 $\mu$m, hereafter $K$) bands. The detectors were sensitive to point sources brighter 
than about 1 mJy at the 10$\sigma$ level, corresponding to limiting magnitudes 
of 15.8, 15.1, and 14.3, respectively (Vega based; for corrections to AB magnitude scale
see below). Point-source photometry is repeatable to better 
than 10\% precision at these limiting magnitudes, and the astrometric uncertainty for these sources is 
less than 0.2$''$. The 2MASS catalogs contain positional and photometric information 
for about 500 million point sources and 2 million extended sources.

\subsection{The Main-Sample Selection}  

\begin{figure}[!t]
\plotone{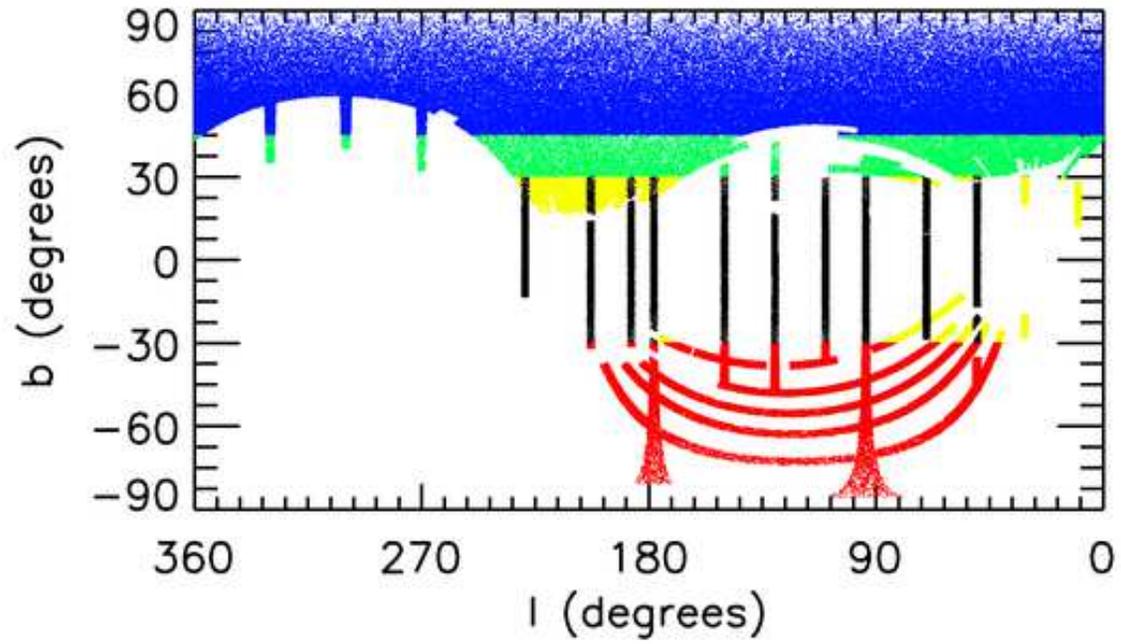}
\caption{The sky coverage for SDSS Data Release 7, used in this study, in Galactic coordinates. The points show a small random subsample of the full sample of 73 million stars analyzed in this paper. The different colors represent the various data file sets (blue, $\textit{b}>45^{\circ}$; green, $45^{\circ}>\textit{b}>30^{\circ}$; black, the 10 SEGUE strips; yellow, $|\textit{b}|<30^{\circ}$, stars not in SEGUE strips; and red, $\textit{b}<-30^{\circ}$). 
}
\label{Fig:sky}
\end{figure}

The main sample is selected from the SDSS Data Release 7 using the 
following two main criteria:
\begin{enumerate}
\item unique unresolved sources: objc\_type=6, binary processing flags
          DEBLENDED\_AS\_MOVING, SATURATED, BLENDED, BRIGHT, and NODEBLEND
          must be false, parameter nCHILD=0, and 
\item the model $r$-band magnitudes (uncorrected for extinction) must satisfy  $rMod < 21$,
\end{enumerate}
These criteria yielded $73$ million stars (for an SQL query used to select the main sample see 
Appendix A).  The distribution of selected sources on the sky is shown in Figure~\ref{Fig:sky}. 

For isolated sources, the $r<21$ condition ensures that photometric errors
are typically not larger than 0.05 mag \citep[see Fig.~1 in][]{Sesar07}. 
For sources with $r<19$, the errors reach their systematic limit
of $\sim$0.02 mag. When reported errors are smaller than 0.02 mag, we 
reset them to 0.02 mag to account for expected photometric zeropoint calibration 
errors \citep{ubercal2008}. 
The behavior of best-fit $\chi^2_{pdf}$ distributions
described in \S\ref{sec:chi2} justifies the reset of errors. Errors can be much larger for sources in complex 
environments, and sometimes reported errors are unreliable 
\citep[e.g., when sources are closer than 3$''$, the photometric errors are overestimated, see Figure 14 in][]{SIJ08}. 
If the cataloged photometric error is larger than 0.5 
mag in the $griz$ bands, or larger than 1.5 mag in the $u$ band, that data point is not used
in the analysis (formally, we reset the magnitudes to 999.9 and their
errors to 9999.9 in publicly available files, see Appendix B).

\subsection{SDSS-2MASS Subsample}

Following \citet{Covey07}, acceptable 2MASS sources must have 
2MASS quality flags $rd\_{flag}$ == 222, $bl\_{flag}$ == 111, and 
$cc\_{flag}$ == 0, and selected 2MASS sources are positionally matched 
to SDSS sources with a distance cutoff of 1.5$''$. 
The combined SDSS-2MASS catalog contains $\sim$23 million sources.
The wavelength coverage of the SDSS and 2MASS bandpasses are shown
in Fig. 3 in \citet{Finlator00}. The distributions of SDSS-2MASS
sources in various color-color and color-magnitude diagrams are 
discussed in detail by  \citet{Finlator00} and \citet{Covey07}. 
We emphasize that practically all sources in an SDSS-2MASS point source 
sample defined by a $K$-band flux limit are sufficiently bright to be
detected in all other SDSS and 2MASS bands. For orientation, main sequence stars selected
by the condition $K<14.3$ are closer than approximately 1-2kpc.

Similarly to the treatment of SDSS photometry, for stars with reported errors
in the $J$, $H$, and $K$ bands greater than 0.5 mag, we reset magnitudes 
and errors to 999.9 and 9999.9, respectively. We also reset photometric errors
to 0.02 mag when reported errors are below this limit \citep[systematic errors in 2MASS photometry 
are 0.02 mag;][]{Skrutskie97}. The Vega-based 2MASS photometry is translated to SDSS-like 
AB system following \citet{Finlator00}.
\begin{eqnarray}
J_{AB} = J_{2MASS}+0.89 \\ \nonumber
H_{AB} = H_{2MASS}+1.37 \\ \nonumber
K_{AB} = K_{2MASS}+1.84 \nonumber
\end{eqnarray}

Note that these corrections have no impact on fitting and results
(because the same corrections are applied to models and observations
and thus cancel, see below), but are convenient when visualizing SEDs.

\subsection{Model Assumptions and Fitting Procedures \label{sec:model}}

There are two empirical results that form the basis of our method. 
First, the stellar locus in the multi-dimensional color space spanned
by SDSS and 2MASS colors is nearly one dimensional (because for most 
stars the effective temperature has much more effect on colors than other physical 
parameters, such as age and metallicity). The locus position reflects basic stellar
physics and is so well defined that it has been used to test the quality 
of SDSS photometry \citep{Ivezic04}, as well as to calibrate new
photometric data \citep{High2009}. 

Second, the {\it shape} of the dust extinction curve can be described as 
a one-parameter family, usually
parametrized by $R_V=A_V/E(B-V)$ \citep{CCM, ODonnell, Fitz99, FM09}. 
Using this parametrization, extinction in an arbitrary photometric bandpass
$\lambda$ is equal to 
\begin{equation}
\label{eq:Clambda}
              A_\lambda = C_\lambda(R_V) \, A_r,
\end{equation}
where $A_r$ is extinction in the SDSS $r$ band, and $C_\lambda(R_V)$ describes
the shape of the extinction curve\footnote{The often used parametrization
of dust extinction curve, $k(\lambda-V)=E(\lambda-V)/E(B-V)$, is related to
$C_\lambda$ via $k(\lambda-V) = R_V\,(C_\lambda \, A_r/A_V -1)$; \citet{SF2010}
give $A_V/A_r=1.200$.}. Hence, the observed colors 
can be fit using only three free parameters: the position along the locus, 
$R_V$, and $A_r$ (eq.~\ref{eq:Clambda} is not the only way to ``close''
the system of equations; for a detailed discussion see Appendix B). Some caveats to this 
statement, such as the fact that not all unresolved sources are found along the locus 
(e.g., quasars and unresolved binary stars), and that even for fixed dust properties $A_r$ 
and $A_\lambda$ depend on the source spectral energy distribution, are  
discussed in quantitative detail further below. We note that it is not mandatory
to adopt an extinction curve parametrization given by eq.~\ref{eq:Clambda}. 
For example, we could simply adopt the $A_\lambda$ values determined for high 
Galactic latitude regions by \citet{Sch2010}. However, large dust extinction observed at
low Galactic latitudes offers a possibility to constrain the shape of the dust extinction
curve, and eq.~\ref{eq:Clambda} provides a convenient one-parameter description 
that works well in practice (but see also \citealt{FM09} for a different functional
parametrization with two free parameters). 

A similar method was recently proposed by \citet{BJ2011}, where a strong prior is obtained
from measured (trigonometric) distances and a requirement that stars must be consistent
with stellar evolutionary track in the Hertzsprung-Russell diagram (as opposed to our
constraint that stellar colors must be consistent with the stellar color locus). Such a prior
has the advantage of being able to easily distinguish giant stars from main sequence 
stars. Unfortunately, trigonometric distances are not available for the vast majority of stars in our sample.

\subsubsection{Fitting Details}
 
The best-fit empirical stellar model from a library described in \S\ref{sec:CoveySEDs},
and the dust extinction according to a $C_\lambda(R_V)$ parametrization 
described in \S\ref{sec:Aparam}, 
are found by minimizing $\chi^2_{pdf}$ defined as 
\begin{equation}
  \chi^2_{pdf} = {1 \over N-k} \,\sum_{i=1}^N \left({ c^{obs}_i - c^{mod}_i  \over \sigma_i } \right)^2,
\end{equation}
where $c^{obs}_i$ are $N$ observed adjacent (e.g., $u-g$, $g-r$, etc.)
colors ($N=4$ for only-SDSS dataset, and $N=7$ for SDSS-2MASS dataset).
The number of fitted parameters is $k=3$ for all parameters, and $k=2$ when a fixed 
value $R_V=3.1$ is assumed (see below). 

The model colors are constructed using extinction-corrected magnitudes
\begin{equation}
       m^{corr}_\lambda = m^{obs}_\lambda - A_\lambda,
\end{equation}
with $\lambda=(ugriz[JHK])$, resulting in 
\begin{equation}
      c^{mod} = c^{lib}(t) + \left[C_{\lambda 2}(R_V)-C_{\lambda 1}(R_V)\right]\, A_r.
\end{equation}
Here $\lambda 1$ and $\lambda 2$ correspond to two adjacent bandpasses
which define colors $c^{mod}$ and $c^{lib}$.  
Hence, by minimizing $\chi^2_{pdf}$, we obtain the best-fit values for three free 
parameters: $R_V$, $A_r$, and the model library index, $t$ (intrinsic stellar
color, or position along the locus). Once these
parameters are determined, the overall flux normalization (i.e. apparent 
magnitude offset) is determined by minimizing $\chi^2_{pdf}$ for the 
fixed best-fit model. 

We minimize $\chi^2_{pdf}$ by a brute force method. All 228 library SEDs (see \S\ref{sec:CoveySEDs})
are tried, with dust extinction values in the range $0 \le A_r \le 10$ 
with 0.02 mag wide steps. This is not a very efficient method, but
the runtime on a multi-processor machine was nevertheless much shorter,
in both human and machine time, than post-fitting analysis of the results. 

We investigate the impact of $R_V$ by producing two sets of best-fit $t$ and $A_r$. First, 
we fixed $R_V=3.1$, and then allow $R_V$ to vary in the range 1-8, with 
0.1 wide steps. The results for the two cases are compared and analyzed in
the next section. 

The errors, $\sigma_i$, are computed from photometric errors quoted
in catalogs, with a floor of 0.02 mag added in quadrature to account
for plausible systematic errors (such as calibration uncertainties), as well as for 
the finite locus width. In principle, $\sigma_i$ could be varied with the
trial library SED to account for the varying width of the stellar locus. We
have not implemented this feature because it does not dominate the 
systematic errors. 

For a given $R_V$ value (whether constant, or a grid value in the free $R_V$
case), once the minimum $\chi^2$, $\chi_{min}^2$, is located, 
an ellipse is fit to the section of the $\chi^2$ surface defined by 
$\chi^2 < \chi_{min}^2 + 6.17$ (i.e., within 2$\sigma$ 
deviation for 2 degrees of freedom):
\begin{equation}
  \chi^2(t,A_r|R_V) = a(t-t^\ast)^2 + b(t-t^\ast)(A_r-A_r^\ast) + c(A_r-A_r^\ast)^2 
\end{equation}
were $t$ is the model index, and $t^\ast$ and $A_r^\ast$ are the best-fit
values corresponding to $\chi_{min}^2$. Using the best-fit parameters
$a$, $b$ and $c$, the (marginalized) model and $A_r$ errors can be
computed from 
\begin{equation}
\label{eq:sig1}
\sigma_t = \left(a - \frac{b^2}{4c}\right)^{-\frac{1}{2}}
\end{equation}

\begin{equation}
\label{eq:sig2}
\sigma_A = \left(c - \frac{b^2}{4a}\right)^{-\frac{1}{2}}
\end{equation}
Note that the $b$ coefficient controls the covariance between $t^\ast$ and $A_r^\ast$. 
The $\chi^2$ surface around the best-fit $t$/$A_r$ combination is described well by 
an ellipsoid, although this error ellipse approximation breaks down far from the best-fit.
The $\chi^2$ surface for stars with $\chi^2_{min}>200$ is not fit with an ellipse and 
such stars, contributing less than two percent of the entire sample, are instead 
marked as bad fits.

\subsection{The Covey et al. Stellar SED Library \label{sec:CoveySEDs} }

\citet{Covey07} quantified the main stellar locus in the $ugrizJHK$
photometric system using a sample of $\sim$600,000 point sources detected 
by SDSS and 2MASS. They  tabulated the locus position and width as a function 
of the $g-i$ color, for 228 $g-i$ values in the range $-0.25 < g-i < 4.50$.  
We adopt this locus parametrization as our empirical SED library. Strictly speaking, 
this is not an exhaustive SED library that includes all possible combinations of 
effective temperature, metallicity and gravity, but rather a parameterization of the 
mean locus and its width in the multi-dimensional SDSS-2MASS color space.
We note that \citet{Covey07} used the SFD map to correct SDSS and 2MASS 
photometry for interstellar dust extinction. Because they only studied 
high galactic latitude regions where typically $A_r<0.1$, errors in derived 
locus parametrization due to errors in the SFD map are at most 0.01 mag, 
and thus smaller or at most comparable to photometric calibration errors. 

This $g-i$ parametrization reflects the fact that the stellar effective temperature, 
which by and large controls the $g-i$ color, is more important than other physical 
parameters, such as age (gravity) and metallicity, in determining the overall SED shape
\citep[for a related discussion and principal component analysis of SDSS stellar spectra see][]{McGurk2010}.  
The adopted $g-i$ range includes the overwhelming majority of all unresolved SDSS
sources, and approximately corresponds to MK spectral types from early A to late M.  
Due to 2MASS flux limits, the stellar sample analyzed by \citet{Covey07} does 
not include faint blue stars \citep[those with $r\ga16$ for $g-r<0.6$; see Fig.~4 in][]{Finlator00}. 
Consequently,  the \citet{Covey07} locus corresponds to predominantly metal-rich main sequence 
stars ($[Fe/H]>-1$) because low-metallicity halo stars detected by SDSS are predominantly
faint and blue (see Fig.~3 in I08). According to
Galfast model (J08), stars detected in SEGUE stripes are dominated by metal-rich main
sequence stars, although we note that the fraction of red giant stars in SEGUE is expected
to be much larger than observed by SDSS at high Galactic latitudes ($\sim20\%$
vs. $\sim5\%$). 

The adopted model library cannot provide a good fit for SEDs of unresolved pairs of 
white and red dwarfs \citep{Smolcic2004}, hot white dwarfs \citep{Eisen06},
and quasars \citep{Richards2001}, whose SEDs can differ from the adopted library by
many tenths of a magnitude. Systematic photometric discrepancies at the level of a few 
hundredths of a magnitude are also expected for K and M giants, especially in the $u$ 
band \citep{Helmi2003}. Similar $u$ band discrepancies are expected for metal-poor 
main sequence stars (I08). Nevertheless, all these populations together never contribute
more than $\sim20\%$ of the full sample \citep{Finlator00, Juric08}, and in most cases 
can be recognized by their large values of $\chi_{min}^2$. At least in principle, 
additional libraries appropriate for those other populations can be used a posteriori to 
fit the observed SEDs of sources that have large $\chi_{min}^2$ when using SEDs of 
main sequence stars. This additional analysis has not been attempted here, though
our results represent the first necessary step: finding sources with large $\chi_{min}^2$.

\subsection{Parametrization of Dust Properties \label{sec:Aparam}}

To implement the fitting method described in \S\ref{sec:model}, the
shape of the extinction curve ($C_\lambda$, see eq.~\ref{eq:Clambda}) must be 
characterized. $C_\lambda$ in the SDSS bands
was initially computed (prior to the beginning of the survey, to enable
spectroscopic targeting) using the standard parametrization of the extinction 
curve \citep{CCM, ODonnell} with $R_V=3.1$.
The resulting values ($C_\lambda$=1.87, 1.38, 0.76, 0.54, with $\lambda=u,g,i,z$) are 
commonly adopted to compute the extinction in the SDSS bands, together with $A_r$
given by the SFD map via $A_r = 2.75 E(B-V)$.

A preliminary analysis of the position of the stellar locus in the SDSS-2MASS color space 
suggested that the above $C_\lambda$ values need to be slightly adjusted \citep{Meyer2005}. 
Further support for this conclusion was recently presented by Sch2010. 
Here we revisit the Meyer et al. analysis using an improved SDSS photometric catalog 
from the so-called stripe 82 region\footnote{Available from 
http://www.astro.washington.edu/users/ivezic/sdss/catalogs/stripe82.html
} \citet{Ivezic07}. SDSS photometry in this catalog is about twice as accurate 
as typical SDSS photometry due to averaging of many observations and various 
corrections for systematic errors. The SDSS-2MASS subset of that catalog includes 102,794 sources 
unresolved by SDSS (out of about a million in the full sample), which also have a 
2MASS source with $K<14.3$ within 1.5$''$. 
The results of our analysis provide an updated set of $C_\lambda$ coefficients, which 
are then used to select a dust extinction model for generating the required
$C_\lambda(R_V)$ dependence. Similarly to a recent analysis by Sch2010,
we find that the \citet{ODonnell} model can be rejected, and adopt the CCM dust 
extinction law \citep{CCM}.

\subsubsection{ Determination of the locus shifts} 

The interstellar extinction reddens the stellar colors and shifts the 
position of the {\it whole} stellar locus at high Galactic latitudes, where 
practically all stars are located behind the dust screen. At high
Galactic latitudes,  distances to an overwhelming majority of stars are 
larger ($\ga 100$ pc)  than the characteristic scale height of the interstellar 
dust layer ($\sim$70 pc, J08). Both the amount of reddening and its wavelength 
dependence can be determined by measuring the locus position and 
comparing it to the locus position corresponding to a dust-free case. 
The latter can be determined in regions with very small extinction ($A_r \sim 0.05$)
where errors in the SFD extinction map as large as 20\% would still be negligible. 

\begin{figure}[!t]
\epsscale{0.9}    
\plotone{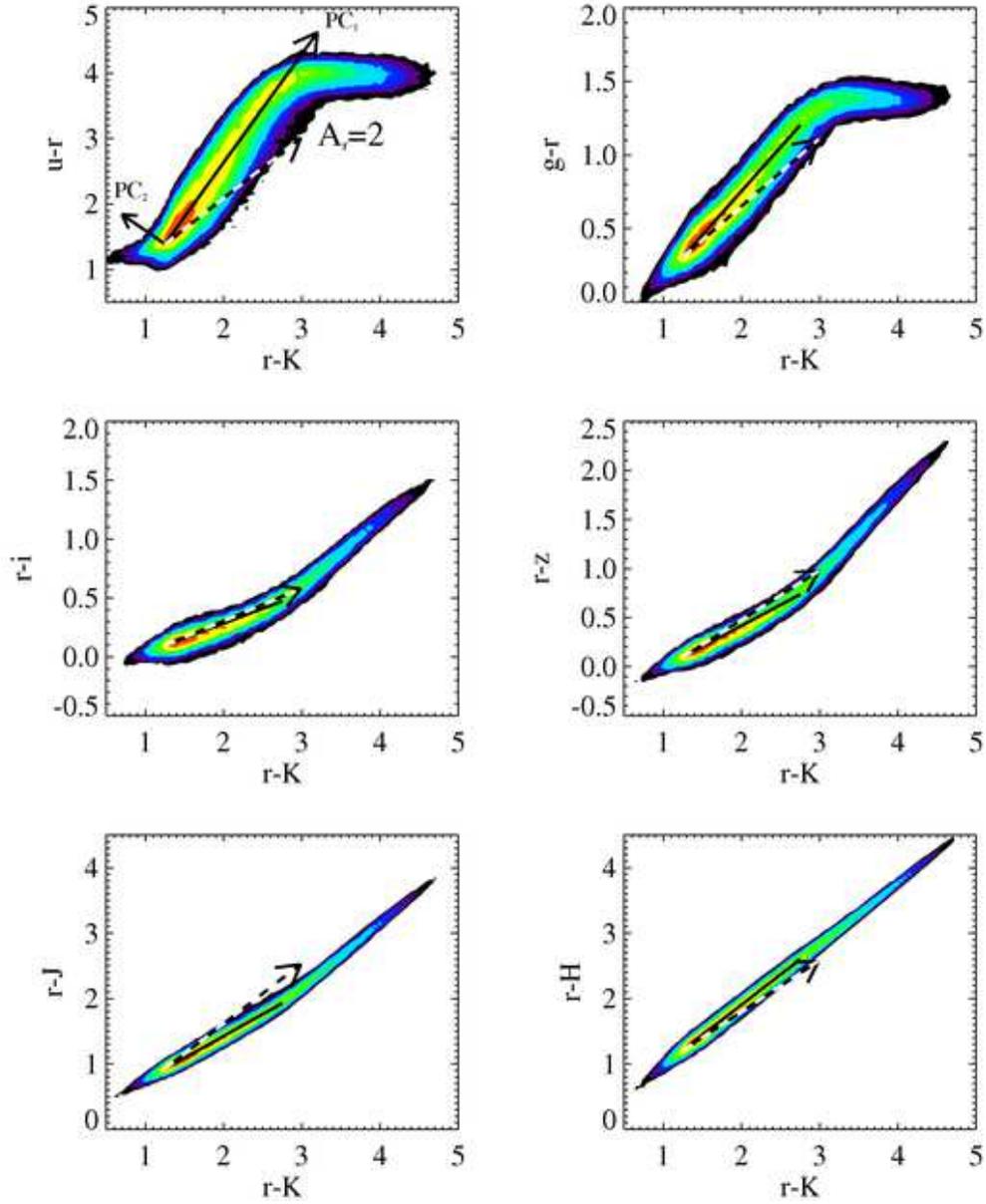}
\epsscale{1.0}
\vskip 0.2in    
\caption{The distribution of unresolved SDSS sources with 2MASS detections
in the $\lambda-r$ vs. $r-K$ color-color diagrams, with $\lambda=u, g, i, z, J$ 
and $H$. The source density is shown as color-coded maps, and it increases from 
black to green to red. The two arrows marked PC$_1$ and PC$_2$ in the top left 
panel illustrate the ``principal color'' axes discussed in text and used to 
track the locus shifts due to interstellar dust reddening. The dashed vector 
in each panel shows the reddening vector for $A_r=2$ and standard $R_V=3.1$ 
dust \citep{CCM}.}
\label{Fig:PCex}
\end{figure}

We measure the locus position in the seven-dimensional SDSS-2MASS
color space using an extended version of the ``principal color'' method 
developed by \citet{Ivezic04} to track the quality of SDSS photometric calibration. 
We utilize six independent two-dimensional projections spanned 
by the $r-K$ and $\lambda-r$ colors, where $\lambda=u, g, i, z, J$ and $H$ (see Fig.~\ref{Fig:PCex}). 
Since the extinction in the 2MASS $K$ band is small and fairly model and $R_V$-independent 
($A_K/A_r$ = 0.133 for $R_V=3.1$, with only a $\sim$10\% variation over the range of 
plausible $R_V$ and dust models, as discussed further below), the locus shifts in the $r-K$ direction 
provide robust constraints for $A_r$. For example, a 10\% uncertainty in the $A_K/A_r$ ratio
results in only 1.5\% uncertainty in $A_r$ determined from a given $A_r-A_K$ value.
We determine these shifts iteratively, starting with $A_r$ given by the 
SFD map, and adjusting $A_r$ until the observed and corrected $r-K$ color distributions agree 
in a maximum likelihood sense.  
This determination of $A_r$ is very similar to the ``blue tip'' method introduced
in Sch2010. The two main differences are due to the addition of 2MASS
data. First, the low-metallicity faint blue stars are not included in the sample analyzed
here. Such stars could systematically influence the locus morphology and reddening 
estimates based on the ``blue tip'' method; nevertheless, our results are in good agreement with the 
Schlafly et al. results, as discussed below. Second, the availability of the $K$ magnitudes enables a 
robust and straightforward determination of $A_r$, {\it without any consideration of the 
SFD map}.  For a detailed discussion of these issues, see Appendix C. 

After $A_r$ is estimated from the $r-K$ color offsets, the locus offsets in the $\lambda-r$ directions 
then provide constraints for the extinction wavelength dependence, $C_\lambda$. We measure 
these offsets using principal colors, $P_1$ and $P_2$, with $P_1$ parallel to the blue part of 
the stellar locus, and $P_2$ perpendicular to it (see the top left panel in Fig.~\ref{Fig:PCex} 
for illustration of the principal axes, and for a comparison of the locus orientation with the 
direction of the standard reddening vectors).  
The blue part of the stellar locus at the probed faint magnitudes ($14<r<17$) 
includes mostly thick disk stars with distances of the order 1 kpc or larger, 
which are thus beyond all the dust. 

We measure the position of the blue part of the locus in 
each $\lambda-r$ vs. $r-K$ diagram using stars with $1.5 < r-K < 2.5$ (approximately; the range
is enforced using the $P_1(\lambda)$ color). The blue part of the locus is parametrized as
\begin{equation} 
       P_1(\lambda) = \cos(\theta_\lambda)\, (r-K) + \sin(\theta_\lambda)\,(\lambda-r) + c_1(\lambda)
\end{equation} 
and
\begin{equation} 
       P_2(\lambda) = -\sin(\theta_\lambda)\, (r-K) + \cos(\theta_\lambda)\,(\lambda-r) + c_2(\lambda).
\end{equation} 
The best-fit angle $\theta_\lambda$ found using stripe 82 dataset is equal to 
(61.85$^{\circ}$, 33.07$^{\circ}$, 14.57$^{\circ}$, 23.47$^{\circ}$, 34.04$^{\circ}$, 43.35$^{\circ}$). for 
$\lambda=(u,g,i,z,J,H)$. The values of $c_1$ and $c_2$ are completely arbitrary; we set $c_{\lambda}=0$, and 
determine $c_2(\lambda)$ by requiring that the median value of $P_2(\lambda)$ color is 0
($c_2$=0.463, 0.434, 0.236, 0.424, $-$0.048, $-$0.019, for $u,g,i,z,J,H$, respectively). 
Given the locus shift $\Delta P_2(\lambda)$, and $A_r$ determined from the $r-K$ color offset 
(or alternatively from the $\Delta P_1$ offsets), the corresponding $A_\lambda$ can be 
determined from 
\begin{equation} 
 C_\lambda \equiv {A_\lambda \over A_r} = 1 + \tan(\theta_\lambda)(1-{A_K\over A_r})+
                {1\over \cos(\theta_\lambda)}\, {\Delta P_2(\lambda) \over A_r}.
\end{equation} 
Assuming a constant $A_K/A_r$ ratio, it is straightforward to compute the error of this estimate.

The locus position must be measured over a sky area where the amount of dust
and dust properties can be assumed constant. The smaller the area, the more
robust is this assumption. However, the chosen area cannot be arbitrarily small
because the error in the locus position, and thus the $C_\lambda$ error, is inversely proportional to the 
square root of the star counts. Within the analyzed stripe 82 region, the counts of 
SDSS-2MASS stars in the blue part of the stellar locus never drop below 70 stars deg$^{-2}$. 
We bin the data using $4^{\circ}$ wide bins of R.A. (with $|Dec|<$1.27 deg., an area of 
$\sim$10 deg$^2$ per bin), which guarantees that random errors in $A_\lambda$ never 
exceed $\sim$2\% (even for the $u$ band, and a factor of few smaller in other 
bands). In addition, we consider four larger regions: the high-latitude northern 
sky with $b>45^\circ$, split into $l<180^\circ$ and $l>180^\circ$ subregions, 
a northern strip defined by $30^\circ < b < 45^\circ$, and a southern strip
defined by $-45^\circ < b < -30^\circ$ (for these regions, we use SDSS DR7 photometry).

%
%

\subsubsection{Interpretation of the locus shifts and adopted dust extinction model} 
\label{sec:Clambda} 

We find that the variations in the shape of the extinction curve 
across the 28 R.A. bins from Stripe 82 region are consistent within
measurement errors. The values of $C_\lambda$ obtained for the whole 
Stripe 82 region are listed in the first row in Table~\ref{Tab:ClambdaConstraints}. 
Practically identical coefficients are obtained for the southern 
strip defined by $-45^\circ < b < -30^\circ$. The extinction curve values 
for the northern sky are consistent with the southern sky, and {\it we recommend 
the entries listed in the first row in Table~\ref{Tab:ClambdaConstraints}
for correcting SDSS and 2MASS photometry for interstellar dust extinction.}
One of the largest 
discrepancies is detected in a region from the northern strip defined  
by $30^\circ < b < 45^\circ$ and $0^\circ < l < 10^\circ$; and these
values are listed in the second row in Table~\ref{Tab:ClambdaConstraints}. 
Nevertheless, the north vs. south differences are not large, and, using
models described below, correspond to an $R_V$ variation of about 0.1. 

Much larger north vs. south differences are detected when comparing 
the best-fit $A_r$ values to the SFD map values. The 
accuracy of the $A_r$ determined here is about 3-10\%, depending on
the amount of dust. We find that the SFD $A_r$ values are consistently larger
by about 20\% than our values determined across the southern 
hemisphere. Interestingly, no such discrepancy is detected across
the northern sky, to within measurement errors of $\sim$5\%. In several
isolated regions, the discrepancies are much larger. For example, 
in a region defined by $-45^\circ < b < -30^\circ$ and $157^\circ < l < 160^\circ$,
the SFD values appear overestimated by 50\% (the median value of 
$A_r$ in that region given by the SFD map is 1.3). These results are similar
to those presented in Sch2010, where the spatial variation of errors in
the SFD map and their possible causes are discussed in more detail. 
The conclusion that the SFD $A_r$ values are consistently overestimated
in the southern hemisphere is also consistent with the results based on galaxy
count analysis by \cite{Yasuda2007}, which is essentially an independent 
method.  

\begin{figure}[!t]
\epsscale{0.9}    
\plotone{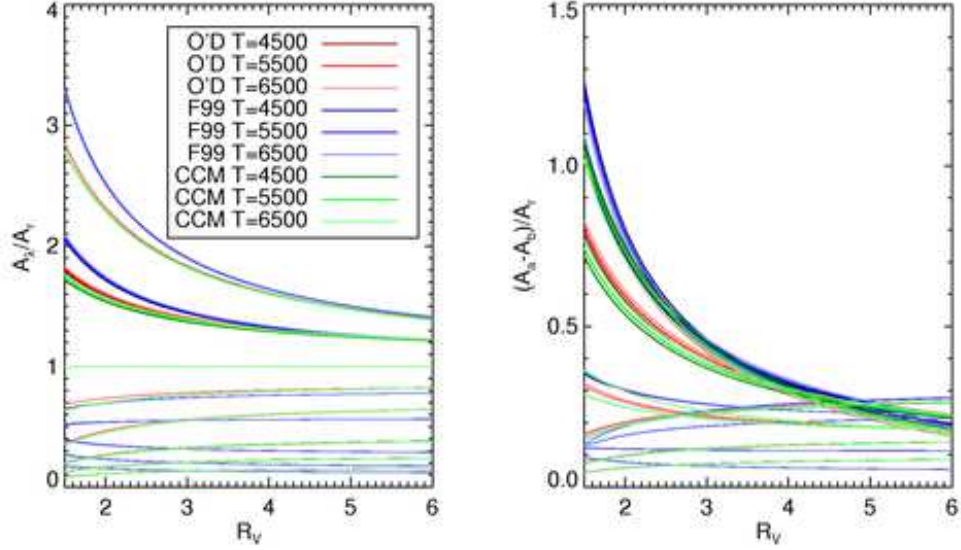}
\epsscale{1.0}
\vskip 0.2in    
\caption{Model predictions for the extinction curve shape as a function 
of $R_V$ for three different models: O'D \citep{ODonnell}, F99 \citep{Fitz99},
and CCM \citep{CCM}, evaluated for stars with three different effective 
temperatures (as listed in the legend, in Kelvin). The left panel shows 
$C_\lambda=A_\lambda/A_r$ for $\lambda=(u,g,r,i,z,J,H,K)$ (top to bottom, 
respectively); the right panel is analogous, except that the ratios
based on colors ($u-g$, $g-r$, $r-i$, $i-z$, $z-J$, $J-H$, and $H-K$) are 
shown. As expected, most of the sensitivity to $R_V$ comes from the blue 
bands ($u$ and $g$).}
\label{Fig:Ed1}
\end{figure}

We adopt the $C_\lambda$ values determined for the Stripe 82 region
(the first row in Table~\ref{Tab:ClambdaConstraints}) to select 
a dust extinction law used in subsequent fitting of SEGUE data. 
Using the same assumptions and code as Sch2010, we compute dust extinction
curves for three popular models, and for three different input stellar
spectral energy distributions. As can be seen in Figure~\ref{Fig:Ed1},
the differences between the models are much larger than the impact 
of different underlying spectra. 

A comparison of the observational constraints and model predictions 
is summarized in Figure~\ref{Fig:Ed2}. Following Sch2010, we use 
ratios of the reddening values for this comparison. The differences in 
the extinction curve shape between the southern and northern sky
determined here are similar to their differences from the Sch2010 
results, and are consistent with estimated measurement uncertainties.
The \citet{ODonnell} model predicts unacceptable values of 
the $(A_r-A_i)/(A_i-A_z)$ ratio for all values of $R_V$. The other
two models are in fair agreement with the data. Due to a slight 
offset of the Sch2010 measurements, they argued that the CCM model
\citep{CCM} is also unsatisfactory, although the discrepancy was
not as large as in the case of the \citet{ODonnell} reddening law. 

Although none of the models shows a perfect agreement with the
data, the discrepancies are not large. To further illustrate the
constraints from different bands, we determine the best-fit 
$R_V$ and its uncertainty in each band using the CCM model.
If a model is acceptable, the constraints from different bands
must be statistically consistent. As shown in Figure~\ref{Fig:RvConstraints},
this is indeed the case, and we obtain the best-fit $R_V=3.01\pm0.05$.
The systematic error of this estimate, implied by the variation
of the extinction curve shape across the analyzed regions, is
about 0.1. The corresponding figure for the F99 \citep{Fitz99} 
reddening law is similar, with the best-fit $R_V=3.30\pm0.1$,
while for the  \citet{ODonnell} model, $R_V=3.05\pm0.05$. However,
for the latter, the predicted extinction in the $i$ band is 
inconsistent with the rest of the bands at about 2$\sigma$ level
(see Figure~\ref{Fig:RvOD}). This inconsistency is the main reason
for rejecting the O'Donnell model both here and by Sch2010. 
A comparison between the CCM and O'Donnell laws is further illustrated 
in Figure~\ref{Fig:ODCCMdiff}; it appears that polynomial fitting adopted
by both CCM and O'Donnell (to the 7$^{th}$ and 8$^{th}$ order, respectively) 
has caused wiggles whose integral over SDSS bandpasses is the largest in 
the $i$ band. The predicted values of the extinction curve for all three models, 
using their individual best-fit values for $R_V$, 
are listed in Table~\ref{Tab:ClambdaConstraints}. 

For the rest of our analysis, we generate $C_\lambda(R_V)$
values using the CCM law and an F star spectral energy 
distribution (7000 K). The adopted curves are shown in 
Figure~\ref{Fig:AlambdaRv}, and a few representative values
are listed in Table~\ref{Tab:ClambdaFreeRv}. For comparison, 
we also list $C_\lambda$ values suggested by Sch2010, and 
the values computed using extinction curve parametrization 
proposed by \citet{FM09}.

\subsection{ Illustration of the Method and Fitting Degeneracies  }
\label{sec:methodSummary}

To summarize, we make two basic assumptions when analyzing observed SEDs of
low-latitude stars (SEGUE stripes). First, we assume that the median
stellar locus in SDSS and 2MASS  bandpasses, as quantified by \citet{Covey07} 
at high Galactic latitudes,  is a good description of stellar colors at all Galactic latitudes. 
Second, we assume that the normalized dust extinction curve, 
$A_\lambda/A_r$, can be described as a function of single
parameter, $R_V=A_V/E(B-V)$. Therefore, for a given set of measured 
colors, four in SDSS-only case, and seven in SDSS-2MASS
case, we fit three free parameters: stellar model (position along the
one-dimensional locus), $t$, dust amount, $A_r$, and $R_V$.  

When the number of measured colors is small, when the color errors are large,  
or when the sampled wavelength range is not sufficiently wide, the best-fit 
solutions can be degenerate.  The main reason for this degeneracy is the similarity 
of the stellar locus orientation and the direction of the dust reddening vector (see Figure \ref{Fig:PCex}).
This degeneracy is especially strong for stars in the blue part of the locus
($g-i<1$) and remains even when SDSS photometry is augmented by 2MASS 
photometry (a photometric band at a wavelength much shorter than the 
SDSS $u$ band is needed to break this degeneracy). 

Figure~\ref{Fig:riVSgr} illustrates an example of degenerate solutions in the $r-i$ vs. $g-r$ 
color-color diagram, and how degeneracies are partially broken when the $i-z$ color is 
added to the data. Because the direction of the reddening vector in the $i-z$ vs. $r-i$ 
color-color diagram is essentially independent of $R_V$, the measured $r-i$ and $i-z$ 
colors provide robust constraints for $t$ and $A_r$, irrespective of $R_V$. The addition
of the measured $g-r$ color to $r-i$ and $i-z$ colors then constrains $R_V$. 

Since the stellar locus in the $i-z$ and $r-i$ color-color diagram and the reddening vector 
are not perpendicular, the covariance between the best-fit $t$ and $A_r$ values does not 
vanish. The addition of other bands, e.g. 2MASS bands to SDSS bands, alleviates this 
covariance, but not completely (and only slightly for blue stars). We quantify 
this effect using simulated observations, as described below.

\subsection{Tests of the Method \label{sec:methodtests}}

To test the implementation of $\chi^2$ minimization algorithm,
and to study the dependence of best-fit parameter uncertainties on 
photometric errors, the amount of extinction, and the intrinsic stellar
color, we first perform relatively simple Monte Carlo simulations and 
analyze the resulting mock catalog based on realistic stellar and dust distributions,
and photometric error behavior. 

\subsubsection{The Impact of Photometric Errors} 

In the first test, we study the variation of best-fit parameters
with photometric errors, where the latter are generated using 
Gaussian distribution and four different widths: 0.01, 0.02,
0.04, and 0.08 mag. The dust extinction curve shape is fixed to $R_V=3.1$,
and we only use SDSS photometry. 
The noiseless ``observed'' magnitudes for a 
fiducial star with intrinsic color $g-i=1.95$ (roughly at the
``knee'' of the stellar locus in the $r-i$ vs. $g-r$ color-color
diagram) and $A_r=1.5$, 
are convolved with photometric noise generated independently
for each band, and the resulting ``observed'' colors are used in fitting.
The errors in best-fit models and $A_r$ are illustrated in 
Figures~\ref{Fig:Dgi} and \ref{Fig:Dar}. 

The median errors in the best-fit stellar SED, parametrized by the $g-i$
color, are about twice as large as the assumed photometric 
errors. When photometric errors exceed about 0.05 mag, the 
best-fit $A_r$ distribution becomes {\it bimodal}, with the additional mode 
corresponding to a solution with a bluer star behind more dust. Therefore, 
even the addition of the red $z$ passband is insuficient to break
the stellar color--reddening degeneracy when the photometry
is inaccurate (this conclusion remains true even when 2MASS bands are
added). {\it Our fitting results should thus be trusted only for stars sufficiently 
bright to have photometric errors smaller than about 0.05 mag in most 
bands.}  With this constraint, the formal best-fit errors are typically within 20\%
of the true errors.

\subsubsection{The Reddening vs. Intrinsic Stellar Color Degeneracy} 

In the second test, we have investigated the covariance between
the best-fit model and $A_r$ values. Here again the dust extinction 
curve shape is fixed to $R_V=3.1$. Figure~\ref{Fig:DgiChi2}
shows the $\chi^2$ surfaces for a blue and a red star, and for 
two values of $A_r$, when only SDSS bands are used in fitting
and Gaussian noise with $\sigma=0.02$ mag is assumed for all bands. The 
best-fit model-$A_r$ covariance is larger for the bluer star, 
in agreement with the behavior illustrated in Figure~\ref{Fig:riVSgr} 
(the angle between the reddening vector and the stellar locus 
is smaller for the blue part of the locus, than for the red part). 
The $A_r$ vs. $g-i$ covariance does not strongly depend on assumed $A_r$.
When the 2MASS bands are added, the morphology of the $\chi^2$ surface 
is essentially unchanged (recall that $R_V$ was fixed in these tests). 

These tests show that our implementation of the $\chi^2$ minimization algorithm
produces statistically correct results, and that the accuracy of SDSS and 2MASS
photometry is sufficient (for most sources) to break degeneracy between the dust 
reddening and intrinsic stellar color in case of a fixed dust extinction curve ($R_V=3.1$). 
Nevertheless, {\it the best-fit results should be interpreted with caution when 
photometric errors exceed 0.05 mag, especially for intrinsically blue stars}.

\subsubsection{Tests Based on a Realistic {\it Galfast} Mock Catalog } 

To quantify the expected fidelity of our best-fit parameters,
including $R_V$,  for a realistic distribution of stellar colors, photometric errors 
and dust extinction, we employ a mock catalog produced by the {\it Galfast}
code (Juri\'{c} et al., in prep). {\it Galfast} is based on the Galactic structure model
from J08 and includes thin-disk, thick-disk, and halo components.
The stellar populations considered here include main sequence and post-main 
sequence subgiant and giant stars. All other populations, such as blue horizontal 
branch stars, brown dwarfs, white dwarfs and quasars, are expected to contribute
only a few percent of the total source count at low Galactic latitudes relevant
here. SDSS and 2MASS photometry is generated using the \citet{Covey07} SED library
(using the $g-i$ color provided by {\it Galfast}). The photometric errors are modeled 
using parametrization given by eq.~5 in \citet{IvezicLSST}, and the best-fit values
for 5$\sigma$ limiting depth derived using cataloged errors for SDSS and 2MASS data
(for SDSS $ugriz$ bands: 21.5, 23.0, 22.8, 22.6 and 20.5, respectively; for 2MASS 
$JHK$ bands: 17.0, 16.0 and 15.5, on Vega scale). The dust extinction along the
line-of-sight to each star is assigned using the three-dimensional dust distribution 
model of \citet{AL2005}. The shape of the dust extinction curve is fixed to the CCM 
model values for $R_V=3.1$. The normalization of the extinction for a given line
of sight is determined by requiring a match to the SFD map at a fiducial distance of 
100 kpc (that is, a complex dust distribution is retained in two out of three 
coordinates).  

The intrinsic absolute magnitude and color distribution of stars in the simulated 
low latitude ($|b|<5^\circ$) sample is very different from distributions seen with
high latitude samples.  The two main differences are much bluer {\it intrinsic} color 
distribution, and a much larger fraction of red giants in the low-latitude dataset. The 
origin of these differences is illustrated in the top two panels in Figure~\ref{Fig:mock2}. 
As shown in the top left panel, the simulated sample is dominated by stars with intrinsic 
$g-i<1.2$, and includes a large fraction of red giants (40\% with $M_r < 2$). These giants 
pass the $r>14$ selection cut due to large dust extinction ($A_r\sim3$ mag for giants 
in the simulated sample). At high Galactic latitudes, most red giants are brighter
than SDSS saturation limit $r\sim14$). 

The distributions of modeled stars in the color-magnitude and color-color magnitude 
diagrams closely match SDSS and 2MASS data (for an illustration see Figure~\ref{Fig:mock}).  
The much redder observed colors of stars in SEGUE stripes, compared to high-latitude
sky, are reproduced with high fidelity. For example, the median $g-i$ color for the
SEGUE $l\sim110^\circ$ stripe moves from 1.0 at $r\sim16$ to 1.7 at $r\sim21$;
only 2\% of stars with $r\sim21$ have $g-i<1$. For comparison, at high Galactic 
latitudes, the median $g-i$ color also becomes redder for fainter stars,
but reaches a value of 1 at $r\sim19.5$, or over 3 mags fainter than at low
Galactic latitudes. 
Although the two sets of diagrams are encouragingly similar, there a few 
detailed differences: the observed diagrams have more outliers, and a few diagrams
(e.g., $J-K$ vs. $i-z$ and $i-z$ vs. $r-i$) imply different reddening vectors than used
in simulations ($R_V=3.1$). We discuss these differences in more detail in the next section. 

The resulting mock catalog is processed in exactly the same way as catalogs with observations. 
Note that the simulated photometry is generated with the same SED model library and dust extinction
curve as used in fitting. We analyze four different fitting methods: we use both only-SDSS (four
colors) and SDSS-2MASS (seven colors) photometric data, and we consider both $R_V=3.1$ (the true value) 
and $R_V$ as a free fitted parameter. Only stars with $r<20$ and $K<13.9$ (Vega) are used in analysis; 
this cutoff results in the median photometric errors of 0.02 mag in the $r$ band and 0.04 mag in the $K$ 
band (and 0.06 in the $u$ band, which is the only band where errors exceed the $K$ band errors).  
There are about 94,000 simulated stars that satisfy these criteria (the simulated area covers 25 deg$^2$). 
We first analyze the fitting results when $R_V$ is fixed to its true value, and then extend our analysis to 
fitting results when $R_V$ is a free parameter. 

When $R_V$ is fixed, the obtained $\chi^2_{pdf}$ distributions closely resemble expected
distributions for 2 and 5 degrees of freedom, with slightly more objects in the tails. 
For example, 86\% and 93\% of the sample are expected to have $\chi^2_{pdf}<2$
for only-SDSS and SDSS-2MASS cases, while we obtained 73\% and 80\%. The latter
fractions remain the same when the $r$ band and $K$ band limits are relaxed by 
1 mag. For further analysis, we only use stars with $\chi^2_{pdf}<2$. 

The bottom left panel in Figure~\ref{Fig:mock2} shows the distribution of simulated stars in 
the intrinsic apparent magnitude vs. color space, where we use only-SDSS best-fit intrinsic 
$g-i$ color and correct ``observed'' $r$ band magnitudes using the best-fit $A_r$. Its overall 
similarity with the top left panel is encouraging. The main difference is at the blue edge,
$g-i<0.3$,  with about 20\% of stars having best-fit $g-i$ color biased blue (simulated
sample essentially does not include stars with $g-i<0.3$ because this is turn-off color for
thick disk stars which contributes stars in that magnitude-color range). The same stars
also have overestimated $A_r$. These biases are the result of the reddening-color degeneracy 
and could be mitigated by adopting a strong prior such as removing SEDs with $g-i<0.3$ from 
the SED model library. 

The bottom right panel compares the best-fit $A_r$ to the input value. The best-fit $A_r$ is 
systematically larger than the input values by about 10\%. This overestimate is due to 
color-reddening degeneracy discussed above: when $A_r$ is overestimated, the best-fit stellar 
color is biased blue. When the full SDSS-2MASS dataset is used, the outliers seen in the bottom 
right panel in  Figure~\ref{Fig:mock2} disappear, and the $A_r$ bias is smaller by a few percent
(10\% vs. 15\% for blue stars). Overall, there is no dramatic improvement resulting from the addition 
of 2MASS photometry (the root-mean-square scatter for the $A_r$ difference decreases from 0.42 mag 
to 0.33 mag when 2MASS photometry is added). 

We find that the best-fit values based on only SDSS data are biased when the $u$ band 
errors are large: $A_r$ by 0.27 (true values are smaller) and $g-i$ by 0.2 mag (bluer) for 
stars with $u$ band errors of $\sim$0.1 mag. When the SDSS-2MASS dataset is used, both 
bias values fall to about 2/3 of only-SDSS values. Therefore, accurate $u$ band photometry 
is  {\it crucial} for obtaining accurate best-fit results. In order to minimize the effects of this 
bias, we further limit the sample to stars with $u$ band errors below 0.05 mag. 
Unfortunately, only 40\% of stars satisfy this cut. 

The true errors in both stellar color and $A_r$ (as determined by comparing the best-fit
and true values) are about twice as large as marginalized errors computed using 
eqs.~\ref{eq:sig1} and \ref{eq:sig2}, both in case of only-SDSS and SDSS-2MASS
fits. This increased scatter is probably due to color-$A_r$ degeneracies and deviations of the 
maximum likelihood contours from a two-dimensional ellipse approximation: the errors in $g-i$ 
color and $A_r$ errors are strongly correlated with a slope of $\delta(g-i)/\delta(A_r) \sim -0.65$ 
(when this correlation is used to ``correct'' the best-fit color by sliding them along this relation, 
the residuals are consistent with photometric errors; in other words, the entire ``additional'' 
color scatter is along this relation).  The root-mean-square (rms) scatter for 
$A_r$ errors is 0.42 mag and 0.33 mag for only-SDSS and SDSS-2MASS fits (20\% and 16\% for relative
errors, i.e., errors normalized by true $A_r$), and the rms for $g-i$ color errors are 0.29 mag
and 0.23 mag, respectively.  We note that these errors are valid for individual stars, which
suffer from the color-reddening degeneracy. When the results are averaged in small pixels on
the sky, the scatter is significantly smaller (because the spread of stars along the color-reddening
degeneracy manifold is fairly symmetric).  For example, the rms error for $A_r$ in 0.2$\times$0.2 deg$^2$
pixels decreases by a factor 3-4, to a level of about 5-10\% (depending on the line of sight 
direction and the median $A_r$).

\subsubsection{``Free-$R_V$'' Case} 
\label{sec:freeRvtests}

The analysis of fits with $R_V$ treated as a free parameter revealed that SDSS data alone are 
insufficient to reliably constrain $R_V$, while SDSS-2MASS dataset produced 
good results. Figure~\ref{Fig:GalfastRv} compares the two resulting distributions of best-fit 
$R_V$ (the input value is $R_V=3.1$).  When SDSS-2MASS photometry is used, $R_V$ can be 
determined with a bias of $<0.1$, and a precision (rms) of 0.10 when all stars from the 
simulated sample with $\chi^2_{pdf}<2$ are considered. The $R_V$ error is not correlated 
with stellar color, nor with distance; $A_r$ is the only parameter that controls the $R_V$ error. 
As expected, the $R_V$ error increases for small $A_r$. A good practical limit is $A_r>1$,
which guarantees bias below 0.1 and an rms of at most 0.3. The $R_V$ error decreases with 
$A_r$, and drops to 0.15 at $A_r=2$ and below 0.1 for $A_r>4$. For $A_r<1$, the precision
of $R_V$ estimate significantly deteriorates; for $0.5<A_r<0.7$, the median best-fit $R_V$ 
becomes biased to 3.2, with an rms of 0.5. 

Unsurprisingly, the $R_V$ error is much larger when using only SDSS photometry; when
considering all stars with $\chi^2_{pdf}<2$, the best-fit $R_V$ is biased to 3.3, with an
rms of 1.2, rendering it practically useless. The main reason for this poor performance 
are the facts that three free parameters are constrained using only four colors, and
that these three parameters are strongly degenerate. The SDSS-2MASS dataset shows
superior performance when $R_V$ is a free parameter not because 2MASS data can 
constrain $R_V$ (using $C_\lambda(R_V)$ parametrization employed here), but because
2MASS data better determine $A_r$ and intrinsic stellar color, which gives
more leverage to SDSS data (mostly the $u$ and $g$ band) to constrain $R_V$. 

As a result of this test, we conclude that {\it only $R_V$ estimates based on SDSS-2MASS 
dataset should be used,} and those only for stars with $\chi^2_{pdf}<2$ and $A_r>1$.

\subsubsection{``Dusty'' Parallax Relation} 
\label{sec:redgiantsmock}
The analysis of the mock {\it Galfast} sample uncovered an interesting possibility for 
identifying candidate red giant stars in SEGUE stripes. Distinguishing red giant stars using 
only SDSS colors is difficult even at high Galactic latitudes (offsets from the main
sequence stellar locus are at most 0.02-0.03 mag; for more details see \citealt{Helmi2003}), 
and seems futile at low Galactic latitudes. However, the best-fit $A_r$ contains information
about distance to a star, and this fact can be used for dwarf vs. giant star separation.

After obtaining the best-fit intrinsic $g-i$ color,
we compute distance to each star using a photometric parallax relation appropriate for main 
sequence stars (I08). For red giants, the resulting distances are grossly understimated 
(for example, a red giant star with $g-i=1$ has $M_r \sim 0$, while main sequence stars
with the same color have $M_r \sim 6$, resulting in a distance ratio of $\sim15$ for the 
same apparent magnitude). However, because red giant stars are much more distant 
away than main sequence stars of the same color, their best-fit values of $A_r$ are also
on average significantly different. The latter difference is a consequence of the fact that $A_r$ is 
proportional to the dust column along the line of sight, which in turn is roughly proportional to distance
(although not exactly because the dust number density varies with position). 

These differences in the best-fit $A_r$ vs. main-sequence distance behavior between main sequence and 
red giant stars are illustrated in the top two panels in Figure~\ref{Fig:DPR}. The dashed lines mark
the region in the $A_r$ vs. distance diagram dominated by simulated stars with $M_r<3$
(as illustrated in the bottom left panel).  Red giant stars are found in the upper left corner 
of this diagram because their (main sequence) distances are too small given their $A_r$:  
it takes about 1 kpc of dust column to produce $A_r\sim1$ mag and thus stars with $A_r>1$ 
should be further than $\sim$1 kpc. 

This separation of red giant and main sequence stars in the $A_r$ vs. distance diagram can 
be elegantly summarized via a relation that we dub ``dusty parallax''. First, using the median 
best-fit $A_r$ in narrow distance bins for stars with best-fit main sequence distances $D<0.5$ kpc 
(see the blob discernible in the lower left corner), we obtained a linear relationship
\begin{equation}
\label{eq:AvsD}  
                        A_r = 1.06 \, {D \over {\rm kpc}}. 
\end{equation} 
The best-fit coefficient of 1.06 mag kpc$^{-1}$ is in good agreement with the coefficient
corresponding to true $A_r$ and distance for stars with $M_r>5$, 1.13 mag kpc$^{-1}$
(and implies that a similar algorithm can be applied to real data). 
This relation can be employed to estimate distance from the best-fit $A_r$ for
all stars, and in turn absolute magnitude $M_r$  via ``dusty parallax'' relation
\begin{equation}
\label{eq:DPR}  
                     M_r^{DPR} = r -5\,\log_{10}(0.94\,A_r) - A_r - 10. 
\end{equation}   
A comparison of true $M_r$ and $M_r^{PDR}$ is shown in the bottom right panel in Figure~\ref{Fig:DPR}. 
The root-mean-square scatter for the ($M_r-M_r^{PDR}$) difference is 1.2 mag. 

The coefficient from eq.~\ref{eq:AvsD} reflects the spatial distribution of dust generated using a 
smooth model from \citet{AL2005}. In reality, localized clumps of dust will result in larger estimated
distances and thus some main sequence stars will be misinterpreted as candidate red giants. 
Nevertheless, the precision of this relation seems sufficient to broadly separate red giant and main 
sequence stars using their best-fit $g-i$ color and $A_r$. 

In many ways, this ``dusty'' parallax relation is similar to the reduced proper motion (RPM) method
(for a detailed discussion, see Appendix B in \citealt{SIJ08}); the main difference is that RPM estimates 
distance using its relationship with proper motion (assuming a fixed true tangential velocity, distance
is inversely proportional to proper motion), while DPR estimates distance using a relationship between 
dust extinction and distance. We return to this relation and the selection of red giants when analyzing 
real data samples in the next section.

To summarize this testing section, the analysis of simulated datasets has revealed important 
limitations of the best-fit results, mostly stemming from the finite photometric precision of 
SDSS and 2MASS surveys. Most notably, the SDSS dataset alone does not have enough power 
to reliably constrain $R_V$, and {\it only $R_V$ estimates based on SDSS-2MASS dataset should 
be used,} and those only for stars with $\chi^2_{pdf}<2$ and $A_r>1$. 
The tests based on a mock {\it Galfast} catalog also demonstrated 
that the fraction of red giant stars in low Galactic latitude samples is much larger than observed 
at high Galactic latitudes. These conclusions are important for the interpretation of results 
described in the next section.

\section{ANALYSIS OF THE RESULTS}
\label{sec:analysis}

We apply the method described in the preceding Section (and summarized in
\S\ref{sec:methodSummary}) in four different ways. 
We fit separately the full SDSS dataset (73 million sources) using 
only SDSS photometry, and the SDSS-2MASS subset (23 million sources)
using both SDSS and 2MASS photometry. We first consider a fixed 
$C_\lambda$ extinction curve determined for Stripe 82 region (the 
coefficients listed in the first row in Table~\ref{Tab:ClambdaConstraints}),
and refer to it hereafter as the ``fixed $R_V=3.1$'' case (although the
best-fit CCM model corresponds to $R_V=3.0\pm0.1$). These fixed-$R_V$ fits are 
obtained for the entire dataset, including high Galactic latitude regions
where dust extinction is too small to reliably constrain the shape of the
extinction curve (i.e., $R_V$) using data for individual stars. To 
investigate the variation of $R_V$ in high-extinction and low Galactic 
latitude regions, we use the CCM $C_\lambda$ curves discussed in 
Section~\ref{sec:Clambda} (and shown in Figure~\ref{Fig:AlambdaRv}). In this 
``free $R_V$'' case, we only consider the ten SEGUE stripes limited to the 
latitude range $|b|<30^\circ$, which include 37 million sources in the full 
SDSS dataset, and 10 million sources in the SDSS-2MASS subset. As discussed
in \S\ref{sec:freeRvtests}, the ``free $R_V$'' results are only reliable when
based on the full SDSS-2MASS photometric dataset. We include the ``free $R_V$'' 
only-SDSS results in the public distribution for completeness, but do not 
discuss them further. 

The resulting best-fit parameter set is rich in content and its full 
scientific exploitation is far beyond the scope of this paper. 
The purpose of the preliminary analysis presented below is to illustrate
the main results and to demonstrate their reliability, as well as to motivate further
work by others -- all the data and the best-fit parameters are made publicly available, 
as described in Appendix B. 

We first analyze ``fixed $R_V$'' fits, and compare results based on only-SDSS
data with those obtained using the full SDSS-2MASS dataset. This comparison 
shows that both datasets result in similar best fits, which adequately 
explain the observed dust-reddened SEDs of most stars in the samples. 
The main conclusion derived from the ``free $R_V$'' fits is the lack of strong evidence
for a significant overall departure from the canonical value of $R_V=3.1$.

\subsection{Fixed $R_V$ Case} 

Two sets of results based on a fixed dust extinction curve (``fixed $R_V=3.1$'' case)
are compared: those based on the full SDSS-2MASS photometric dataset whose seven
colors provide better fitting constraints, and those for a larger and fainter only-SDSS 
sample which includes only four colors. We begin with a basic statistical analysis of
the best-fit $\chi^2_{pdf}$ distributions.

\subsubsection{The best-fit $\chi^2_{pdf}$ distributions}
\label{sec:chi2} 

The distribution of the best-fit $\chi^2_{pdf}$, separately for low-extinction and 
high-extinction regions, and for low-SNR and high-SNR sources (bright and faint), 
is shown in Figure~\ref{Fig:Chi2Slices110}.  As evident, there is no strong dependence of
the shape of the best-fit $\chi^2_{pdf}$ on SNR. In low-extinction regions (top
two panels) the obtained $\chi^2_{pdf}$ distributions closely resemble
theoretical $\chi^2_{pdf}$ distributions with 2 and 5 degrees of freedom. This
agreement is not too surprising because the empirical model library was derived 
using the same dataset, and essentially demonstrates that cataloged photometric
errors for SDSS and 2MASS are reliable. 

In the high-extinction regions (although we discuss here only a single SEGUE stripe, 
we have verified that our conclusions are valid for all ten stripes), the core of the 
observed $\chi^2_{pdf}$ distributions is still similar to theoretically expected 
distributions (computed for Gaussian error distributions, and assuming that SEDs of all 
stars in the sample are well described by the model library), but tails are more extended
than in low-extinction high-latitude regions. For comparison, about 70\% of a
sample is expected to have $\chi^2_{pdf}<1.2$ (valid for the low number of degrees
of freedom considered here), while we obtained about 50\% for the observed 
distributions. The increased fraction of red giants at low Galactic latitudes, 
increased but unrecognized photometric errors (e.g., due to crowding), more
complex dust extinction curve behavior than captured by the adopted CCM model, 
as well as increased metallicity of disk stars, may all contribute to the tails of the 
observed $\chi^2_{pdf}$ distributions. 

For further analysis, we use subsamples of stars with $r<19$, $K<14$ (Vega scale), 
and $\chi^2_{pdf}<2$, unless noted otherwise. These criteria select stars with relatively small 
photometric errors (typically $<0.05$ mag in most bands) and whose reddened SEDs are well
described by the model SED library and the CCM extinction curve. About 50-60\%
of stars in only-SDSS sample, and 70-80\% stars in SDSS-2MASS subsample, 
are typically selected by the adopted $\chi^2_{pdf}<2$ cut (for theoretical
$\chi^2_{pdf}$ distributions with 2 and 5 degrees of freedom, 86\% and 93\% of stars 
would satisfy this $\chi^2_{pdf}$ cut).

\subsubsection{The Northern Galactic Cap Region \label{sec:ngp}} 

Due to small $A_r$ for the $b>30^\circ$ sky region (the median $A_r$ from the SFD map is $\sim$0.08 mag), 
the errors for best-fit $A_r$ for individual stars can be as large as best-fit $A_r$ 
itself when using only-SDSS fits (fixed $R_V$ case). Both the formal $A_r$ errors, and the
differences between best-fit and SFD values for $A_r$ begin to increase rapidly for $r>18$ and
become unreliable for $r>19$. This behavior is in agreement with tests described in \S\ref{sec:methodtests}
and the behavior of SDSS photometric errors as a function of magnitude (even for blue stars, 
the median $u$ band error is already 0.05 mag at $r=19$, and 0.2 mag when stars of all colors are considered). 

Nevertheless, by taking a median value for typically several hundred stars per $\sim$1 deg$^2$, 
a map can be constructed that reproduces the features seen in the SFD map (see the top left panel 
in Figure~\ref{panelsLambertNGP}). Quantitative analysis of the median differences between the best-fit 
$A_r$ and the SFD $A_r$ values shows that the former are larger by about 50\% on average, with a scatter 
of about 20\%. This bias is probably due to color-reddening degeneracy and small extinction at high Galactic 
latitudes which is only a factor 2-3 larger than photometric errors. An additional effect contributing to this 
bias are zeropoint calibration errors in SDSS photometry: the median differences between the best-fit $A_r$ 
and the SFD $A_r$ values show a structure reminiscent of the SDSS scanning pattern (see the top right
panel in Figure~\ref{panelsLambertNGP}). These coherent residuals imply problems with the transfer of SDSS 
photometric zeropoints across the sky. 

The median differences between observed and best-fit model 
magnitudes show deviations of up to 0.01 mag, and are largest in the $i$ band,  as illustrated in the 
bottom left panel in Figure~\ref{panelsLambertNGP}). Therefore, these relatively small local calibration 
errors (each of the six scanning strips in an SDSS scan, i.e., the ``camera columns'', is independently 
calibrated) are mis-interpreted as a local extinction variation at the level of a few times 0.01 mag. 

With the addition of 2MASS photometry, the agreement with the SFD map improves.  The best-fit  $A_r$
values are overestimated, relative to SFD values, by only $\sim$0.02 mag (25\% on average), and the
median differences do not show structure resembling  the SDSS scanning pattern (see the bottom right
panel in Figure~\ref{panelsLambertNGP}). We note that $r<18$ selection limit (and $K<14$ in 2MASS case)
results in about one star per the resolution element of SFD map. Therefore, to significantly improve the 
spatial resolution of extinction map at high Galactic latitudes, a sample several magnitudes deeper than
SDSS-2MASS sample is required.

\subsubsection{The SEGUE Stripes} 

The main goal of this work is to determine extinction at low Galactic latitudes. 
We consider ten $\sim2.5^{\circ}$. wide SEGUE stripes with $|b|<30^\circ$. The full SDSS
sample includes 37 million sources, with 10 million sources in the SDSS-2MASS subset.
We find that results based on the two datasets are similar, though the latter is expected
to produce more reliable results. We first illustrate the behavior
of best-fit $A_r$ as a function of distance for all stripes, and then provide more quantitative
discussion of the differences in best-fit results in the next section, which is focused on a single
stripe ($l\sim110^\circ$). We also provide a comparison to the SFD extinction maps further below. 

A visual summary of the best-fit $A_r$ using only-SDSS fits for the ten SEGUE stripes, 
in the range $|b|<5^\circ$ and for three distance slices ranging from 0.3 kpc to 2.5 kpc, 
is shown in Figure~\ref{Fig:ArSDSSdSlicesAll}. Distances to stars are determined by assuming that 
all sources are main sequence stars, and using photometric parallax relation from I08 
with [Fe/H]=$-0.4$ (with the best-fit intrinsic colors). An expected scatter in metallicity 
of 0.2-0.3 dex for disk stars corresponds to about 10-15\% uncertainty in distance. 
Although not all sources are main sequence stars (such as red giants, which have grossly 
underestimated distances, see \S\ref{sec:redgiantsdata} below for discussion), the 
fraction of main sequence stars in the samples is sufficiently large that the median $A_r$ 
is not strongly biased. Furthermore, sources whose SEDs are significantly
different from the main sequence SEDs are not included: the figures are constructed
only with sources that have the best-fit $\chi^2_{pdf}<2$. We also excluded red giant
candidates, as described below. 

It is easily discernible from  Figure~\ref{Fig:ArSDSSdSlicesAll} that the extinction 
along the line-of-sight (that is, $A_r$) increases with distance. On average, the stripes
towards the Galactic center have more large-extinction ($A_r>1$) regions. In several 
directions, $A_r$ exceeds several magnitudes and practically no stars are detected by SDSS.

\subsubsection{Selection Function Differences for only-SDSS and SDSS-2MASS subsamples} 

Another projection of the sky position--distance--$A_r$ space is shown in 
Figures~\ref{Fig:DAcounts_onlySDSS} (only-SDSS case) and \ref{Fig:DAcounts_SDSS2MASS}
(SDSS-2MASS case). As evident, the morphology of these $A_r$ vs. distance diagrams differs
significantly between the two subsamples. The main reason for these changes is different sample 
selection functions in the flux-color space -- and not differences in the best-fit $A_r$ or 
distance values which agree well on a star by star basis (see the next section). 

For only-SDSS case, the main selection criterion (in addition to $\chi^2_{pdf}<2$ in both cases) is 
$r<19$ and the $u$ band error limit of 0.05 mag.  The latter condition is necessary to assure 
reliable fitting results when only four colors are used, and results in a strong bias towards the
blue end of the observed color distribution. In SDSS-2MASS case,  it is sufficient to require 
$K<15$ (Vega) to obtain reliable fitting results because there are seven colors, and because
this condition limits the $K$ band and $u$ band errors to about 0.1 mag (with much smaller errors 
in other bands). 
This selection condition results in  a strong bias towards the red end of the observed color 
distribution. Due to their selection functions, the effective $r$ band limiting magnitude for 
reliable only-SDSS samples varies from $r\sim17$ at $g-i=1$ to $r\sim15$ at $g-i=3$, while 
for SDSS-2MASS samples it varies from $r\sim17$ at $g-i=1$ to $r\sim20$ at $g-i=3$. As a 
result, SDSS-2MASS samples contain many more nearby red dwarfs at distances below 500 pc, 
while only-SDSS sample extends further than SDSS-2MASS sample, to about 2.5 kpc. 
On average, about twice as many stars survive the quality cuts for SDSS-2MASS sample
as for only-SDSS sample (although the latter typically contains about four times as
many stars at $|b|<5^\circ$ before any selection). 

In the $A_r$ vs. distance diagram, the selection cutoff for SDSS-2MASS sample is nearly
vertical, and limits the sample to distances below about 1.5 kpc (assuming $A_r<5$ and 
main sequence stars). For only-SDSS sample, the upper limit on $u$ band error introduces
a diagonal selection boundary that excludes stars in the upper right corner. With the
selection criteria adopted above, the sample becomes limited to $A_r < 2$ at a distance
of about 1 kpc, with an overall distance limit of about 2.5 kpc. 

The slopes of $A_r$ vs. distance relations along main-sequence locus seen in 
Figures~\ref{Fig:DAcounts_onlySDSS} and \ref{Fig:DAcounts_SDSS2MASS}
constrains the local (within 1 kpc) extinction per unit distance normalization 
to the range $A_r/D=0.7-1.4$ mag kpc$^{-1}$, with larger values corresponding
to smaller angular distances from the Galactic center. The variation of this
normalization with Galactic longitude is consistent with the exponential scale length 
for thin disk stars obtained by J08 ($L_1=2.6\pm0.5$ kpc, see their Table 10). 
Nevertheless, the variation fo $A_r/D$ with longitude observed here is more complex
than predicted by simple axially symmetric dust distribution model.

\subsubsection{ The Selection of Candidate Red Giant Stars }
\label{sec:redgiantsdata}

The $A_r$ vs. distance diagrams based on SDSS-2MASS data (see Figure~\ref{Fig:DAcounts_SDSS2MASS})
show an excess of sources in the top left corner (the effect is not as strong for only-SDSS case because
the selection effects due to the $u$ band error limit, discussed in the previous section, remove most of 
these sources). Based on a mock catalog discussion in \S\ref{sec:redgiantsmock}, these sources are 
consistent with red giant stars. Informed by their distribution, and clear separation from the locus of
main sequence stars, we adopted the following criteria for the selection of candidate red giants:
\begin{enumerate} 
\item Best-fit main sequence distance below 1 kpc, $D_{kpc} < 1$, 
\item Best-fit extinction, $A_r > 1.5 + 1.5\,D_{kpc}$, and 
\item Best-fit intrinsic color, $0.4 < g-i < 1.4$. 
\end{enumerate} 
The first two criteria are based on the morphology observed in the $A_r$ vs. distance diagrams, 
and the third criterion removes outliers whose best-fit intrinsic colors are inconsistent with 
the color distribution for the majority of sources selected by the first two criteria. 

We applied these criteria to all ten SEGUE stripes and found that the fraction of selected
stars varies significantly with Galactic longitude, from $\sim15$\% for stripes at $l=50^\circ$
and $l=70^\circ$ to $\sim2$\% for stripes within 20$^{\circ}$. from the Galactic anticenter. 
The inclination of the main sequence stellar locus in the $A_r$ vs. distance diagrams
also varies with Galactic longitude, with its slope (determined for distances up to 1 kpc)
decreasing from about 2.0 mag kpc$^{-1}$ for the $l=50^\circ$ stripe to 0.6 mag kpc$^{-1}$ for the 
$l=187^\circ$ stripe. Hence, our selection criterion \#2 above could be improved by taking this
variation into account (for the same reason, the proportionality ``constant'' in 
eq.~\ref{eq:AvsD} varies with longitude). 

The observed variation of the fraction of candidate red giants with Galactic longitude
represents a strong constraint for the Galactic structure models, and the change of 
$A_r$ vs. distance slope reflects the variation of dust number volume density in the 
Galactic disk. Hence, the data presented here can be used to improve Galactic stellar
population models such as {\it Galfast} and TRILEGAL, and dust distribution models, 
such as the \citet{AL2005} model employed by {\it Galfast}. The required detailed analysis
is beyond the scope of this work.

\subsubsection{Detailed Analysis of the $l\sim110^\circ$ SEGUE stripe}

For a detailed analysis of the best-fit results, we select a single fiducial SEGUE stripe
with $l\sim110^\circ$. A simple but far-reaching 
conclusion of the work presented here is that fits to intrinsic stellar SEDs and dust 
extinction on per star basis are capable of reproducing the morphology of observed 
color-color diagrams in highly dust-extincted regions.
This success is illustrated in Figure~\ref{Fig:CMD_SEGUEl110_dataVSfixedRv}, where
six characteristic color-color diagrams constructed with observed SDSS-2MASS photometry
are contrasted with analogous diagrams constructed using best-fit results. We reiterate
that the observed morphology in these diagrams at low Galactic latitudes is {\it vastly 
different} than at high latitudes (the latter is illustrated in the figure by the Covey et al. locus).  

When considering SDSS-2MASS sample, fits based on the full seven-color set and those
restricted to the four SDSS colors produce quantitatively similar, though not 
identical results. The root-mean-square (rms) scatter of the difference in best-fit intrinsic 
colors is 0.04 mag, and rms for best-fit $A_r$ difference is 0.07 (the median $A_r$ is 
1.9). For $A_r\sim5$, the values based on only-SDSS photometry become biased (larger)
by about 4\% relative to SDSS-2MASS values. 
A star-by-star comparison presented in Figure~\ref{Fig:compareArGI_onlySDSSvsSDSS2MASS}
shows a few regions (e.g. $g-i\sim1.5$ and small $g-i$) where results can
differ substantially; nevertheless, the fraction of affected sources is small and
negligible when results are averaged over many stars. The latter point is illustrated in
Figure~\ref{Fig:ArSDSSvsSDSS2MASS}, which compares the two $A_r$ maps for stars
at a limited range of distances. The two maps agree to better than 0.05 mag even
in regions where $A_r > 4$. This agreement demonstrates that SDSS data alone are
sufficient to obtain the best-fit intrinsic color and extinction along the line-of-sight
for the majority of stars (when $R_V$ is fixed). In the rest of analysis we use SDSS-2MASS results, 
except in a few cases where we explore distances beyond 2 kpc. 

A cross-section of the three-dimensional $A_r$ map, based on only-SDSS sample from 
the $l\sim110^\circ$ is shown in Figure~\ref{Fig:ArSDSSdistSlicesSEGUE110}. As evident, 
the best-fit $A_r$ increases with the stellar distance between 0.3 kpc and 2.5 kpc. It is 
noteworthy that the two quantities are determined independently (distance is computed 
a posteriori, from the best-fit apparent magnitude). A closer look at distances
below 1 kpc using SDSS-2MASS dataset is shown in Figure~\ref{Fig:ArSDSSdSlices110zi}.  
An impressive feature is the abrupt jump in $A_r$ towards $b\sim2^\circ$ for stars
with distances above 0.9 kpc, thus providing a robust and fairly precise lower distance limit for that dust 
cloud! 

Differences between best-fit $A_r$ values determined here and the SFD map are illustrated
in Figure~\ref{Fig:ArSDSScompSFD}. Since the latter corresponds to extinction along the line 
of sight to infinity, our values are systematically smaller in regions with large $A_r$ and 
similar at large Galactic latitudes, as expected. A detailed analysis of these $A_r$ differences, when combined 
with stellar distance estimates, can provide valuable constraints for various ISM studies. 
For example, in Figure~\ref{Fig:KeiraL110} we demonstrate good correspondence between 
the $A_r$ differences and the distribution of molecular (CO) emission; our results imply
that those molecular clouds must be more distant than $\sim$1 kpc, and that the
substructure seen around $b\sim-2.5^\circ$ is more distant that the one at $b\sim2^\circ$
(see also Figure~\ref{Fig:ArSDSSdSlices110zi}, and a more quantitative discussion in 
\S\ref{sec:SpatDistDust}). Other SEGUE strips contain more examples 
where such ``bracketing'' of distances to molecular clouds can be attempted. 

We note that the SFD map is expected to sometimes fail at low Galactic latitudes not just 
because of stars being embedded in dust, but also because its construction relied upon 
accurate point source subtraction (which is only performed for $|b|<4.7^\circ$) and dust
 having a single temperature along each line of sight. 
These assumptions might be violated in the Galactic plane and may be responsible for regions 
where SFD values are smaller than in our maps (e.g., at $b\sim 10-20^{\circ}$ in Figure 26).

\subsection{Free $R_V$ Case} 
\label{sec:freeRv}

\begin{figure}[!t]
\epsscale{0.9}    
\plotone{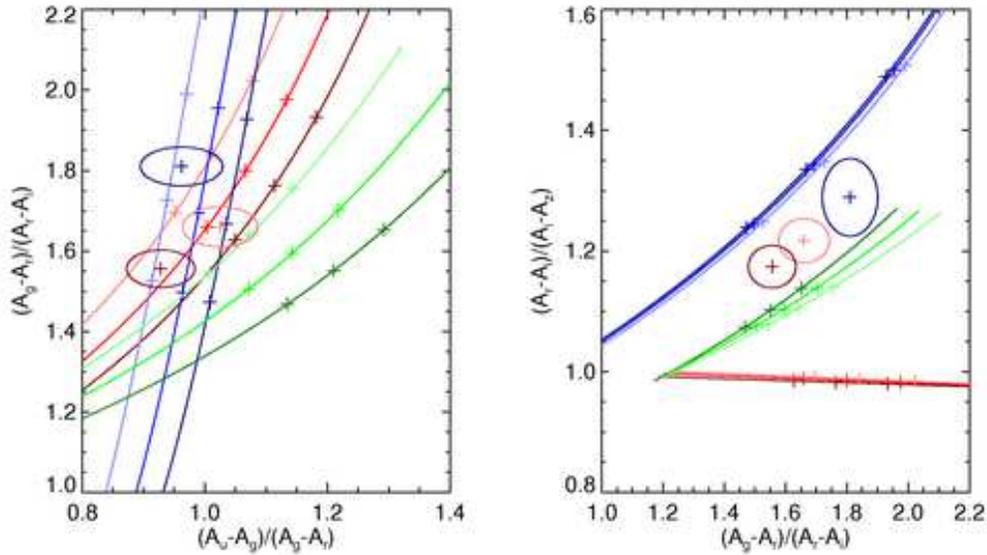}
\epsscale{1.0}
\vskip 0.2in    
\caption{A comparison of the constraints on the extinction curve shape
(the three plus symbols, with approximate 1$\sigma$ uncertainty limits 
shown as ellipses) and three model predictions (see Figure~\ref{Fig:Ed1} 
for legend; the three crosses along the curves correspond to $R_V$=2.6, 3.1,
and 3.6). The pink symbol corresponds to the Stripe 82 region (southern
Galactic hemisphere), the brown symbol to the northern Galactic hemisphere,
and the blue symbol is the constraint from the Schlafly et al. (2010) analysis.
The blue \citep{Fitz99} and green \citep{CCM} models are in fair agreeement with 
the data, while the red model \cite{ODonnell} predicts unacceptable values of the 
$(A_r-A_i)/(A_i-A_z)$ ratio for all values of $R_V$ (see also 
Figure~\ref{Fig:RvConstraints}).} 
\label{Fig:Ed2}
\end{figure}

\begin{figure}[!t]
\epsscale{0.9}    
\plotone{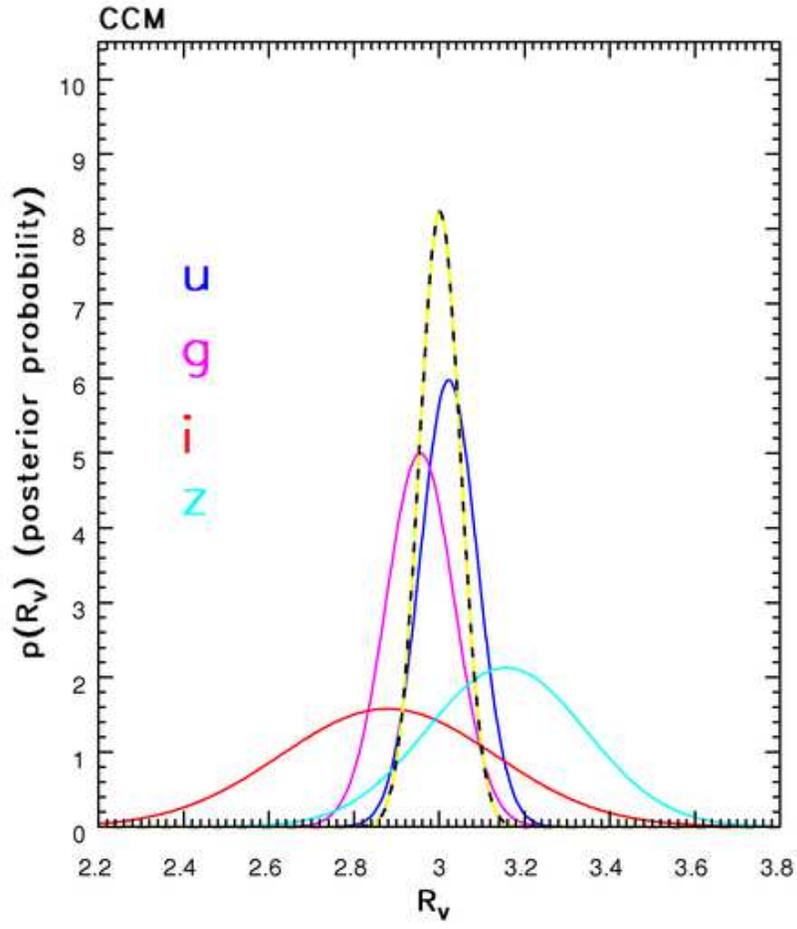}
\epsscale{1.0}
\vskip -1.3in    
\caption{Constraints on $R_V$ based on the CCM \citep{CCM} dust reddening
law. Only the SDSS bands, which provide the strongest constraints on $R_V$,
are shown (see the legend). The dashed line shows the overall constraint
on $R_V$ (posterior probability distribution for a flat prior), with 
the best-fit value of $R_V=3.01\pm0.05$.}
\label{Fig:RvConstraints}
\end{figure}

\begin{figure}[!t]
\epsscale{0.9}    
\plotone{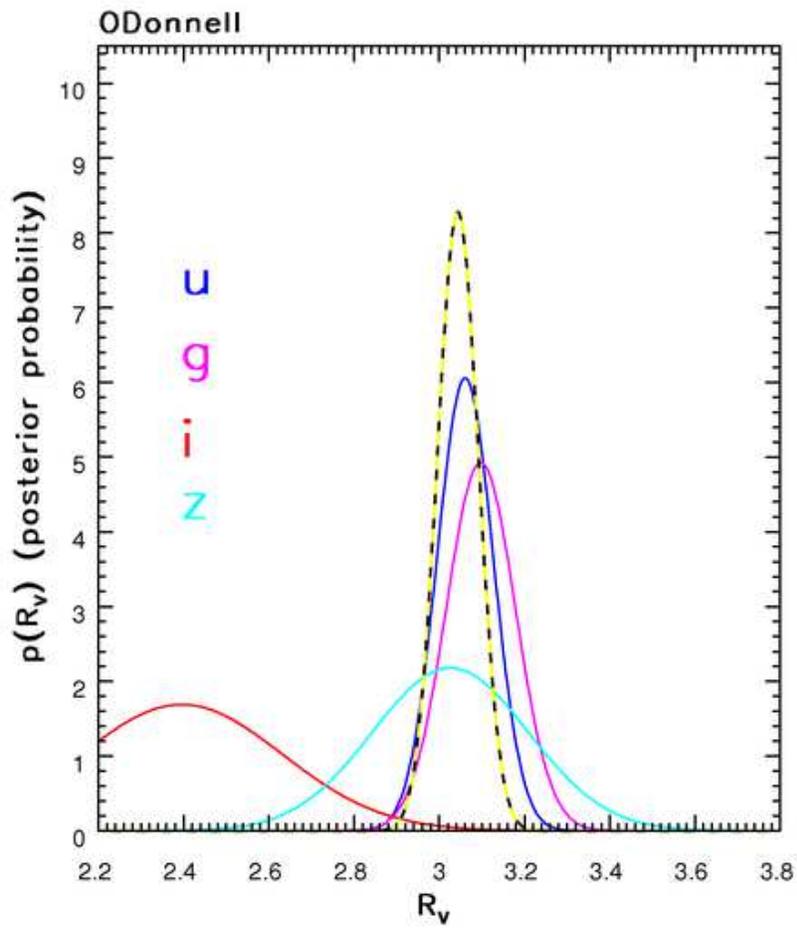}
\epsscale{1.0}
\vskip -1.3in    
\caption{Analogous to Figure~\ref{Fig:RvConstraints}, except that 
\citet{ODonnell} dust reddening law is used. The 
predicted extinction in the $i$ band is inconsistent with 
constraints from other bands.}
\label{Fig:RvOD}
\end{figure}

\begin{figure}[!t]
\epsscale{1.0}
\plotone{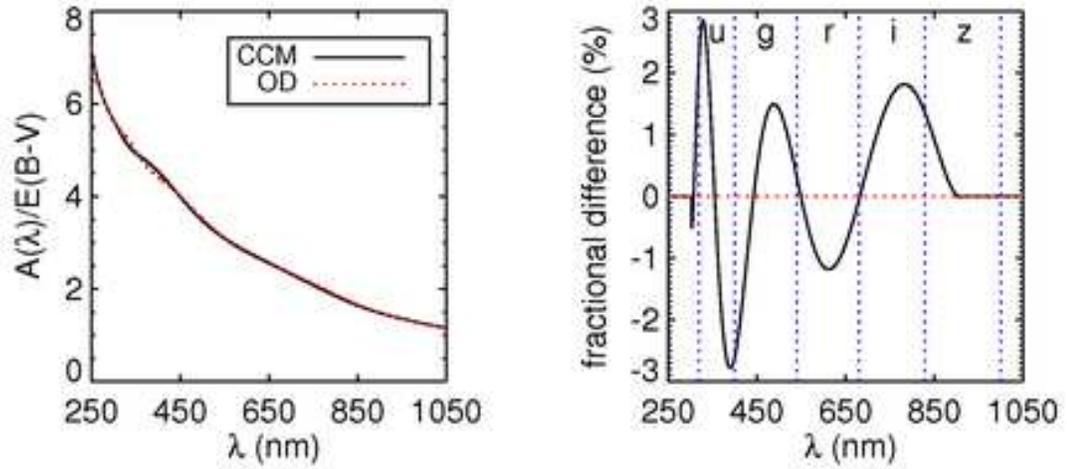}
\vskip -0.1in
\caption{The left panel shows the CCM (black, solid line) and the O'Donnell (red, dashed line) 
dust reddening laws as function of wavelength. The right panel displays the fractional difference 
between the two dust extinction models with the SDSS filter transmission regions overlaid (vertical, 
blue, dashed lines). Given the filter transmission regions, the largest integrated difference is 
expected in the $i$ band, which is what we observe.}
\label{Fig:ODCCMdiff}
\end{figure}

\begin{figure}[!t]
\plotone{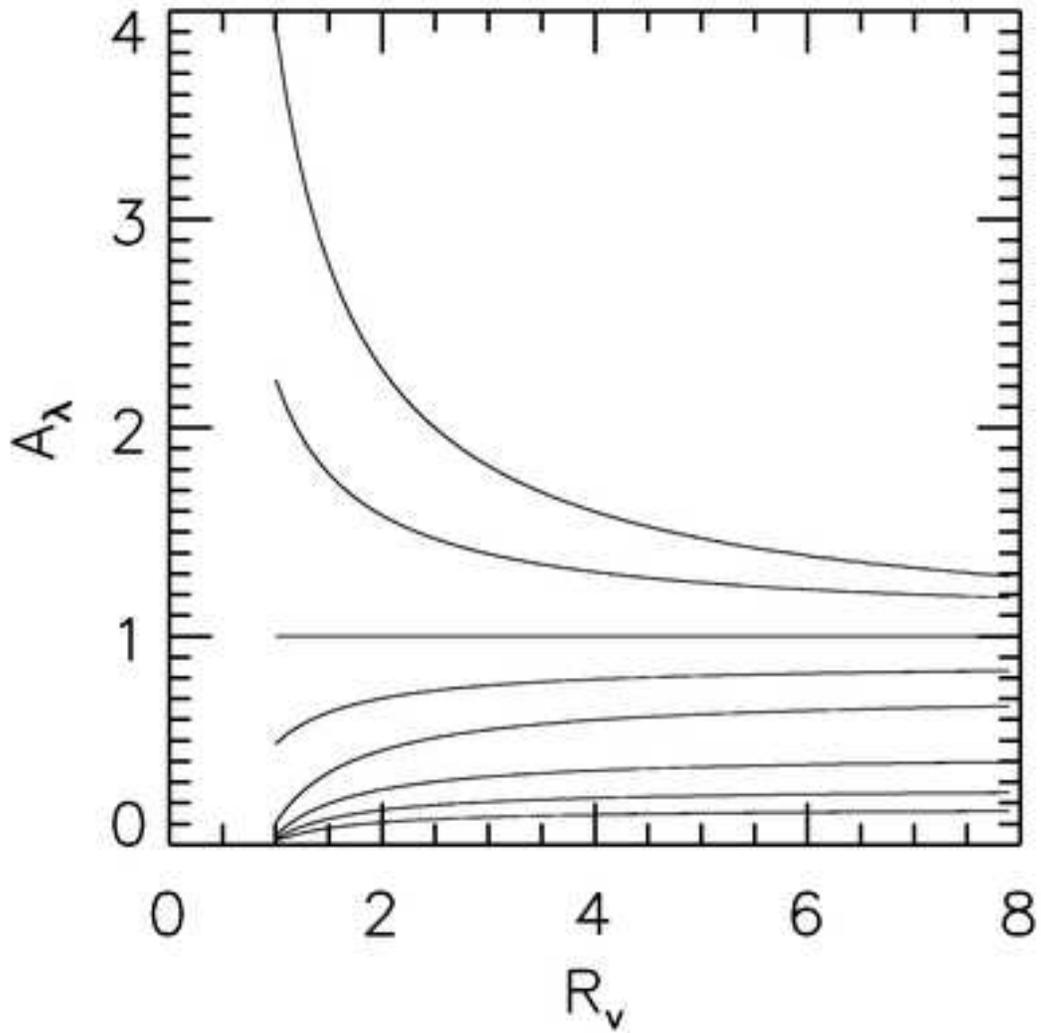}
\caption{The adopted $A_{\lambda}/A_r$ ratio, shown as a function of 
$R_V$, for $\lambda=(ugrizJHK)$, from top to bottom ($A_r=1$).  The curves are computed
for an F star using the CCM \citep{CCM} dust reddening law.}
\label{Fig:AlambdaRv}
\end{figure}

\begin{figure}[!b]
\epsscale{1.06}        
\hskip -0.5in
\plotone{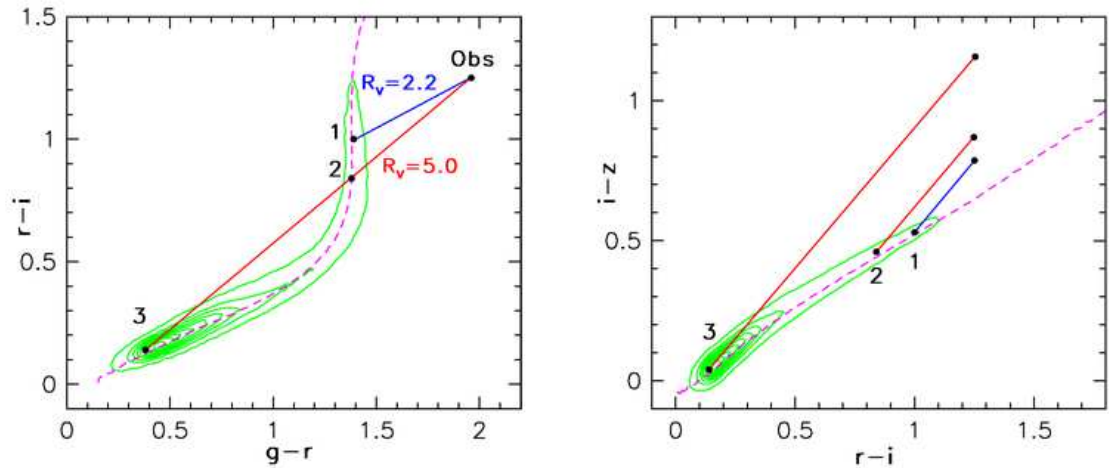}
\epsscale{1.0}        
\vskip -1.5in
\caption{
An illustration of the constraints on intrinsic stellar colors,
extinction in the $r$ band, $A_r$, and the ratio of total to selective
extinction, $R_V$. In both diagrams, the linearly-spaced 
contours show the main stellar locus as observed at high Galactic
latitudes. The dashed lines mark the median stellar locus from 
\citet{Covey07}. In the left panel, the dot marked ``Obs''
represents a hypothetical observation. Depending on the adopted 
$R_V$, as marked, different combinations of intrinsic stellar 
colors (i.e., the position along the stellar locus) and $A_r$ 
are consistent with the observed $g-r$ and $r-i$ colors. Multiple
solutions are possible even for a fixed value of $R_V$. The
three solutions marked 1-3 correspond to ($R_V$,$A_r$)=
1:(2.2,1.0), 2:(5.0,2.2), and 3:(5.0,6.0). As shown in the right 
panel, these degeneracies are broken if the $i-z$ color is 
also available: the three ($R_V$,$A_r$) combinations have 
different reddened $i-z$ colors which breaks the degeneracy between
the intrinsic stellar color and $A_r$. The degeneracy is broken
because the reddening vectors in the right panel are nearly parallel 
despite very different $R_V$ values.}
\label{Fig:riVSgr}
\end{figure}

\begin{figure}[!t]
\plotone{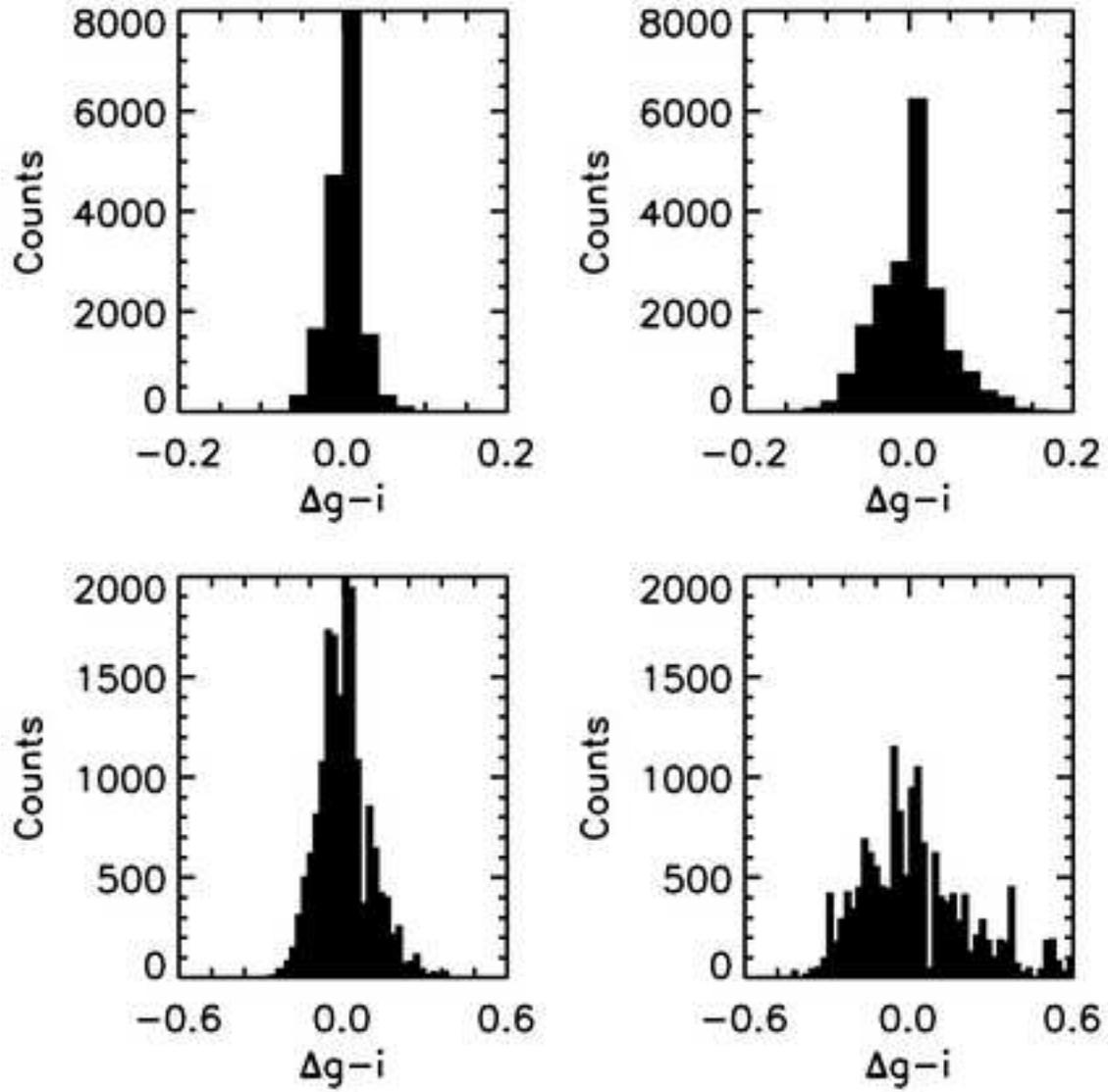}
\caption{A Monte Carlo study of best-fit stellar model errors (parametrized by the 
$g-i$ color) as a function of photometric errors, for a fiducial star with $g-i$=1.95 
and $A_r$=1.5 (the abscissa, $\Delta$$g-i$=true - fit). The photometric errors are generated 
from Gaussian distributions with widths equal to 0.01 mag (top left), 0.02 mag (top right), 
0.04 mag (bottom left) and 0.08 mag (bottom right). The errors in the best-fit
$g-i$ are about twice as large as assumed photometric errors.}
\label{Fig:Dgi}
\end{figure}

\begin{figure}[!t]
\plotone{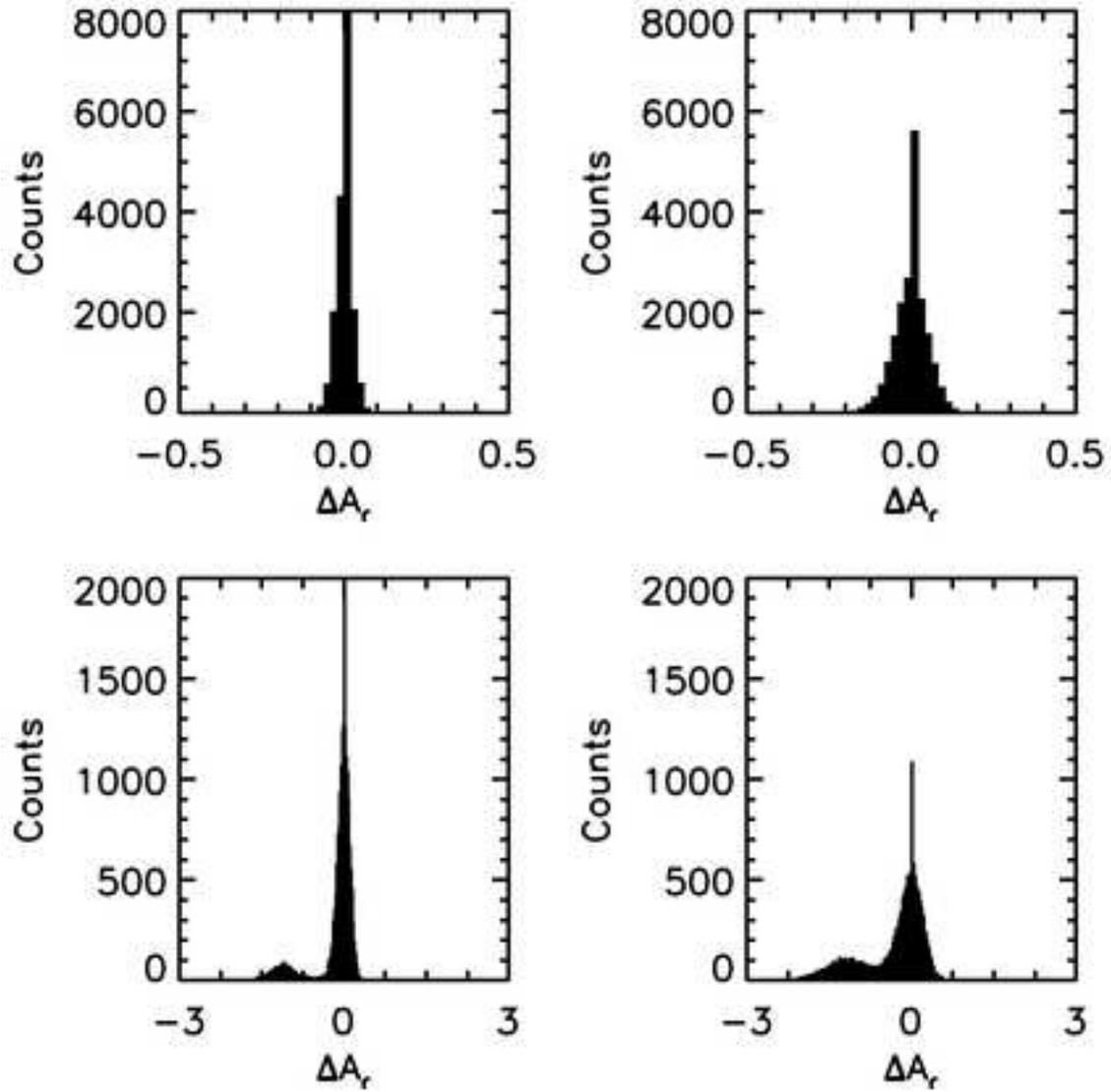}
\caption{Analogous to Fig.~\ref{Fig:Dgi}, except that the
errors in the best-fit $A_r$ are shown ($\Delta$$A_r$)=true - fit). 
Note that for large photometric errors (the bottom two panels),
the $A_r$ error distribution becomes bimodal; the additional mode corresponds
to a solution with a bluer star with more reddening.}
\label{Fig:Dar}
\end{figure}

\begin{figure}[!t]
\plotone{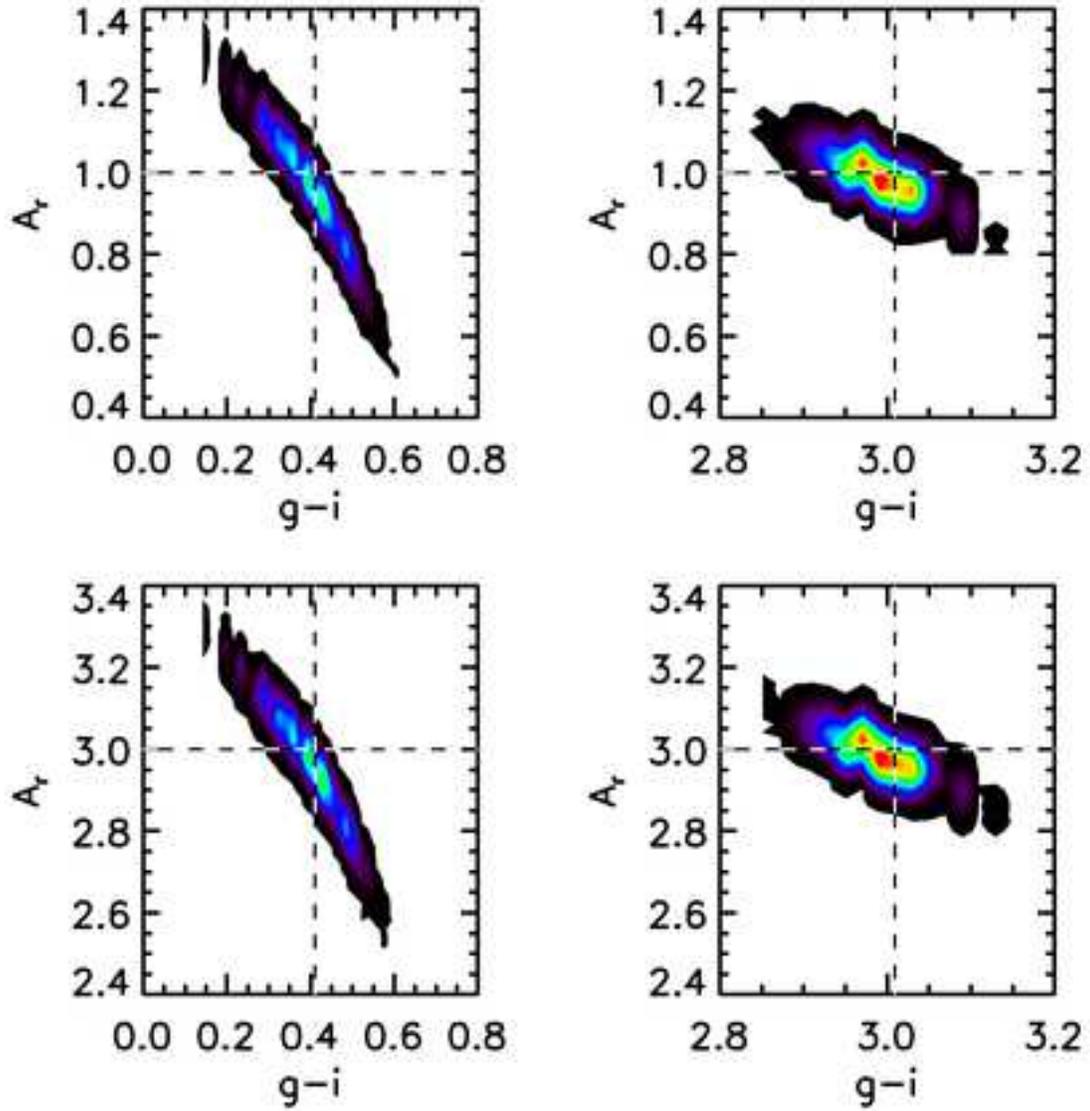}
\caption{Analysis of the covariance in the best-fit values for
$A_r$ and $g-i$ using a simulated dataset. The panels show the distributions
of the best-fit values for $A_r$ and $g-i$ for two different fiducial stars 
(left column: a blue star with true $g-i$=0.4; right column: a red star 
with true $g-i$=3.0), and two different extinction values (top panels: 
$A_r=1$; bottom panels: $A_r$=3). Photometric errors in the $ugriz$ bands
are generated using Gaussian distributions with $\sigma$=0.02 
mag (uncorrelated between different bands). Note that the 
$A_r$ vs. $g-i$ covariance is larger for the blue star, and does not
strongly depend on assumed $A_r$.}
\label{Fig:DgiChi2}
\end{figure}

\begin{figure}[!t]
\epsscale{0.8}
\plotone{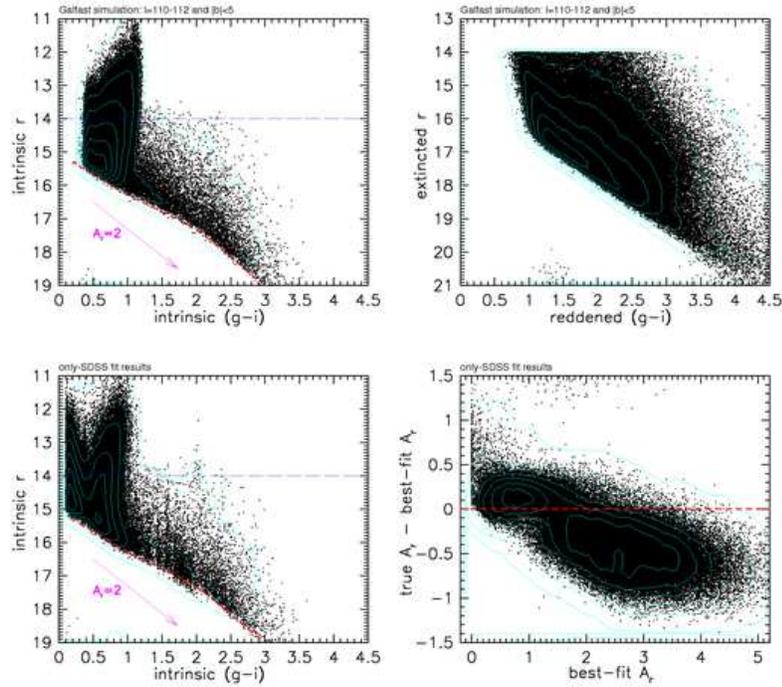}
\vskip -1.4in
\caption{Analysis of a {\it Galfast} simulated SDSS-2MASS sample from a SEGUE strip 
($l\sim 110^\circ$ and $|b|<5^\circ$). The $r$ vs. $g-i$ color-magnitude diagrams in the 
top two panels explain why the fraction of giants is much larger than observed at high Galactic 
latitudes. The same simulated sample, defined by {\it observed extincted} magnitude cuts 
$14<r<21$ and $K<14.3$ (Vega)  is shown in both panels. The left panel is constructed using 
{\it un-extincted} magnitudes, and the right panel with ``observed'' magnitudes (note the offset of the 
y axis by 2 mag). The horizontal dashed line in the left panel shows the SDSS saturation limit; 
stars above this line are dominated by red giants (the ``plume'' towards towards $g-i\sim1$). 
The diagonal dashed line shows the magnitude limit for main sequence stars with $K<14.3$ and 
{\it no dust extinction}. The reddening arrow corresponds to $A_r=2$ and $R_V=3.1$ CCM extinction 
curve. The bottom left panel is analogous to the top left panel, except that the SDSS-based best-fit 
values for $g-i$ and $A_r$ are used. The bottom right panel shows the difference between the input 
value of $A_r$ and the best-fit values, as a function of the latter. The dashed line is added to 
guide the eye.  The root-mean-square scatter for the $A_r$ difference (rms for y axis) is 0.33 mag,
and the bias for large $A_r$ is about 15\% (an overestimate of $A_r$ due to color-$A_r$ degeneracy, 
see text). 
}
\label{Fig:mock2}
\end{figure}

\begin{figure}[!t]
\epsscale{0.8}
\vskip 0.3in
\plotone{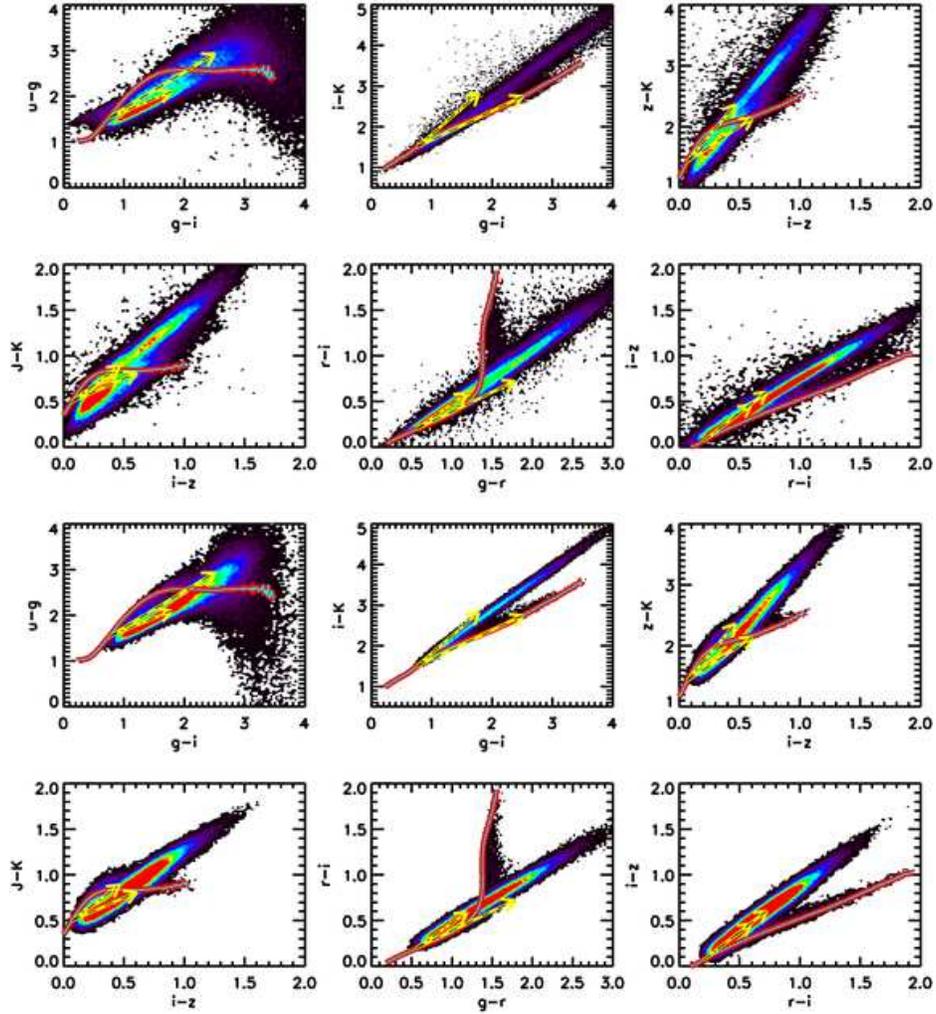}
\vskip 0.2in
\caption{A comparison of six SDSS-2MASS color-color diagrams for data from the SEGUE 
$l\sim 110^\circ$ strip (the top six panels), and for a mock catalog produced with the {\it Galfast} 
code (the bottom six panels). The color-coded contours shows the source counts on a linear 
scale. The two dashed arrows show reddening vectors for $A_r=2$ and $R_V=2$ and 4. The 
locus of circles shows the \citet{Covey07} empirical SED library and illustrates the morphology 
of the same diagrams observed at high Galactic latitudes (and corrected using the SFD map; 
typically $A_r\sim0.1$). The two sets of diagrams are encouragingly similar, with a few detailed 
differences: the observed diagrams have more outliers, and a few diagrams (e.g., $J-K$ vs. $i-z$ 
and $i-z$ vs. $r-i$) imply different reddening vectors than used in simulations ($R_V=3.1$).}
\label{Fig:mock}
\end{figure}

\begin{figure}[!t]
\vskip 0.2in
\epsscale{0.95}
\plotone{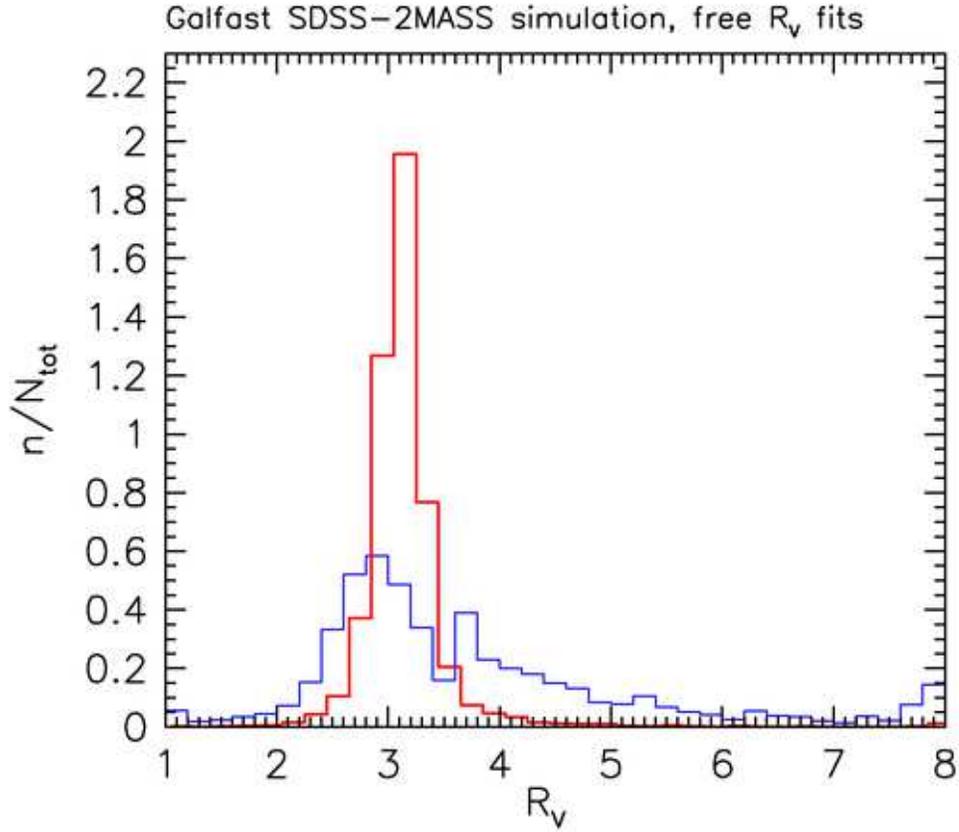}
\vskip -3.0in
\caption{A comparison of the best-fit $R_V$ values for SDSS-2MASS (narrow histogram) and only-SDSS 
(broad histogram) cases, using a simulated {\it Galfast} mock catalog. The input value is fixed to $R_V=3.1$. 
The equivalent Gaussian widths determined from the interquartile range are 0.1 and 1.2, respectively. }
\label{Fig:GalfastRv}
\end{figure}

\begin{figure}[!t]
\vskip 0.in
\epsscale{0.8}
\plotone{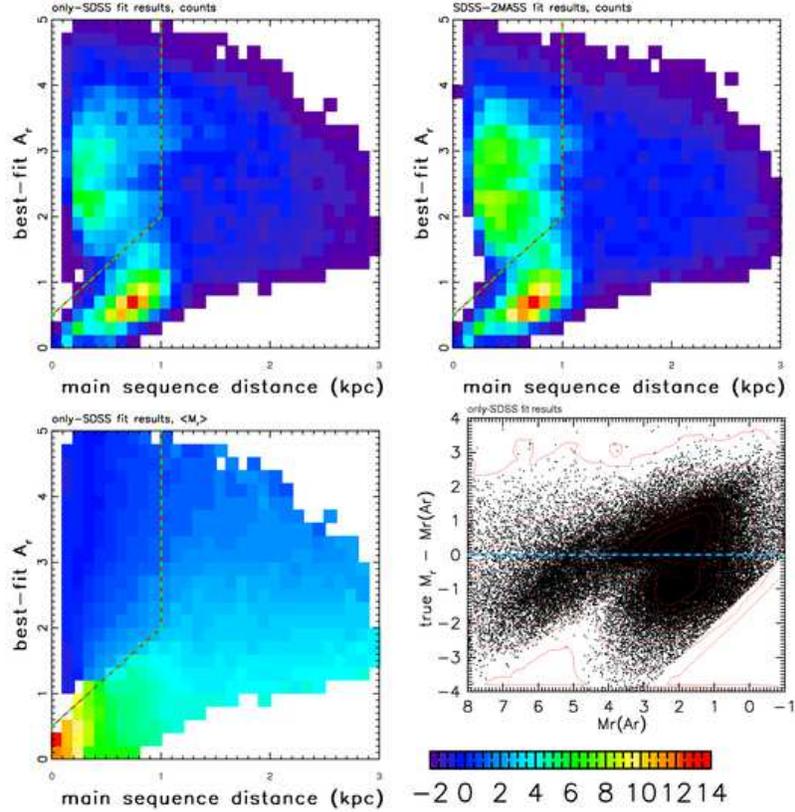}
\vskip -0.6in
\caption{The left panel shows the relationship between best-fit $A_r$ and distance computed using 
the SDSS-based best-fit stellar color and a photometric parallax relation appropriate for main sequence 
stars, for the same simulated sample as in Figures~\ref{Fig:mock2} and \ref{Fig:mock}. The color-coded
map shows the counts of stars on a linear scale. The dashed lines isolate candidate red giant stars that 
have small distances and large $A_r$. The top right panel is analogous to the top left panel, except 
that the best-fit values correspond to SDSS-2MASS data. The bottom left panel shows the median input
absolute magnitude ($M_r$) for stars in each pixel, color coded according to the legend in the lower
right corner. The ``red giant region'' in the top two panels is dominated by giants ($M_r<3$). 
The bottom right panel shows the difference between true absolute magnitude and an estimate obtained 
from the ``dusty parallax relation'' (see eq.~\ref{eq:DPR}), as a function of the latter. The dashed line 
has is added to guide the eye. The root-mean-square scatter between the two magnitudes 
(rms for y axis) is 1.1 mag.}
\label{Fig:DPR}
\end{figure}

\begin{figure}[!t]
\epsscale{0.6}
\vskip 0.2in
\plotone{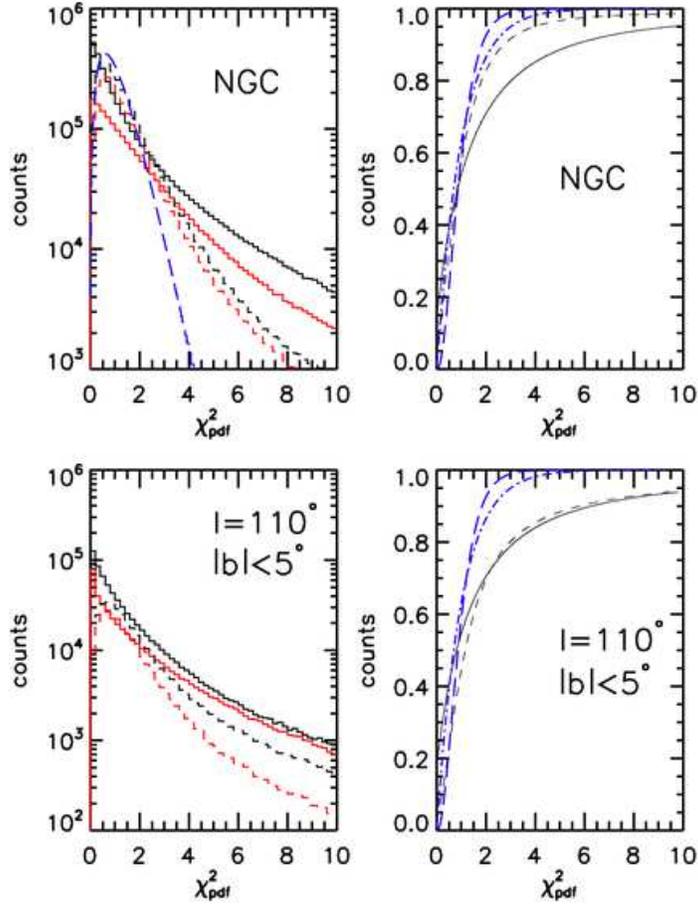}
\vskip 0.2in
\caption{The distribution of the best-fit $\chi^2_{\mathrm{pdf}}$ ($R_V=3.1$), with differential
distributions in the left two panels, and cumulative distributions in the right two
panels. The top two panels correspond to the north Galactic cap region ($b>45^\circ$)
and the bottom two panels to the SEGUE $l\sim110^\circ$ strip, limited to $|b|<5^\circ$
(a high-extinction region). The solid lines are used for SDSS-only fits, and the dashed 
lines for fits to SDSS-2MASS data. In the two left panels, the top solid line corresponds 
to subsamples of stars with $r<20$, and the bottom solid line to stars with $20<r<21$. 
The top dashed line corresponds to the full SDSS-2MASS sample, and the bottom dashed 
line to subsamples with $K<13.9$  (Vega scale, approximately corresponding to $K$ band 
errors up to 0.05 mag). The solid lines in the right panels correspond to the full SDSS sample, 
and the short-dashed lines to the full SDSS-2MASS sample. The dot-dashed and long-dashed
lines correspond to $\chi^2_{\mathrm{pdf}}$ distributions with 2 and 5 degrees of freedom.  The 
long-dashed line in the top left panel corresponds to the $\chi^2_{\mathrm{pdf}}$ distribution 
with 5 degrees of freedom for the full SDSS-2MASS sample.
}
\label{Fig:Chi2Slices110}
\end{figure}

\begin{figure}[!t]
\epsscale{0.9}
\phantom{x}
\vskip 0.5in
\plotone{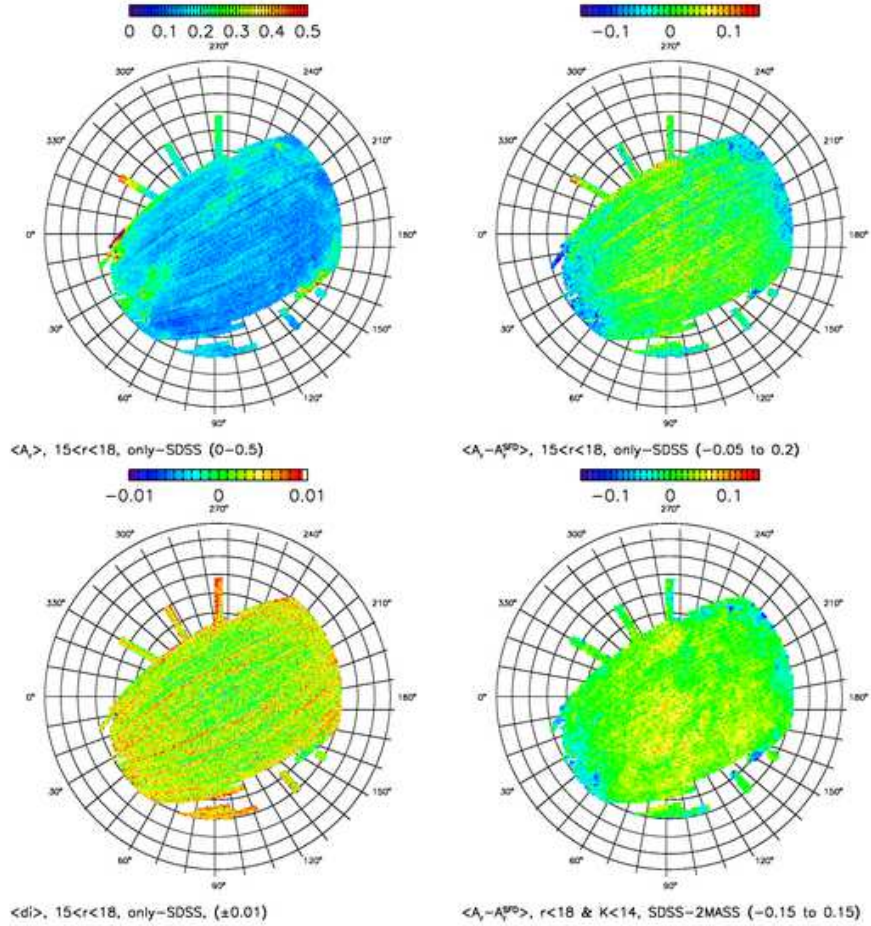}
\vskip -1.2in
\caption{Analysis of the best-fit results for $A_r$ in the low-extinction region with $b>30^\circ$.
The top left panel shows the median $A_r$ in 0.6 deg$^2$ pixels in Lambert projection.
The values are linearly color-coded according to the legend.
Stars with $15<r<18$ and $\chi^2_{\mathrm{pdf}}<2$ from only-SDSS sample with fixed $R_V$
are used for the plot. The median difference between $A_r$ and the values given by the SFD map
are shown in the top right panel. Note the striping reminiscent of the SDSS scanning pattern. 
The bottom left panel shows the median difference between observed and best-fit model magnitudes
in the $i$ band. The bottom right panel is analogous to the top right panel, except that only the
subset of stars also detected by 2MASS ($K<14$) and with full SDSS-2MASS fits (fixed $R_V$) 
are used. Note the much better agreement with the SFD values than in the top right panel. For
more details, please see \S\ref{sec:ngp}.}
\label{panelsLambertNGP}
\end{figure}

\begin{figure}[!t]
\epsscale{0.8}
\vskip 0.1in
\plotone{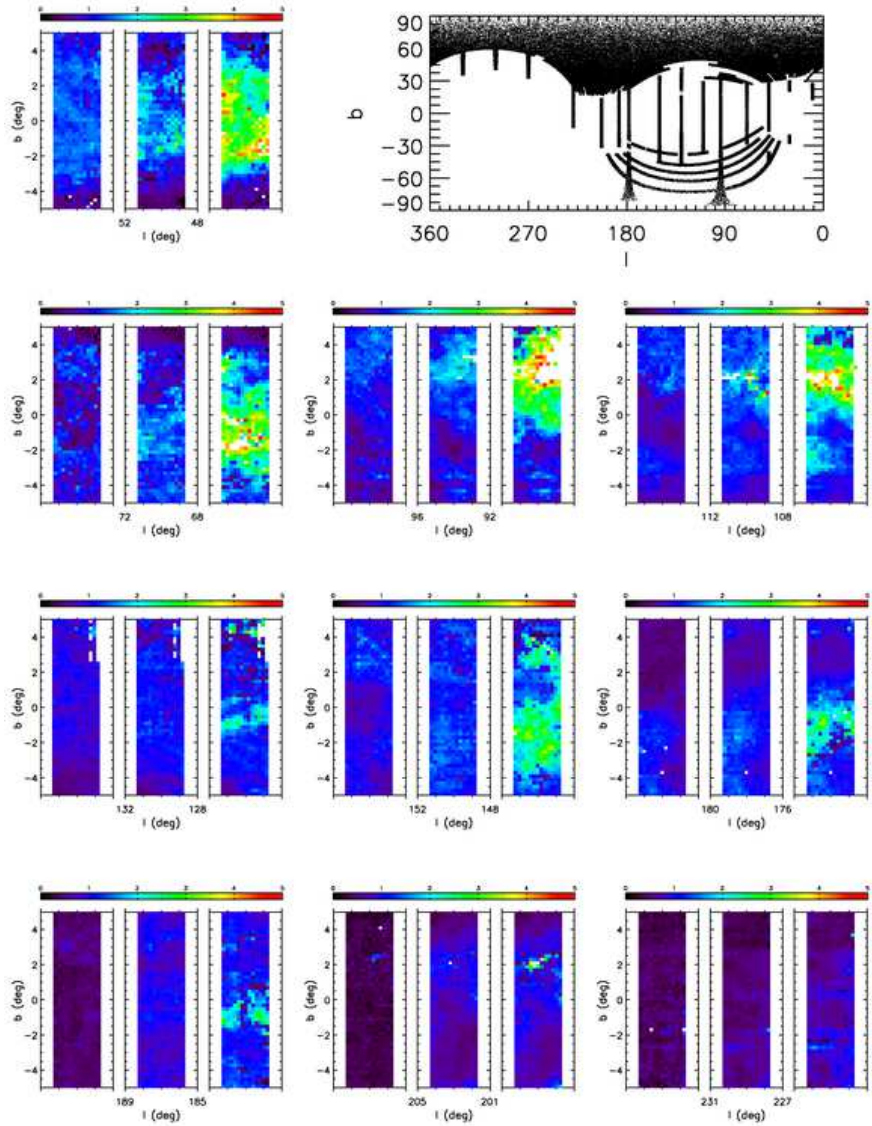}
\vskip 0.2in
\caption{The color-coded maps show the best-fit $A_r$ based on SDSS 
data for the ten analyzed SEGUE stripes. Each stripe is limited to the range 
of $|b|<5^\circ$. A fixed $R_V=3.1$ is assumed. 
The legend above each panel shows the color scale, and each 12$\times$12 
arcmin$^2$ pixel shows the median $A_r$. For each stripe, three 
distance ranges are shown: 0.3-0.6 kpc (left), 1-1.5 kpc (middle) 
and 2-2.5 kpc (right). It is assumed that all stars are on 
main sequence when estimating distances. Only stars with best-fit 
$\chi^2_{pdf}<2$ and outside the red giant region (selected here by 
$A_r<1.5+1.5\,D_{\rm kpc}$) are used for the plot. The top right panel 
shows the sky coverage of the full analyzed dataset.}
\label{Fig:ArSDSSdSlicesAll}
\end{figure}

\begin{figure}[!t]
\vskip 0.in
\epsscale{0.8}
\plotone{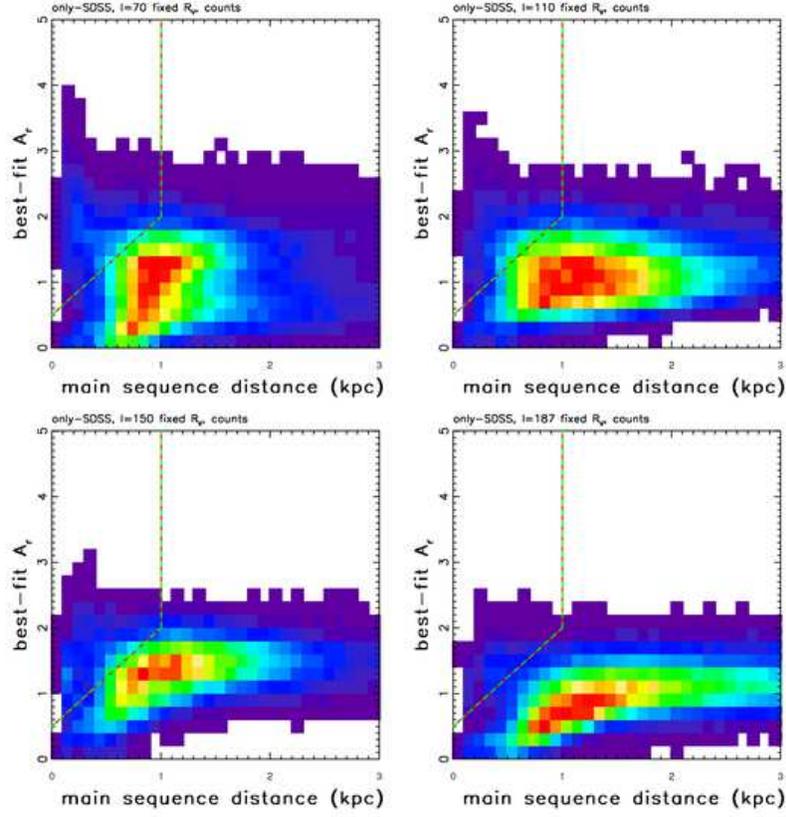}
\vskip -0.7in
\caption{The counts of stars in the only-SDSS case best-fit $A_r$ vs. best-fit main 
sequence distance diagram for four SEGUE strips (top left: $l=70^\circ$, top right: $l=110^\circ$, 
bottom left: $l=150^\circ$, bottom right: $l=187^\circ$; for all panels $|b|<5^\circ$). Only stars 
with $\chi^2_{\mathrm{pdf}}<2$, $r<19$ and error in the $u$ band below 0.05 mag are used. Counts 
are normalized to the maximum value and color coded on the same linear scale, from blue (low) to 
red (high). The two dashed lines mark a region dominated by red giant stars (the top left corner).}
\label{Fig:DAcounts_onlySDSS}
\end{figure}

\begin{figure}[!t]
\vskip 0.2in
\epsscale{0.8}
\plotone{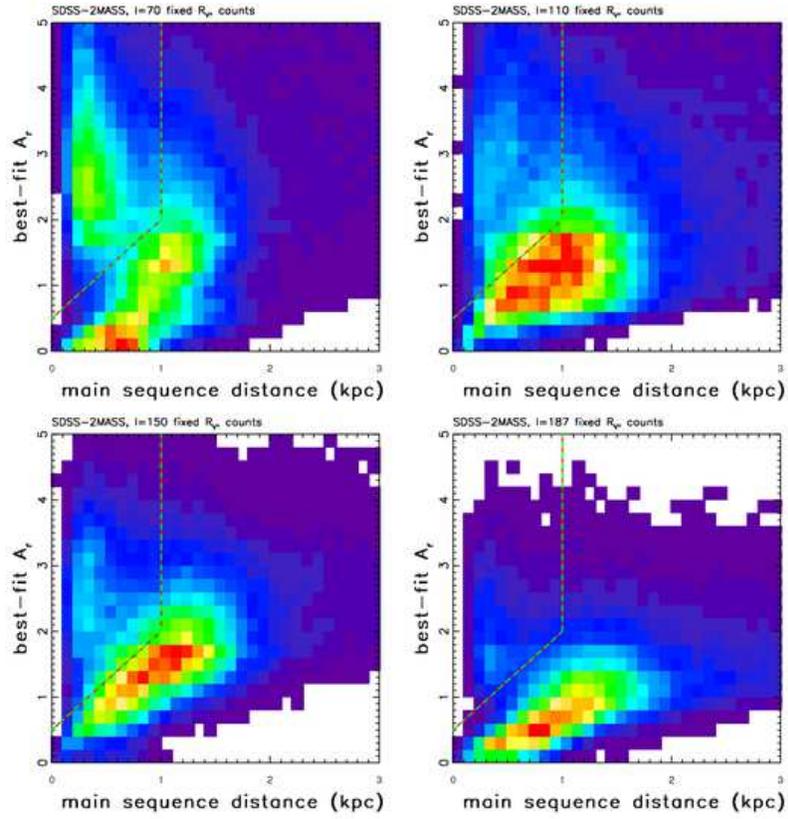}
\vskip -0.7in
\caption{
Analogous to Figure~\ref{Fig:DAcounts_onlySDSS}, except for best-fits based on SDSS-2MASS sample
(only stars with $\chi^2_{\mathrm{pdf}}<2$ and $K<15$ on Vega scale are used). Note the larger 
fraction of red giant stars in the top left corner, and a smaller distance limit, compared 
to Figure~\ref{Fig:DAcounts_onlySDSS} and that the fraction of giants decreases with 
Galactic longitude.}
\label{Fig:DAcounts_SDSS2MASS}
\end{figure}

\begin{figure}[!t]
\epsscale{0.8}
\plotone{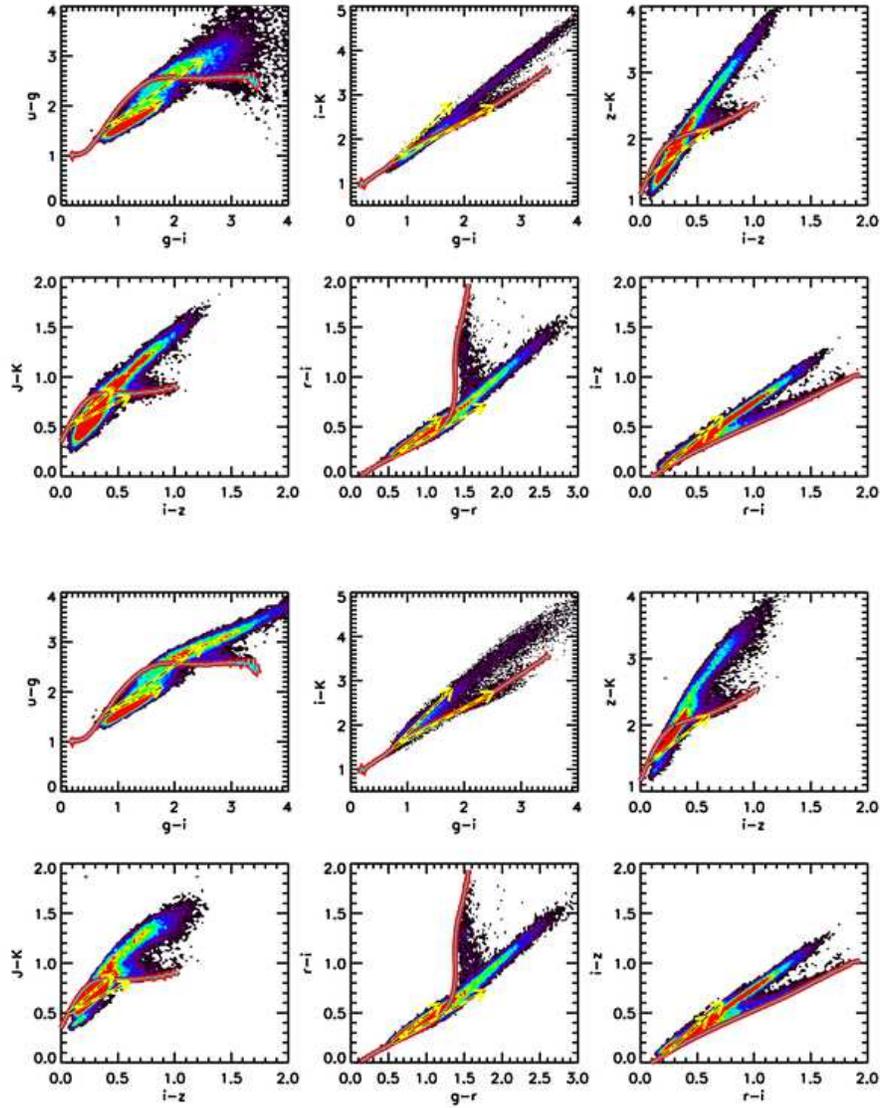}
\vskip 0.2in
\caption{A comparison of six SDSS-2MASS color-color diagrams using data from the SEGUE 
$l\sim 110^\circ$ strip (the top six panels; same as the top six panels in Figure~\ref{Fig:mock}, except
that here only stars with $\chi^2_{\mathrm{pdf}}<2$ are used), and the best-fit model colors based on  
SDSS-2MASS dataset (the bottom six panels, in the same order). The thick lines show the 
\citet{Covey07} empirical SED library and illustrate the 
morphology of the same diagrams observed at high Galactic latitudes. The two sets of diagrams are 
encouragingly similar: fits to intrinsic stellar SED and dust extinction on per star basis are
capable of reproducing the morphology of observed diagrams in highly dust-extincted regions.}
\label{Fig:CMD_SEGUEl110_dataVSfixedRv}
\end{figure}

\begin{figure}[!t]
\epsscale{1.0}
\plotone{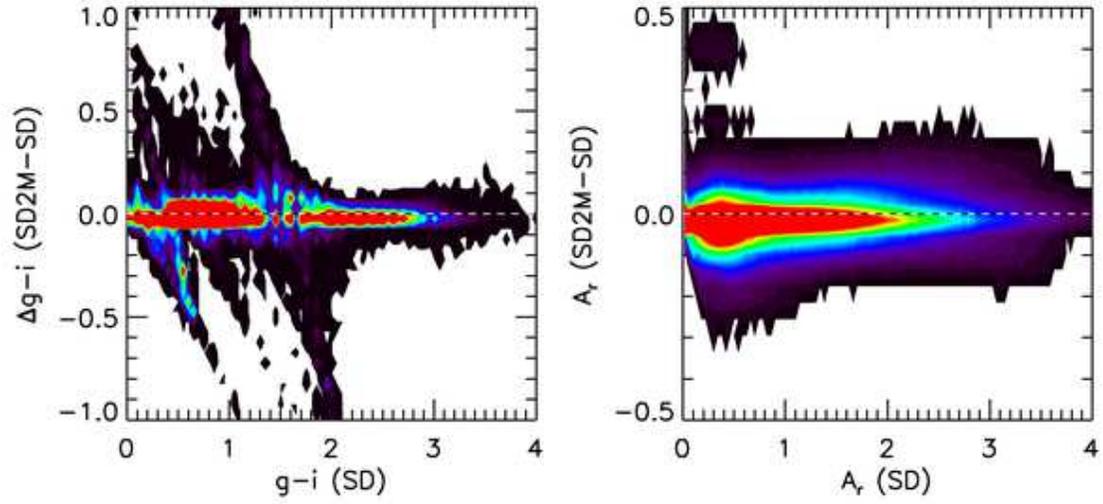}
\vskip -0.1in
\caption{A comparison of the best-fit $g-i$ (left panel) and $A_r$ (right panel)
values obtained with a fixed $R_V=3.1$ for SDSS-2MASS sample from the SEGUE 
$l\sim110^\circ$ stripe, using two different fitting methods. 
The abscissae show the best-fit values obtained using only-SDSS dataset (four fitted
colors), and the ordinates correspond to the residuals of the SDSS-2MASS (seven fitted colors) 
minus the only-SDSS datasets. The number density of stars increases linearly from black to blue 
to red. The dashed lines are added to guide the eye. }
\label{Fig:compareArGI_onlySDSSvsSDSS2MASS}
\end{figure}

\begin{figure}[!t]
\epsscale{0.6}
\plotone{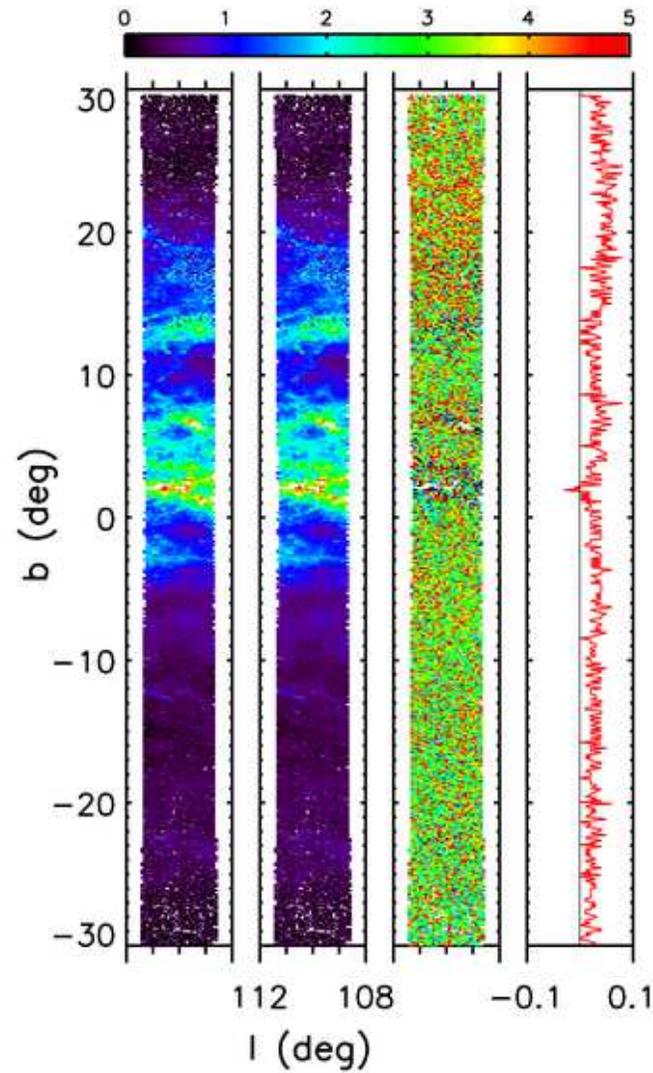}
\vskip 0.in
\caption{Analysis of the differences in the best-fit $A_r$ between fits based on 
SDSS-2MASS dataset (first panel from the left) and those based on only-SDSS data
(second panel). Only stars with best-fit $\chi^2_{pdf}<2$, $r<20$ and main-sequence
distance $0.5-1$ kpc are used for the plot. The top legend shows the coloring code for 
these two panels, and each 6$\times$6 arcmin$^2$ pixel shows the median $A_r$ for stars with 
$\chi^2_{\mathrm{pdf}}<2$. The third panel shows the median difference between the two best-fit 
$A_r$ values (the second panel minus the first panel), with the color coding using the same 
palette, {\it except that the limits are $\pm0.1$ mag}. The fourth panel shows the median 
difference in $A_r$ (i.e., the third panel) for 0.2$^\circ$ wide bins of Galactic latitude.}
\label{Fig:ArSDSSvsSDSS2MASS}
\end{figure}

\begin{figure}[!t]
\epsscale{0.55}
\plotone{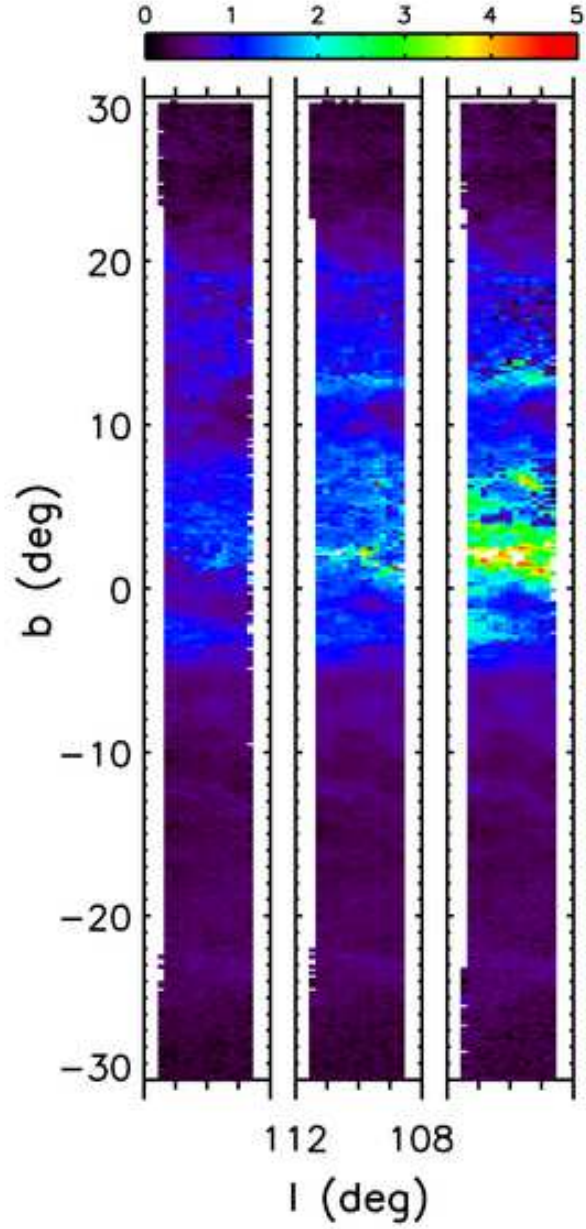}
\epsscale{1.0}
\vskip 0.1in
\caption{The color-coded maps show the best-fit $A_r$ based on only-SDSS 
dataset for the SEGUE $l\sim 110^\circ$ strip. The legend shows the color scale, 
and each 12x12 arcmin$^2$ pixel shows the median $A_r$. The three panels correspond
to main-sequence distance range: $0.3-0.6$ kpc (left), $1-1.5$ kpc (middle) and 
$2-2.5$ kpc (right). Only stars with best-fit $\chi^2_{pdf}<2$ and outside the red giant
region (selected here by $A_r<1.5+1.5\,D_{\rm kpc}$) are used for the plot.}
\label{Fig:ArSDSSdistSlicesSEGUE110}
\end{figure}

\begin{figure}[!t]
\epsscale{0.8}
\plotone{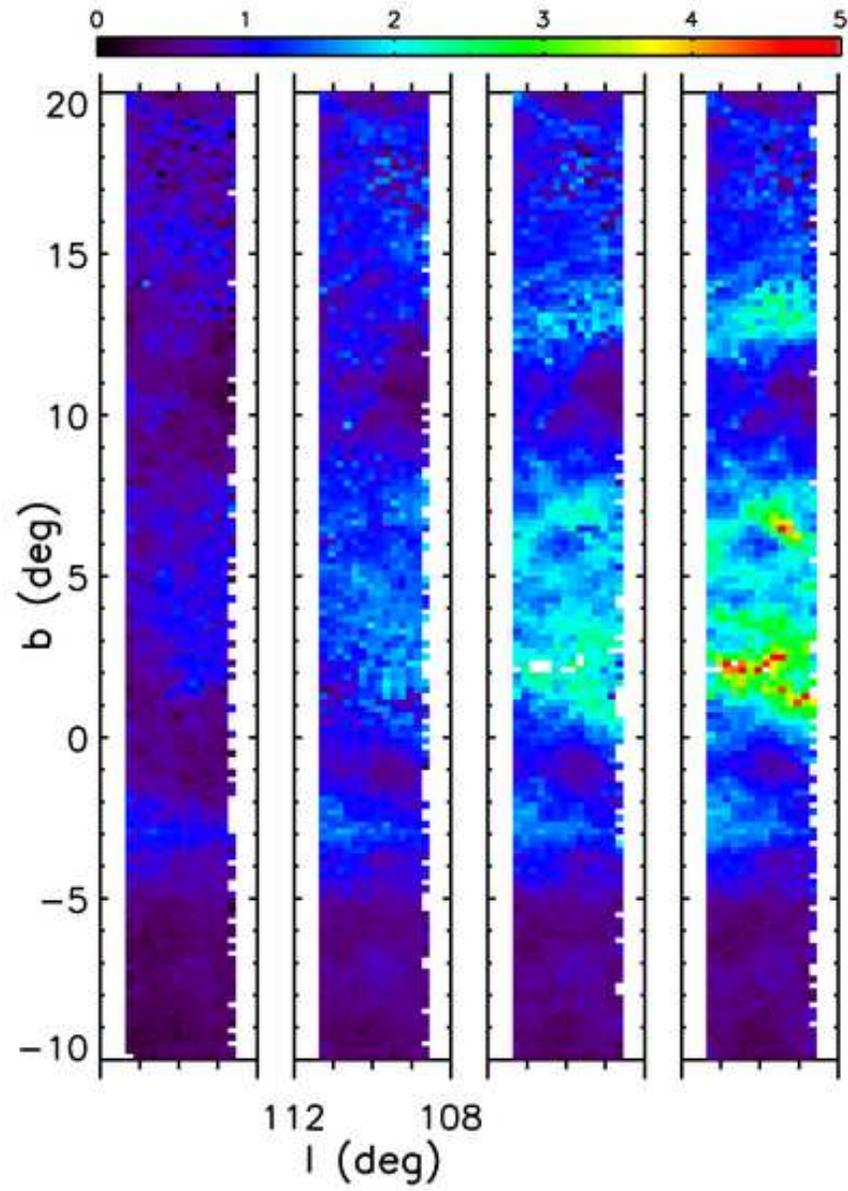}
\vskip 0.1in
\caption{Analogous to Figure~\ref{Fig:ArSDSSdistSlicesSEGUE110}, 
except using the SDSS-2MASS dataset and different distance slices (left to 
right: $0.1-0.5$ kpc, $0.5-0.7$ kpc, $0.7-0.9$ kpc, $0.9-1.1$ kpc). Note the
abrupt increase in $A_r$ for stars towards $b\sim+2^\circ$ that are more distant than 0.9 kpc.}
\label{Fig:ArSDSSdSlices110zi}
\end{figure}

\begin{figure}[!t]
\epsscale{0.6}
\plotone{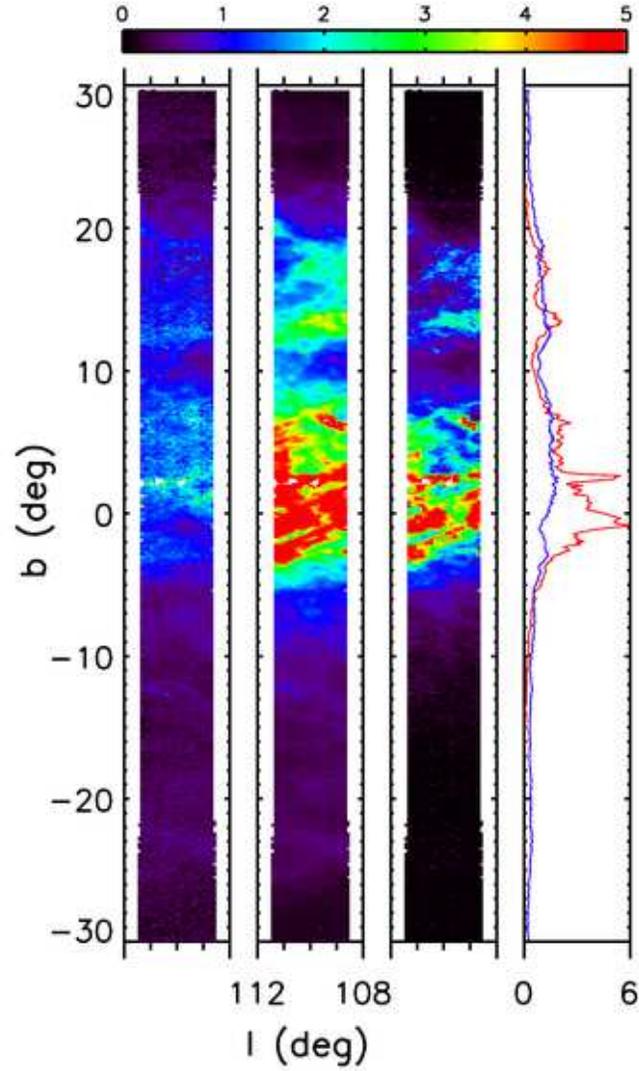}
\epsscale{1.0}
\vskip 0.2in
\caption{Analysis of the differences between best-fit $A_r$ values (left 
panel, based on only-SDSS data; SDSS-2MASS version looks similar) and 
the SFD values (second panel) for stars with $\chi^2_{pdf}<2$ and main-sequence
distance in the range $0.8-1.2$ kpc. The third panel shows the difference of the two 
$A_r$ values (the second panel minus the first panel). Each 6$\times$6 arcmin$^2$ pixel 
in the first three panels is color coded according to the top legend. The fourth panel 
shows the median best-fit $A_r$ (blue line) and the median SFD value (red line) for 
0.2$^\circ$ wide bins of Galactic latitude. If the SFD maps are correct, then the dust 
structures discernible in the two right panels at $b\sim0^\circ$ and $b\sim+2^\circ$ 
must be more distant than $\sim$1 kpc. This conclusion is independently confirmed
for the latter dust cloud in Figure~\ref{Fig:ArSDSSdSlices110zi}.}
\label{Fig:ArSDSScompSFD}
\end{figure}

\begin{figure}[!t]
\epsscale{0.75}
\hskip 0.0in
\plotone{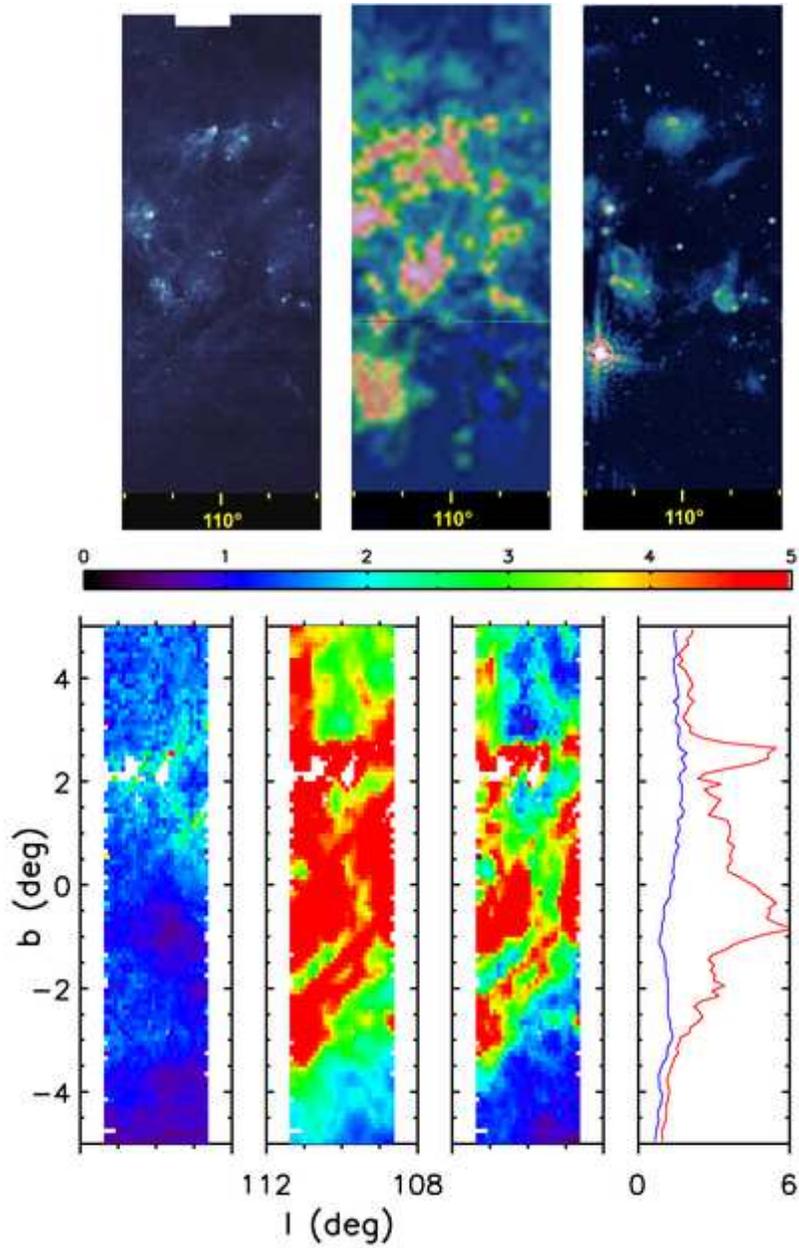}
\vskip 0.1in
\caption{The bottom four panels show the $|b|<5^\circ$ subregion 
of the panels shown in Fig.~\ref{Fig:ArSDSScompSFD}. The top three
panels show the mid-IR (left), CO (middle) and radio continuum (right)
maps on approximately the same scale (obtained using ``The Milky Way 
Explorer'' by Kevin Jardine). The few small irregular white regions in the bottom three
maps do not contain any stars with good photometry. Assuming that the SFD map is 
not grossly incorrect, the dust extinction determined here implies that most of the molecular 
cloud structures seen in the top middle panel must be more distant than $\sim$1 kpc.}
\label{Fig:KeiraL110}
\end{figure}

If there is a significant discrepancy between the {\it shape} of assumed CCM extinction curve for 
$R_V=3.1$ and that required by SEGUE data, photometric residuals between observed and best-fit 
magnitudes should show a correlation with best-fit $A_r$. Indeed, the failure to pass this test 
has revealed that our first instance of fitting erroneously used the O'Donnell extinction curve (due to an error in
``metadata management''). In this case, the photometric residuals (data ``minus'' model) in the 
$i$ band showed a highly statistically significant correlation $\Delta i = -0.015 \, A_r$, which implied 
that the adopted $C_\lambda$ value in the $i$ band was too large by 0.015 (the results for other bands 
did not require a change of $C_\lambda$). The analysis of used $C_\lambda$ values clearly placed
them on top of the O'Donnell model in the right panel in Figure~\ref{Fig:Ed2}, while the revised
value moved the constraint towards the CCM model curves. After our second fitting iteration
that correctly incorporated the CCM model, we regressed photometric residuals and best-fit $A_r$ 
again and found much smaller residuals: $\Delta i = -0.005 \, A_r$ and $\Delta z = 0.003 \, A_r$.
For no other bands were the slopes larger than statistical measurement errors of at most 0.001.
These two relatively small corrections of $C_\lambda$ in the $i$ and $z$ bands result in a shift in the 
right panel in Figure~\ref{Fig:Ed2} {\it away from the CCM model curves, and to a point between the
constraints obtained using stellar locus method for Stripe 82 and the northern Galactic hemisphere!}
That is, the required $C_\lambda$ modifications cannot be accomplished by adopting a CCM model
curve for a different $R_V$ (nor using any of the other two considered models). Hence, SEGUE data 
``knew'' that (independent) empirical constraints on the shape of dust extinction curve from the 
high-latitude sky are better than the CCM model for  $R_V=3.1$! 

The above analysis of photometric residuals shows that there is no a priori reason to expect a 
significant departure
from the canonical $R_V=3.1$ value when $R_V$ is considered a free fitting parameter. Nevertheless,
it is possible that {\it localized} regions in the Galactic disk have a different $R_V$ distribution,
and given the unique nature of our sample, such a study is worthwhile. The analysis of fitting 
results for a mock catalog described in \S\ref{sec:redgiantsmock} showed that only the SDSS-2MASS 
dataset can be expected to provide useful constraints on $R_V$, and this is the fitting case
analyzed here (for completeness, public data distribution includes also only-SDSS case). 

\begin{figure}[!t]
\epsscale{1.0}
\plotone{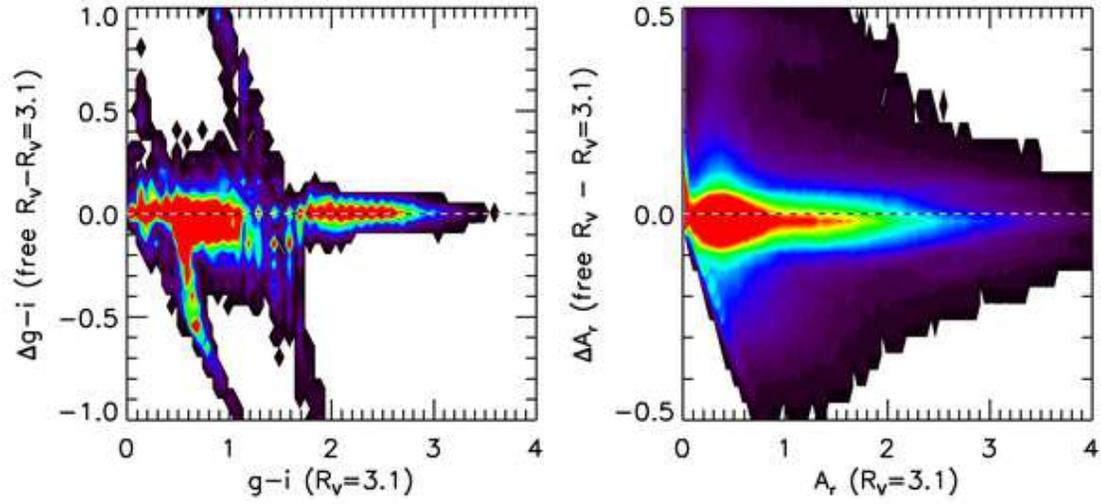}
\vskip -0.1in
\caption{A comparison of the best-fit $g-i$ (left panel) and $A_r$ (right panel)
values for two different treatments of $R_V$, for stars in the $l=110^\circ$ SEGUE strip
(using SDSS-2MASS data). Only stars with best-fit $\chi^2_{pdf}<2$, $r<20$ and
$K<13.9$ (Vega) are used for the plot. The abscissae show the best-fit values obtained for a fixed 
$R_V=3.1$ and the ordinates correspond to the residuals of the differences in the best-fit values 
when $R_V$ is treated as a free-fitting parameter (``free $R_V$'' $-$ ``$R_V$=3.1''). The number density of 
stars increases from black to blue to red. The dashed lines are added to guide the eye. }
\label{Fig:compareArGI_freeVSfixedRv}
\end{figure}

\begin{figure}[!t]
\epsscale{0.8}
\plotone{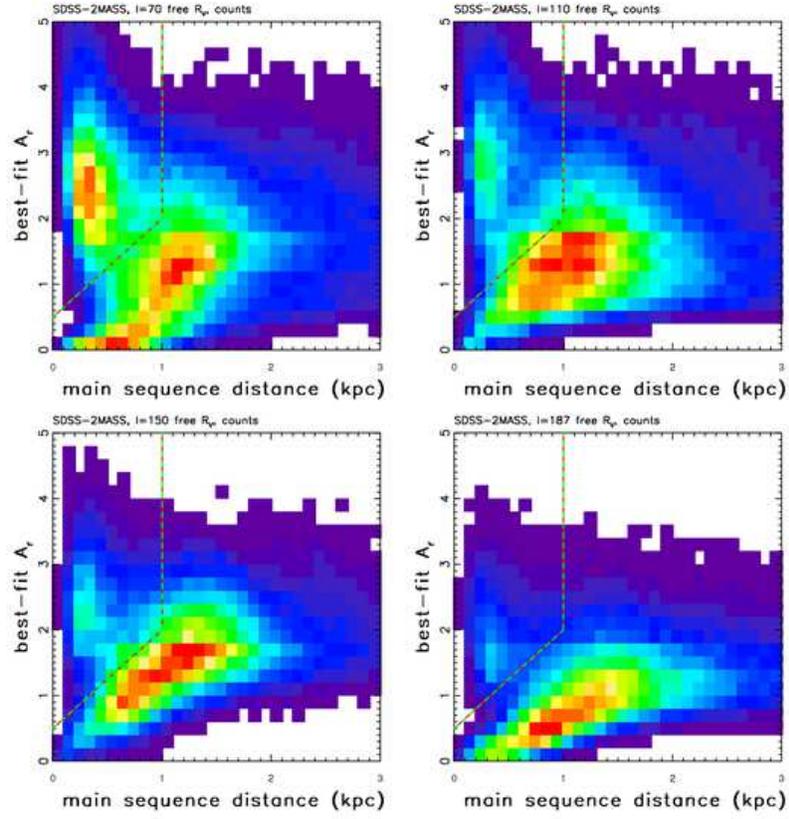}
\vskip -0.7in
\caption{The $A_r$ map analogous to Figure~\ref{Fig:DAcounts_SDSS2MASS}, except that $R_V$ is
treated as a free parameter. Only stars with $|b|<5^\circ$, 
$\chi^2_{\mathrm{pdf}}<2$, $r<19$, and $K<15$ (on Vega scale) are used. 
}
\label{Fig:DAcounts_SDSS2MASS_freeRv}
\end{figure}

\begin{figure}[!t]
\epsscale{1.0}
\plotone{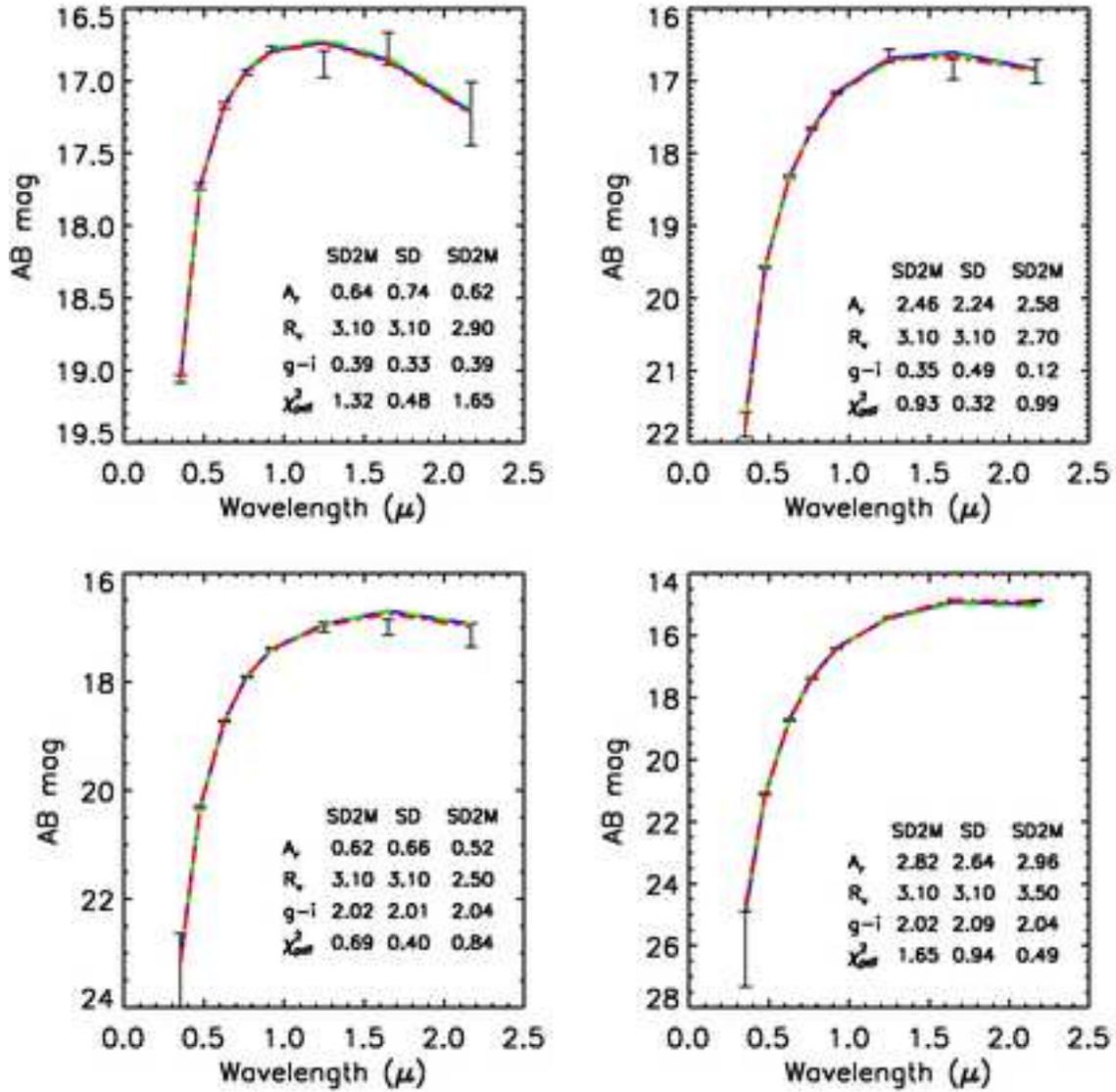}
\vskip -0.0in
\caption{A comparison of three different types of best-fit
SEDs: using only SDSS data with fixed $R_V=3.1$ (blue
line), and using joint SDSS-2MASS dataset with fixed $R_V$ (green 
line) and with free $R_V$ (red line).  As demonstrated by the 
similarity of best-fit lines, the differences in best-fit parameters,
listed in each panel, are due to degeneracies between intrinsic
stellar color, amount of dust and $R_V$. The shown cases
correspond to blue and red stars (top row vs. bottom row), and
small and large $A_r$ (left column vs. right column).}
\label{Fig:FRVSED}
\end{figure}

A comparison of best-fit intrinsic colors and $A_r$ between fixed-$R_V$ and free-$R_V$ cases
is shown in Figure~\ref{Fig:compareArGI_freeVSfixedRv}. While for some sources results can differ 
substantially, the fraction of discrepant sources is small. The resulting distribution of sources in 
the $A_r$ vs. distance diagram, shown in Figure~\ref{Fig:DAcounts_SDSS2MASS_freeRv} for free-$R_V$ 
case, is similar to that based on fixed-$R_V$ case (compare to Figure~\ref{Fig:DAcounts_SDSS2MASS}). 
A comparison of best-fits results for fixed-$R_V$ and free-$R_V$ cases shown in Figure~\ref{Fig:FRVSED} 
reveals that fit residuals are not significantly smaller when $R_V$ is free. 

\begin{figure}[!t]
\epsscale{0.8}
\plotone{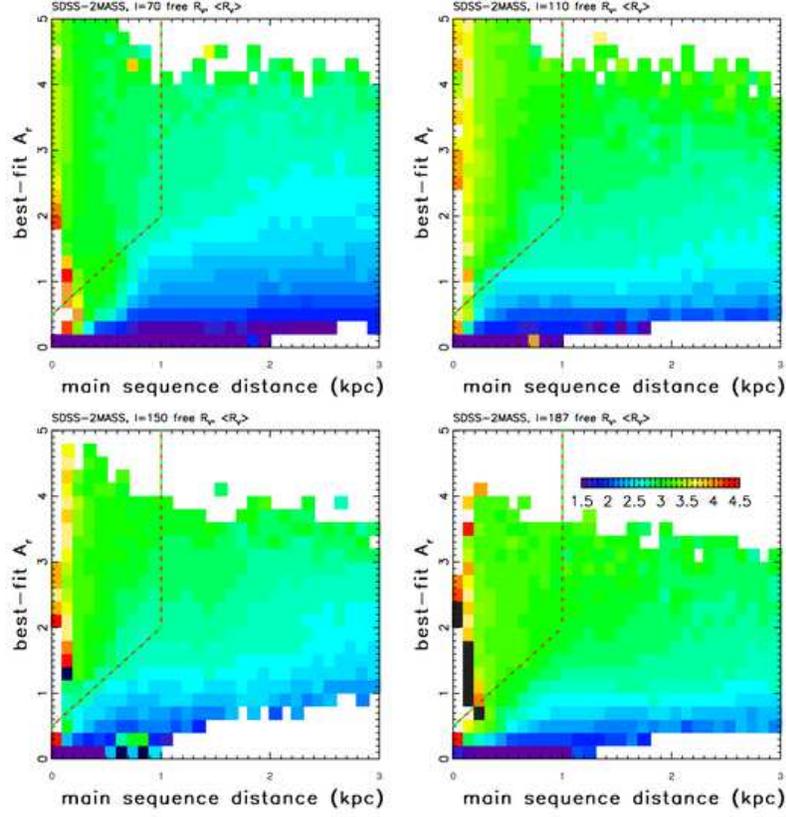}
\vskip -0.7in
\caption{Similar to Figure~\ref{Fig:DAcounts_SDSS2MASS}, except that $R_V$ is
treated as a free parameter, and the color-coded map shows the median value of
$R_V$ (ranging from blue for $R_V=1.5$ to red for $R_V=4.5$, green corresponds 
to $R_V=3$; see the legend in the bottom right panel). Only stars with $|b|<5^\circ$, 
$\chi^2_{\mathrm{pdf}}<2$, $r<19$, and $K<15$ (on Vega scale) are used. 
Note that red giant stars (top left corner) have consistently larger values of $R_V$, and
that consistently $R_V<3$ when $A_r < 1$ for main sequence stars. In other
regions in this diagram where $R_V$ is determined robustly, $R_V=3.1$ cannot be 
ruled out in any of the ten SEGUE stripes at a precision level of $\sim0.1-0.2$.  
}
\label{Fig:DAmedRv_SDSS2MASS_freeRv}
\end{figure}

\begin{figure}[!t]
\epsscale{0.95}
\plotone{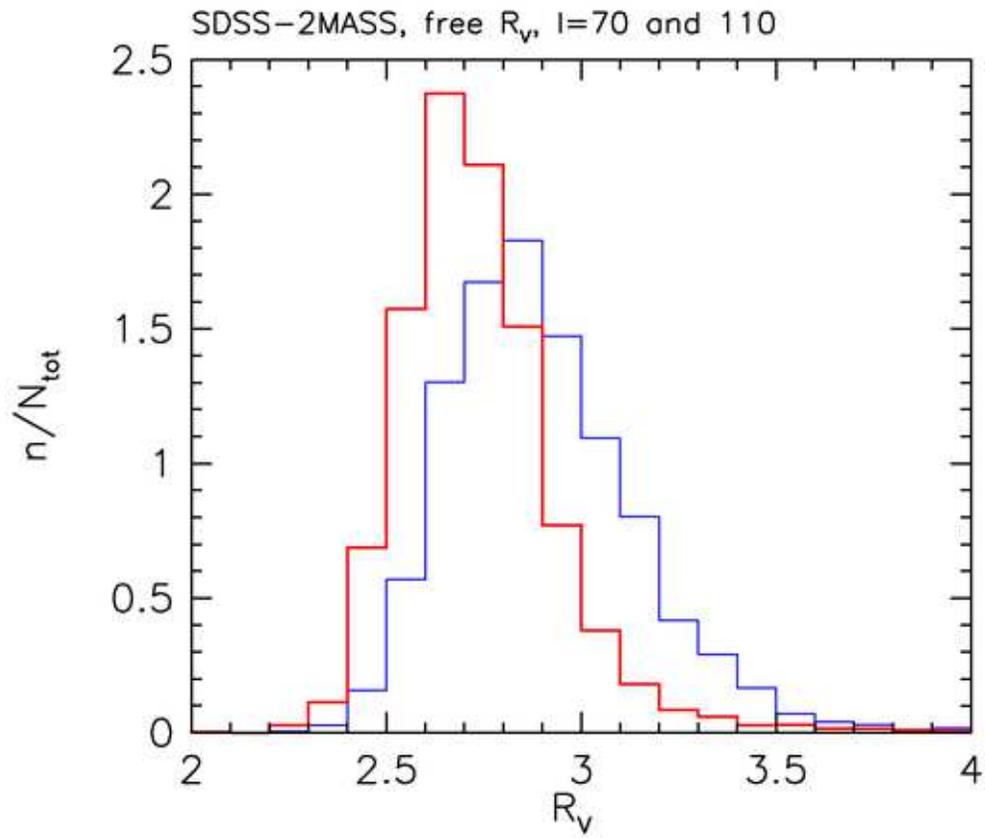}
\vskip -3.0in
\caption{A comparison of the best-fit $R_V$ values for SDSS-2MASS free-$R_V$ case
and stars with distances in the 1.0-2.5 kpc range and $A_r>2.5$, selected from 
$l=70^\circ$ (red, left histogram) and $l=110^\circ$ (blue, right histogram) stripes 
(other selection criteria are the same as for stars plotted in 
Figure~\ref{Fig:DAmedRv_SDSS2MASS_freeRv}).}
\label{Fig:RvHist_SEGUE_l110}
\end{figure}

The median $R_V$, as a function of the position in the $A_r$ vs. distance diagram, is shown in 
Figure~\ref{Fig:DAmedRv_SDSS2MASS_freeRv} for four representative SEGUE stripes. As concluded 
in \S\ref{sec:redgiantsmock}, the $R_V$ results for $A_r<1$ are expected to be biased low. For $A_r>2$, 
and outside the red giant region, the median $R_V$ does not deviate appreciably from its canonical 
value. A more quantitative description of this behavior is shown in Figure~\ref{Fig:RvHist_SEGUE_l110}.
For stars selected by 1 kpc $<D<$ 2.5 kpc and $A_r>2.5$ from $l=110^\circ$ stripe, the median $R_V$ is 
2.90, with a mean of 2.95 and an rms of 0.22 (determined from the interquartile range, the sample 
size is $\sim$9,000 stars). Given various systematic uncertainties that cannot be smaller than 0.1-0.2,
as well as expected random errors ($\sim$0.1), the median $R_V$ is consistent with the canonical
value of 3.1. We note that the width of the $R_V$ histogram is about twice as large as the width
of $R_V$ determined using a fixed-$R_V$ mock sample. Assuming that both widths are reliable, which
may not be strictly quantitatively true, the implied instrinsic scatter in $R_V$ for $l=110^\circ$ stripe 
is $\sim0.2$. Results from other stripes are similar, with the median $R_V$ showing a scatter of 
about 0.1. To illustrate this $R_V$ variation, Figure~\ref{Fig:RvHist_SEGUE_l110} also shows the 
$R_V$ distribution for $l=70^\circ$ stripe, which has a median $R_V$ of 2.80, and an rms of 0.15. 
We have tested for a possibility that the variation of the $R_V$ distribution among stripes is due 
to calibration problems by comparing the median residuals between observed and best-fit magnitudes 
for each SDSS camera column (twelve per stripe) and filter (including 2MASS filters). We did not 
find any evidence for photometric calibration errors larger than 0.01 mag.

As shown in Figure~\ref{Fig:DAmedRv_SDSS2MASS_freeRv}, candidate red giant stars (top left
corner) have consistently somewhat larger values of $R_V$ (by about 0.2-0.4) than the typical star in the 
sample. Given that they are expected to be at much larger distances than main sequence stars, it is possible 
that they sample different types of dust. However, given fairly large range of longitudes sampled by
SEGUE stripes, this conclusion would imply that the dust in the Solar neighborhood (within
1-2 kpc) has anomalously low $R_V$. A more plausible explanation for increased $R_V$ is 
a bias due to slight differences in SEDs between red giants and main sequence stars. 
A preliminary analysis of the SDSS spectroscopic sample has revelead that spectroscopically-confirmed 
giants show an offset from the Covey et al. locus in the seven-dimensional 
SDSS-2MASS color space. Such an offset is, at least in principle, capable of inducing a bias
in best-fit $R_V$. A detailed analysis of this bias and differences in SEDs between main
sequence stars and red giants will be presented elsewhere. For the remainder of analysis 
presented here, we simply exclude candidate red giant stars. 

\begin{figure}[!t]
\epsscale{0.7}
\plotone{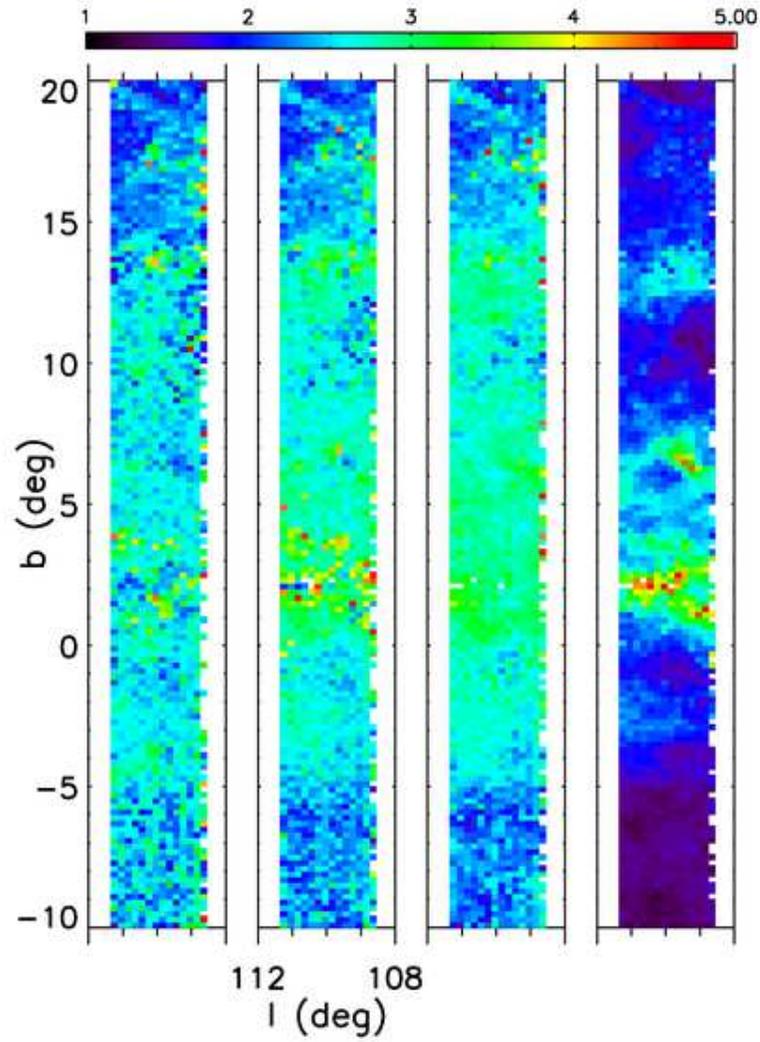}
\epsscale{1.0}
\vskip 0.2in
\caption{The first three panels show the median $R_V$ obtained using SDSS-2MASS 
sample for the SEGUE $l=110^\circ$ strip, and for distance range $0.5-0.7$ kpc (first), 
$0.7-0.9$ kpc (second), and $0.9-1.1$ kpc (third). Only stars outside the 
``red giant'' region, see Figure~\ref{Fig:DAcounts_SDSS2MASS}, and with 
$\chi^2_{pdf}<2$, $r<21$ and $K<14.3$ (Vega) are used for the plot. The pixel size 
is 6$\times$6 arcmin$^2$, and the $R_V$ coloring scheme is shown at the top. 
The fourth panel shows for reference the best-fit $A_r$, for the distance slice $0.9-1.1$ kpc.
Note that $R_V$ is not reliable for $A_r<2$ (black and blue regions in the fourth panel).}
\label{Fig:sd2m_rvmap}
\end{figure}

A cross-section of the three-dimensional $R_V$ map is shown in Figure~\ref{Fig:sd2m_rvmap}
(recall that the $R_V$ values are not reliable in regions of small $A_r$; see the rightmost panel
for reference). For most of high-$A_r$ regions, the median values are consistent with the
canonical values.

\section{The Three-Dimensional Distributions of Dust and Stars} 

Best-fit stellar distance and extinction along the line of sight, $A_r$, determined
here can be used to infer the three-dimensional distributions of dust and stars. 
The determination of these distributions is not straightforward. In case of 
stars, complicated flux-color-extinction selection effects must be taken into
account in order to obtain unbiased distributions. This analysis is best done
with the aid of mock catalogs, such as those produced by {\it Galfast}. In case
of dust, the complexity is further increased because the {\it integral} of
dust volume density along the line of sight is constrained, and not the density itself. 
To translate these constraints into a positive dust volume density 
(more precisely, extinction per unit length as a function of position in the Galaxy), 
a careful statistical treatment of all errors and selection effects is mandatory. 
Since the full analysis is obviously far beyond this preliminary investigation, 
we illustrate the potential of our dataset with two simplified analysis examples.

\subsection{The Spatial Distribution of Dust} 
\label{sec:SpatDistDust} 

\begin{figure}[!t]
\epsscale{0.6}
\plotone{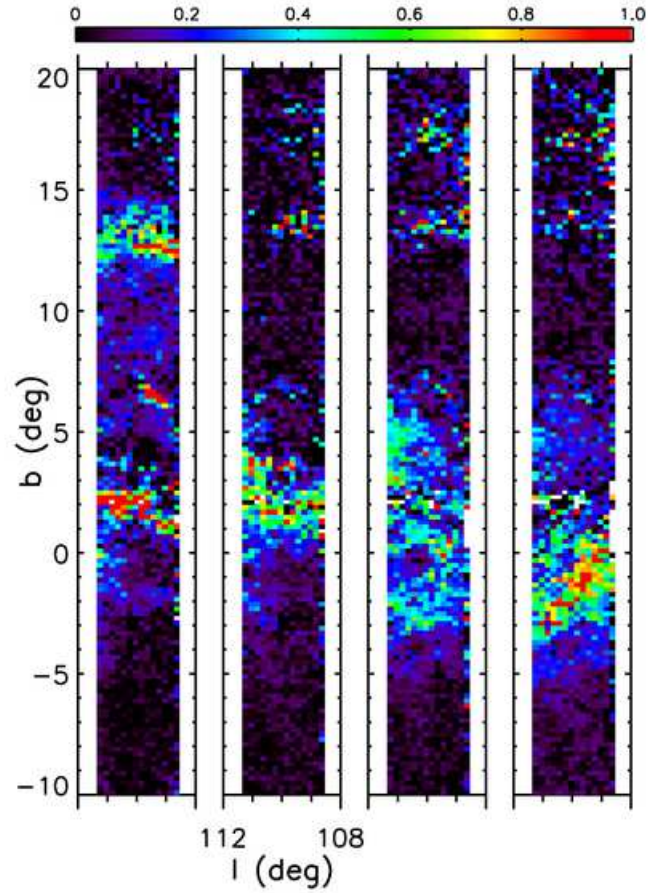}
\epsscale{1.0}
\vskip 0.1in
\caption{Illustration of the three-dimensional dust distribution for SEGUE stripe $l\sim110^\circ$
at mean distances of 1.0, 1.5, 2.0 and 2.5 kpc, using only-SDSS sample and fixed-$R_V$ fits.
Unlike other figures that show the median $A_r$ {\it along the line of sight}, this figure shows the 
{\it differences} in the median $A_r$ (per 12$\times$12 arcmin$^2$ pixel) for samples at distances 
between the quoted distance and limiting distances 0.5 kpc larger and smaller than the mean distance 
(e.g., the first panel shows the difference between the median $A_r$ for 0.5-1.0 kpc and 1.0-1.5 kpc 
subsamples). It is easily discernible that the dust structures observed at $b\sim2^\circ$ and 
$b\sim13^\circ$ are confined to 1-1.5 kpc distance range, while the structure seen at $-3^\circ<b<0^\circ$ 
is due to dust at a distance of $\sim2.5$ kpc (an analogous panel for a mean distance of 3.0 kpc 
shows that this structure is mostly confined to smaller distances). Note that the linear extent 
perpendicular to the line of sight of a given angular size is 2.5 times larger in the last than in 
the first panel.}
\label{Fig:sd2m_diffArDslices}
\end{figure}

A coarse approximate map of the spatial distribution of dust in a given distance range can be obtained 
by subtracting two median $A_r$ maps corresponding to the distance limits of the chosen
range. This method is not statistically optimal, but it suffices for simple visualization.
Figure~\ref{Fig:sd2m_diffArDslices} shows the result of such analysis for mean bin distances 
of 1.0, 1.5, 2.0 and 2.5 kpc, with limiting bin distances 0.5 kpc larger and smaller than the 
mean distance. 

It is easily discernible that the dust structures observed at $b\sim2^\circ$ and $b\sim13^\circ$ are 
confined to 1-1.5 kpc distance range, while the structure seen at $-3^\circ<b<0^\circ$ is due to 
dust at a distance of $\sim2.5$ kpc and subtends $<1$ kpc along the line of sight (an analogous 
panel for a mean distance of 3.0 kpc shows that this structure is mostly confined to smaller distances). 
As discussed earlier, this ability to ``bracket'' distances to dust clouds, and in turn to molecular 
clouds, is an important feature of our dataset. 

\begin{figure}[!t]
\epsscale{0.8}
\plotone{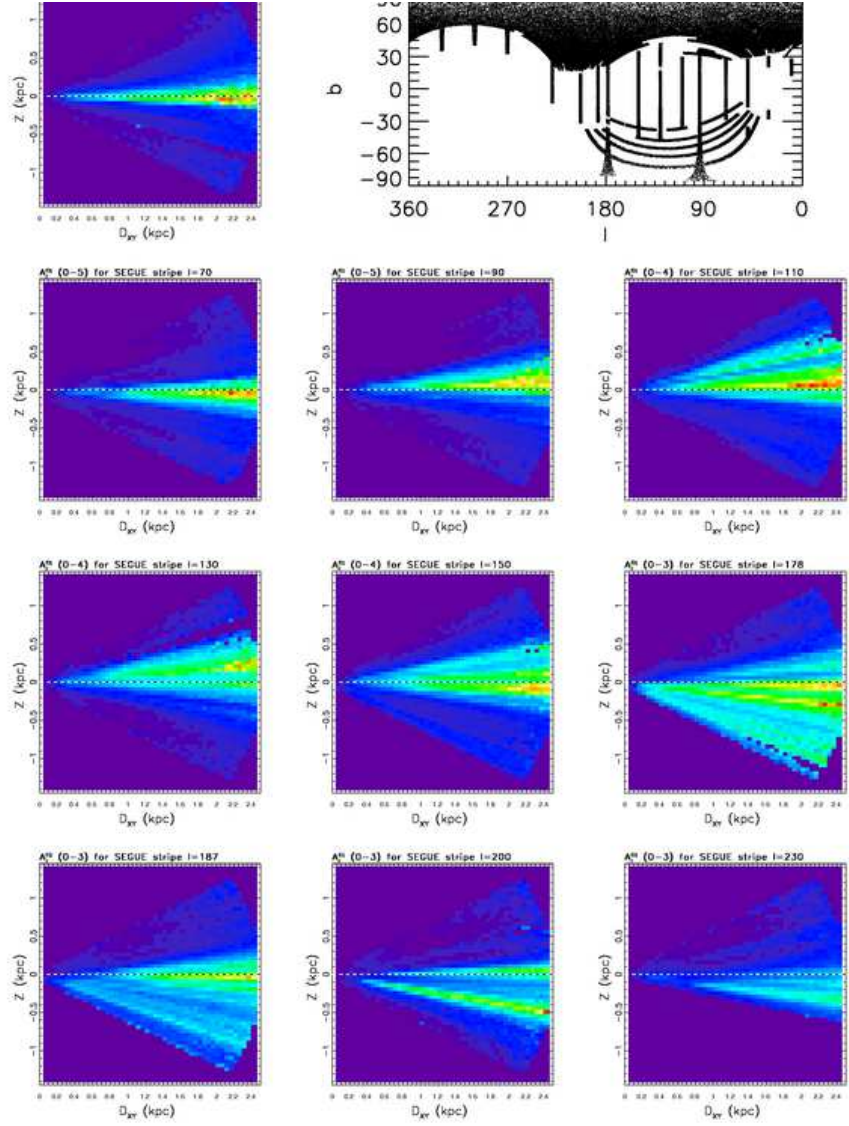}  
\vskip -0.2in
\caption{The median best-fit $A_r$ (extinction along the line of sight) is shown as 
a function of distance from the Galactic plane, $Z$, and distance along the plane, 
$D_{xy}$, for 10 SEGUE stripes (this is {\bf not} a cross-section of three-dimensional
dust distribution!). The best-fit $A_r$ are based on the SDSS-2MASS dataset and 
fixed-$R_V$ fitting case, for stars with $\chi^2_{pdf}<2$ and $K<15$ (Vega).
Each pixel is 50$\times$50 pc$^2$ and subtends 2.5 deg wide stripe in the perpendicular 
(longitude) direction. The color scheme increases linearly from blue to red with a varying 
maximum value: 5 for the first three panels, 4 for the next three, and 3 for the last four panels.}
\label{Fig:ZDcutArPanels}
\end{figure}

Another projection of our dataset, the median $A_r$ as a function of spatial coordinates, is 
shown in Figure~\ref{Fig:ZDcutArPanels} for all ten SEGUE stripes. Aside from the fact that
data for each stripe also resolve the third direction (Galactic longitude), this projection illustrates
the integral constraint on the spatial distribution of dust. For each pixel, or a star in general
case, the measured $A_r$ contains (noisy) information about the dust distribution along 
the line connecting this pixel/star and the observing point (the origin in this figure). 
With an appropriate model description of dust distribution, either parametric or non-parametric,
these $A_r$ maps can be used to constrain the model (for an example of similar analysis, 
see \citealt{JWF2011}). We point out that latitudes with most dust in a given stripe vary with 
the stripe longitude. For example, for the $l\sim130^{\circ}$ stripe, the highest-extinction regions are found 
at positive latitudes, while for the $l\sim230^{\circ}$ stripe, most dust is found at negative latitudes.

\subsection{The Spatial Distribution of Stars} 

\begin{figure}[!t]
\epsscale{0.8}
\plotone{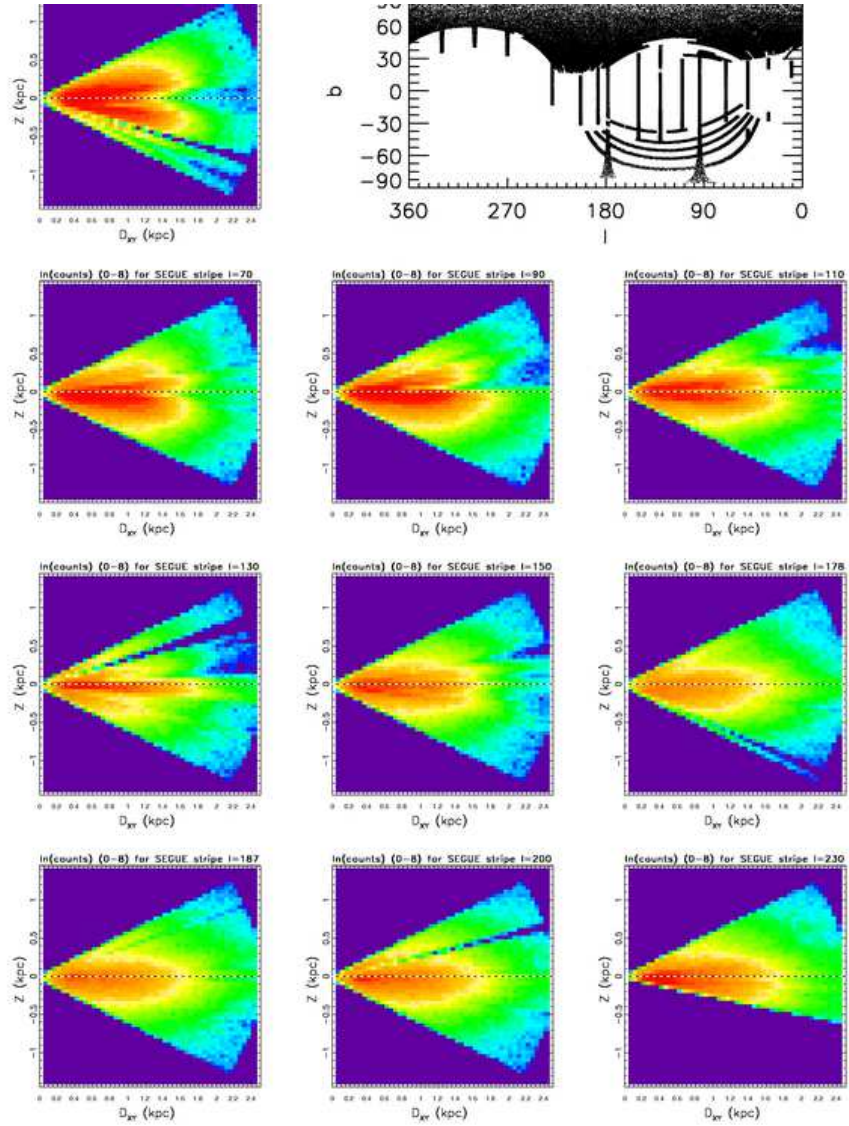}
\vskip -0.2in
\caption{The local volume number density of stars is shown as a function of 
distance from the Galactic plane, $Z$, and distance along the plane, $D_{xy}$,
for the same samples as shown in Figure~\ref{Fig:ZDcutArPanels}. The color scheme 
shows the counts on log scale with the same arbitrary normalization for all stripes. 
The fall-off of the stellar volume  number density at distances beyond $\sim$1 kpc 
is due to the stellar color-dependent sample distance limit and does not reflect the 
disk structure. Note the variation of counts with Galactic longitude (the top four
panels are closer to the Galactic center and contain more stars per unit volume).}
\label{Fig:ZDcutRhoPanels}
\end{figure}

The spatial distribution of stars (the number volume density) is shown in Figure~\ref{Fig:ZDcutRhoPanels}.
We have accounted for the change of volume with distance, but the variable distance limit due 
to faint flux cutoff and variable $A_r$ is not taken into account and is clearly visible in the figure. To 
fully exploit these data for constraining Galactic structure models, a three-dimensional dust map must 
first be derived from $A_r$ constraints (or at least carefully considered to mask 
high-extinction regions), and then one must apply color-dependent distance corrections.

Nevertheless, several encouraging features are 
already discernible in Figure~\ref{Fig:ZDcutRhoPanels}. First, the sample seems fairly
complete for distances below 1 kpc, corresponding to vertical distances from the plane
of up to $|Z|\sim$0.5 kpc. This volume is poorly explored by SDSS high-latitude data 
(e.g. see Figure~15 in J08) and the dataset presented here will enable detailed studies 
of the disk stellar number density profile for small $|Z|$ (e.g., is the exponential profile 
valid within 100 pc from the disk mid-plane?). Second, the stellar number density at
a fiducial location (say, at a distance of 0.5 kpc and $Z=0.3$ kpc) significantly varies
with Galactic longitude. This is expected behavior for an exponential disk profile in 
the galactocentric radial direction, and these data can be used to improve the 
exponential scale length estimates for thin and thick disks (for more details, please 
see \S4 in J08).

\section{ SUMMARY AND DISCUSSION}

This is the first analysis based on SDSS data that simultaneously estimates intrinsic 
stellar color and dust extinction along the line of sight for several tens of millions of 
stars detected in the low Galactic latitude SEGUE survey. The fitting method and various
assumptions are described in \S\ref{sec:methodology}. Our main results are:
\begin{enumerate}
\item The wavelength range spanned by the SDSS photometric system and
the delivered photometric accuracy are sufficient to constrain the intrinsic 
stellar SED and dust extinction along the line of sight. The minimum 
required photometric accuracy of $\sim$0.03 mag prevents non-unique 
solutions in most cases, and the accuracy of best-fit parameters scales 
roughly linearly with smaller errors. At the same time, this accuracy requirement
effectively limits the sample to about $r<19$. 
\item Using the joint SDSS-2MASS photometry for stars at high Galactic latitudes,  
we confirmed the SDSS-based result from Sch2010 that the \citet{ODonnell} reddening 
law can be rejected. We adopted the \citet{CCM} reddening
law in this work, which is similar to the \citet{Fitz99} reddening law adopted by Sch2010. 
Formally, both models are mildly inconsistent with the 
SDSS-2MASS data, but in practice photometric implications of these differences
are minor ($\sim0.01$ mag when $A_r=1$). {\it We recommend the coefficients listed
in the first row in Table 1 for correcting SDSS and 2MASS photometry for interstellar
dust extinction.}
\item For stars detected by both SDSS and 2MASS, and when $R_V$ is not a free
fitting parameter, the best-fit intrinsic stellar color and $A_r$ for only-SDSS (four 
colors) and SDSS-2MASS (seven colors) fitting cases are similar. Although SDSS samples reach 
much further than SDSS-2MASS samples at high Galactic latitudes (the distance limits for blue
stars differ by about a factor of 10), this is not the case at low Galactic latitudes
because observed sources are much redder due to dust. The main benefit of
only-SDSS samples is about a factor of 2 larger distance limit for blue main 
sequence stars; however, the limiting distance for red stars is smaller than for 
SDSS-2MASS case due to a necessary limit on the $u$ band photometric errors. 
\item The SDSS photometry is not sufficient to reliably estimate $R_V$,
with a realistic mock catalog implying errors of about 1. However, the addition 
of 2MASS photometry significantly improves the accuracy of $R_V$ estimates, 
with realistic mock catalogs implying errors of about 0.3 when $A_r \sim 1$, 
and as small as 0.1 for $A_r>4$.  When $R_V$ can be reliably estimated, we find that 
$R_V=3.1$ cannot be ruled out in any
of the ten SEGUE strips (at a systematics-limited precision level of $\sim0.1-0.2$). 
Our best estimate for the intrinsic scatter of $R_V$ in the regions probed by SEGUE
stripes is $\la0.2$. 
\item Simultaneous fits for the intrinsic stellar SED and dust extinction along the line 
of sight allow for efficient recognition of candidate red giant stars in the disk. The selection 
method, which we dub ``dusty parallax relation'', utilizes the increase of dust extinction 
with distance, and identifies candidate giants as stars with anomalously large best-fit $A_r$ for
their best-fit main-sequence distance.
\item The SDSS-2MASS photometric dataset allows robust mapping of the three-dimensional
spatial distributions of main sequence stars and dust to a distance of about 2 kpc (and 
$A_r \la 2-3$). To extend this distance limits, deeper optical and infrared data are needed. 
With LSST and Wide-Field Infrared Survey Explorer (WISE) datasets (see below), the distance 
limit could be extended by close to a factor of 10. 
\item The three-dimensional spatial distributions of stars and dust can be readily 
analyzed with the datasets discussed here, which we make public (see Appendix B). 
{\bf For most scientific applications, we recommend the use of the SDSS-2MASS dataset
with fits based on all seven colors, and with $R_V$ fixed to its canonical value.} 
For studies exploring the $R_V$ variations, the use of
the full SDSS-2MASS dataset is mandatory, with best-fit $R_V$ trustworthy only for 
stars with $\chi^2_{pdf}<2$ and $A_r>1$. Our fits represent a ``stress test'' for
both SDSS and 2MASS photometry, and {\it we emphasize that careful quality cuts must be
applied to avoid unreliable results!} 
\end{enumerate}

Given the results presented by Sch2010, \citet{PG2010}, \citet{JWF2011} and here, 
it is confirmed beyond doubt that there are some systematic problems 
with the normalization of SFD extinction map. Nevertheless, at high Galactic latitudes 
with small extinction these errors do not dominate over the photometric zeropoint 
calibration errors in SDSS data (0.01-0.02 mag), and at low Galactic latitudes 
most stars are embedded in dust and thus the SFD map is of limited use. 

Analysis described at the beginning of \S\ref{sec:freeRv} shows that the datasets
analyzed here can robustly distinguish predictions made by the three popular models
for the shape of dust extinction curve. The O'Donnell model is clearly excluded, and
the other two models do not provide a perfect fit to data either. On the other hand, the 
differences are very small and not much larger than systematic errors in photometry. 
The systematic photometric and other errors translate to a systematic uncertainty in $R_V$ of
about 0.1-0.2. We did not detect any deviations from the canonical value $R_V=3.1$
at this precision level. We reach the same conclusion by \citet{JWF2011}, 
but here we obtained several times smaller errors due to a much wider wavelength range 
of utilized photometry. This uniformity of dust properties within a fairly large volume
(distance limit of the order 1 kpc) probably implies that the ISM dust is well mixed
during its lifetime \citep{draineBook}. 

Last but not least, it will be very informative to directly compare the results presented
here with those obtained by other methods, such as near-infrared color excess 
method \citep{Lombardi2001,Lombardi2011,Majewski2011},  H$\alpha$-based method
\citep{Sale2009}, and Wolf method that is sensitive to gray dust \citep{Yasuda2007,GB2010},
to uncover and quantify various systematic errors that are likely to exist in all methods.


The results presented here will be greatly improved by several upcoming large-scale, 
deep optical surveys, including the Dark Energy Survey \citep{Flaugher08}, Pan-STARRS 
\citep{Kaiser02}, and the LSST \citep{IvezicLSST}.  These 
surveys will significantly extend the faint limit of the sample analyzed here (in case
of LSST by $\sim5$~mag) and are likely to provide more reliable photometry due 
to multiple observations and the use of photometric methods designed for crowded fields. 
Although 2MASS is too shallow to fully complement these new optical surveys, it
will still provide very useful constraints in high-extinction regions. Furthermore, the 
recently released WISE data \citep{WrightWISE} will provide supplemental constraints with
its W1 band at 3.4 \mic, which reaches about 2 mag deeper than 2MASS $K$ band. 
These new datasets are thus certain to provide valuable new information about the dust
and stellar distribution within the Galactic disk beyond the current limiting distance of 
a few kpc.

\acknowledgements
 
\v{Z}. Ivezi\'{c} and B. Sesar acknowledge support by NSF grants AST-615991 
and AST-0707901, and by NSF grant AST-0551161 to LSST for design and
development activity. 
\v{Z}. Ivezi\'{c} thanks the University of Zagreb, where portion of this work was completed,
for its hospitality, and acknowledges support by the Croatian National Science Foundation 
grant O-1548-2009.  M. Berry, \v{Z}. Ivezi\'{c} and B. Sesar acknowledge hospitality by 
the Institute for Astronomy, University of Hawaii. 
D. Finkbeiner and E. Schlafly acknowledge support of NASA grant NNX10AD69G.
This work was supported by the Director, Office of Science, Office of High Energy Physics, of the U.S. Department of Energy under Contract No. DE-AC02-05CH11231.
T.C. Beers acknowledges partial support from
PHY 08-22648: Physics Frontier Center/Joint Institute for Nuclear
Astrophysics (JINA), awarded by the U.S. National Science Foundation.
We acknowledge the hospitality of the KITP at the University of California, 
Santa Barbara, where part of this work was completed (supported by NSF grant 
PHY05-51164). Fermilab is operated by Fermi Research Alliance, LLC under Contract 
No. DE-AC02-07CH11359 with the United States Department of Energy.

Funding for the SDSS and SDSS-II has been provided by the Alfred
P. Sloan Foundation, the Participating Institutions, the National
Science Foundation, the U.S. Department of Energy, the National
Aeronautics and Space Administration, the Japanese Monbukagakusho, the
Max Planck Society, and the Higher Education Funding Council for
England. The SDSS Web Site is http://www.sdss.org/.
The SDSS is managed by the Astrophysical Research Consortium for the
Participating Institutions. The Participating Institutions are the
American Museum of Natural History, Astrophysical Institute Potsdam,
University of Basel, University of Cambridge, Case Western Reserve
University, University of Chicago, Drexel University, Fermilab, the
Institute for Advanced Study, the Japan Participation Group, Johns
Hopkins University, the Joint Institute for Nuclear Astrophysics, the
Kavli Institute for Particle Astrophysics and Cosmology, the Korean
Scientist Group, the Chinese Academy of Sciences (LAMOST), Los Alamos
National Laboratory, the Max-Planck-Institute for Astronomy (MPIA),
the Max-Planck-Institute for Astrophysics (MPA), New Mexico State
University, Ohio State University, University of Pittsburgh,
University of Portsmouth, Princeton University, the United States
Naval Observatory, and the University of Washington.

\appendix

\section{\bf {\tt SQL} Query Example}

The following {\tt SQL} query was used to select and download data
for all SDSS stars with spectroscopic and proper-motion measurements
(see http://casjobs.sdss.org/CasJobs). 

\begin{verbatim}
SELECT
  round(p.ra,6) as ra, round(p.dec,6) as dec, 
  p.run, p.camcol, p.field,           --- comments are preceded by ---
  round(p.extinction_r,3) as rExtSFD, --- r band extinction from SFD
  round(p.modelMag_u,3) as uRaw,   --- N.B. ISM-uncorrected model mags
  round(p.modelMag_g,3) as gRaw,   --- rounding up 
  round(p.modelMag_r,3) as rRaw,  
  round(p.modelMag_i,3) as iRaw, 
  round(p.modelMag_z,3) as zRaw, 
  round(p.modelMagErr_u,3) as uErr, 
  round(p.modelMagErr_g,3) as gErr, 
  round(p.modelMagErr_r,3) as rErr, 
  round(p.modelMagErr_i,3) as iErr,
  round(p.modelMagErr_z,3) as zErr,
  (case when (p.flags & '16') = 0 then 1 else 0 end) as ISOLATED, 
  ISNULL(round(t.pmL,3), -9999) as pmL, --- proper motion data are set to 
  ISNULL(round(t.pmB,3), -9999) as pmB, --- -9999 if non-existent (NULL)
  ISNULL(round(t.pmRaErr,3), -9999) as pmErr  --- if pmErr < 0 no pm data
INTO mydb.dustSample
FROM phototag p LEFT OUTER JOIN propermotions t ON 
  (p.objID = t.objID and t.match = 1 and t.sigra < 350 and t.sigdec < 350)
         --- quality cut on pm
WHERE
  p.type = 6 and              --- select unresolved sources
  (p.flags & '4295229440') = 0 and --- '4295229440' is code for no 
                             --- DEBLENDED_AS_MOVING or SATURATED objects
  p.mode = 1  --- PRIMARY objects only, which implies 
              --- !BRIGHT && (!BLENDED || NODEBLEND || nchild == 0)]
  p.modelMag_r < 21  --- adopted faint limit
--- the end of query
\end{verbatim}

\section{Data Distribution}

All data files, as well as a detailed description of their content, are available from a public data 
repository\footnote{http://www.astro.washington.edu/users/ivezic/r\_datadepot.html}.
Due to the large data volume, we separate our catalogs into four groups. We fit stellar SEDs twice 
for all 10 SEGUE strips: once with selective extinction fixed at $R_V=3.0$, and a second time with 
$R_V$ as a free fitting parameter (limited to the range $1-7.9$). Similarly, we present the only-SDSS 
and SDSS-2MASS datasets separately. For the $R_V=3.0$ case, the data files in each dataset (only-SDSS 
and SDSS-2MASS) are defined by Galactic coordinates, and are designed to contain fewer than 10 million 
stars each. For the free-$R_V$ case, we distribute only the data from SEGUE strips with $|b|<30^\circ$ 
because $R_V$ is poorly constrained at higher Galactic latitudes with small extinction. This data organization 
allows users to download data for a relatively small region of sky without the burden of downloading the 
whole dataset. These datasets are made available in two formats: as FITS tables, and as plain ASCII text files.

All of the data files contain SDSS astrometry and photometry (and proper motions), the SFD value for $A_r$, 
and best-fit model parameters (including a best-fit distance estimate).  Additionally, the SDSS-2MASS data 
files also contain 2MASS astrometry and photometry, the only-SDSS best-fit parameters, and the SDSS-2MASS 
best-fit parameters.

We emphasize that our fits represent a ``stress test'' for both SDSS and 2MASS photometry, and thus 
{\bf careful quality cuts must be applied to avoid unreliable results!}

\section{\bf Discussion of the Methodology}

Here we provide a more detailed discussion of two aspects of methods discussed in
\S\ref{sec:methodology}.

\subsection{ Closing the System of Equations}

Stellar colors constrain reddening due to dust, e.g., $A_{ug}=A_u-A_g$, rather than
dust extinction, here $A_u$ and $A_g$. Therefore, when inferring the amount of 
dust extinction, both in case of single stars that are projected onto the unreddened 
stellar locus in the multi-dimensional color space, and in case of color offsets of the 
whole stellar locus at high Galactic latitudes, there is always one constraint 
{\it fewer} than the number of photometric bands. A convenient way of thinking about 
this ``missing'' equation is that dust extinction is described by its ``scale'' $A_r$ and 
four (or seven in SDSS-2MASS case) measures of the {\it scaleless shape} of the extinction
curve, $C_\lambda=A_\lambda/A_r$. Three different approaches can be used to ``close'' this 
system of equations, and to break ``reddening-extinction'' degeneracy (we do not discuss
the best approach, based on {\it known} distance modulus, $DM$, and absolute magnitude, 
$M_r$, which directly constrains $A_r$ via $r = M_r + DM + A_r$, because our dataset 
does not include a parallax distance to the vast majority of stars).  

The first approach assumes that $A_r$ is provided as an additional input, for example,
from the SFD map as $A_r^{SFD}$. In this case, $A_{ug}=(C_u-C_g)\,A_r^{SFD}$,
and it is easy to show that 
\begin{eqnarray}
        C_u = 1  + {A_{ug}+A_{gr} \over A_r^{SFD}}   \\
        C_g = 1  + {A_{gr} \over A_r^{SFD}}                \\
        C_i  = 1  - {A_{ri} \over A_r^{SFD}}                  \\
        C_z = 1  - {A_{ri}+A_{iz} \over A_r^{SFD}}. 
\end{eqnarray}

If there are systematic errors in $A_r^{SFD}$, they will be propagated to $C_m$. 
Such effects can be tested for by tracing the variation of resulting $C_m$ across
the sky, and by correlating deviations with $A_r^{SFD}$. In particular, given many
lines of sight, it is possible to fit a spatially invariant model for errors in $A_r^{SFD}$ 
(e.g., an additive and a multiplicative error). 

To illustrate the impact of errors in $A_r^{SFD}$ on $C_m$ determined with this method, 
we computed $A_{color}$ using the CCM model with $R_V=3.1$ and true $A_r=1$, and 
we assumed a multiplicative error in $A_r^{SFD}$. A correction factor of 0.95 produces the 
overall best-fit $R_V=3.28$, with the $i$ and $z$ band constraints on $R_V$ biased to 
even higher values. For a correction factor of 0.9, the best-fit $R_V=3.48$, with the
$i$ and $z$ extinction values barely consistent with the CCM extinction curve. For 
additive errors such that the true $A_r = A_r^{SFD}+0.05$, the best-fit $R_V$ varies 
from 2.20 for$A_r^{SFD}=0.1$, to 2.93 at for $A_r^{SFD}=1$. Therefore, this
method is quite sensitive to systematic errors in $A_r^{SFD}$ and should be used
with caution. 

The second approach uses a model-based extinction curve to predict $C_m$ 
as a function of {\it single} parameter $R_V$. Given that there are three 
``spare'' constraints, model predictions can be tested for self-consistency (and 
perhaps used to select the ``best'' model). This method
results in estimates for $A_r$ and can be used to test external maps, such 
as SFD, though {\it only in a model-dependent way}. 

The third approach, adopted here, is to assume (fix) one value of $C_m$ and solve 
for $A_r$ and all remaining $C_m$. While at first this approach sounds arbitrary,
it becomes sound when SDSS data are augmented with 2MASS data. The reason is that the effective
wavelength for 2MASS $K$ band is 2.2 \mic, which in this context is almost as 
large as infinity\footnote{Please do not take this statement {\it out of this context}!}.
When using both SDSS and 2MASS, $A_r$ is estimated using offsets of the $r-K$
color distribution. The main reason why this approach works is the fact that $A_K/A_r$
is small (0.132) and varies little with $R_V$, and among all plausible dust extinction
models. For example, for $R_V>2$ all models predict variations of $A_K/A_r$ not
exceeding 20\%. This variation translates to only about a 3\% error when the $r-K$ shift
is interpreted as $A_r-A_K=(1-A_K/A_r)\,A_r=0.868\,A_r$. When using this approach, 
there are seven colors constructed with eight photometric bands, and the result is
estimates for $A_r$ and six $C_m$. 

\subsection{Methods for Quantifying Color Offsets for the Stellar Locus }

What is the optimal method for measuring $A_{ug}$, $A_{gr}$, etc., using the stellar 
locus?  If we think of the stellar locus in a two-dimensional (2D) color-color diagram
as of an ``image'', then we essentially ``slide'' the image of the reddened 
sample to perfectly align with the image of the ``intrinsic'' dereddened locus. 
This alignment can be performed in each 2D color-color diagram, or alternatively
all four color shifts can be determined simultaneously in the 4D color space. 
At the other extreme, the color shifts can be determined using 1D projections
of each color, as in the ``blue tip'' method proposed by \citet{Sch2010}.

If there were no astrophysical systematics and measurement error distributions
were fully understood, these methods should produce identical results (e.g., the 
$g-r$ offsets estimated from the $g-r$ vs. $u-g$ and $r-i$ vs. $g-r$ diagrams 
would be statistically consistent). However, there are astrophysical systematics,
such as distance, age and metallicity effects, that may introduce various biases.
For example, M dwarf stars in SDSS sample can be as close as 100 pc and thus be within dust layer, 
and the ``blue tip''  is sensitive to age and metallicity of turn-off stars that define 
it. The idea behind the principal colors method employed here is to avoid distance 
effects by considering only stars bluer than M dwarfs, and to mitigate age and metallicity 
effects by measuring shifts perpendicular to the locus. The reason for the latter is that 
age variation ``extends'' or ``shortens'' the locus (i.e., shifts the ``blue tip''), 
but does not strongly affect its position in the perpendicular ($P_2$) direction. 
When considering metallicity, systematic effects are a little bit more complicated, but are 
mostly confined to the $u$ band. For blue stars, the $g-r$ color is essentially a 
measure of effective temperature with negligible dependence on metallicity
\citep{Ivezic08}. At a given $g-r$ color, the $u-g$ color depends on metallicity 
(it becomes bluer as metallicity decreases, see the top right panel in fig.~2 in
\citealt{Ivezic08}). For example, at $g-r$=0.3, the $u-g$ color varies by about 
0.2 mag as the metallicity varies from the median thick-disk value ($-0.5$) to
the median halo value ($-1.5$). This shift is not parallel to the locus in the
$g-r$ vs. $u-g$ color-color diagram, so it does have some effect on the $P_2$ 
distribution. However, already at $g-r=0.5$, the fraction of halo stars in SDSS
sample is sufficiently small that this effect becomes negligible (because such red
halo stars are too faint to be detected by SDSS). Hence, in the range
$0.5 < g-r < 1.2$, only the dust reddening (and photometric calibration errors, of course!) 
can significantly shift the locus perpendicularly to its blue part (even in the $u-g$ 
vs. $g-r$ diagram). An added benefit from the signal-to-noise ratio viewpoint is 
that the $P_2$ distributions are very narrow, an advantage that mitigates the fact that 
the reddening vectors are measured only along $P_2$ directions.

\bibliographystyle{apj}
\bibliography{apj-jour,tomoIV}

\newpage

\begin{table}[h]
\caption{Observational Constraints and Model Values for the Extinction Curve, $C_\lambda\equiv A_\lambda/A_r$}
\centering
\vskip 0.2in
\begin{tabular}{r c c c c c c c}
\hline \hline
Region &  $u$  &  $g$  &  $i$  &   $z$  &  $J$ & $H$ & $K_s$  \\ 
\hline
S82      & 1.810  & 1.400  &  0.759  & 0.561  & 0.317 &  0.200 & 0.132  \\ 
North   & 1.750  & 1.389  &  0.750  & 0.537  & 0.297 &  0.180 & 0.132  \\ 
CCM     & 1.814  & 1.394  &  0.764  & 0.552  & 0.327 &  0.205 & 0.132  \\  
F99      & 1.795  & 1.415  &  0.748  & 0.554  & 0.308 &  0.194 & 0.132  \\ 
OD       & 1.813  & 1.406  &  0.783  & 0.562  & 0.325 &  0.205 & 0.132  \\ 
\hline
\end{tabular}
\label{Tab:ClambdaConstraints}
\tablecomments{The first two rows list observational constraints for the
shape of the extinction curve, $C_\lambda\equiv A_\lambda/A_r$. The value
of $C_\lambda$ in the $K$ band was {\it assumed to be 0.132}. The first
row corresponds to the so-called SDSS Stripe 82 region (defined by 
$300^\circ < {\rm R.A.} < 60^\circ$ and $|Dec|<1.27^\circ$), and the
second row to a northern region defined by $30^\circ < b < 45^\circ$ and 
$0^\circ < l < 10^\circ$. The last three
rows list model predictions computed for an F star spectrum and the 
best-fit value of $R_V$ (CCM=Cardelli et al. 1989: $R_V=3.01$; F99=Fitzpatrick 
1999: $R_V=3.30$; OD=O'Donnell 1994: $R_V=3.05$).}
\end{table}

\begin{table}[h]
\caption{Adopted Extinction Coefficients, $C_\lambda(R_V)$}
\centering
\vskip 0.2in
\begin{tabular}{c c c c c c c c c}
\hline \hline
$R_V$ &  $u$  &  $g$  &  $i$  &   $z$  &  $J$ & $H$ & $K_s$ & Source \\ 
\hline
 2.0 & 2.280  & 1.579  &  0.702  & 0.453  & 0.264 &  0.166 & 0.107 & CCM \\ 
 2.5 & 1.998  & 1.467  &  0.740  & 0.513  & 0.302 &  0.190 & 0.122 & CCM \\ 
 3.0 & 1.817  & 1.395  &  0.764  & 0.552  & 0.326 &  0.205 & 0.132 & CCM \\
\hline
 3.1 & 1.788  & 1.384  &  0.768  & 0.558  & 0.330 &  0.208 & 0.134 & CCM \\
3.1 & 1.855  & 1.446  &  0.743  & 0.553  &  \dots &  \dots & \dots & Sch2010 \\ 
 3.1 & 1.857  & 1.439  &  0.725  & 0.517  &  0.250 & 0.131 & 0.068 & F99 \\
\hline
 4.0 & 1.598  & 1.308  &  0.793  & 0.598  & 0.356 &  0.224 & 0.144 & CCM \\
 5.0 & 1.470  & 1.257  &  0.810  & 0.625  & 0.373 &  0.234 & 0.151 & CCM \\
\hline
\end{tabular}
\label{Tab:ClambdaFreeRv}
\tablecomments{An illustration of the dependence of the adopted extinction
curve, $C_\lambda\equiv A_\lambda/A_r$ on $R_V$ ($C_r=1$ by definition; see 
also Figure~\ref{Fig:AlambdaRv}).
The second line with $R_V=3.1$ lists the values suggested by \citealt{SF2010},
and the third line with $R_V=3.1$ lists the values computed using eq.~5 from \citealt{FM09}
with $\alpha=2.50$ (constrained by $C_r=1$), and using $A_V/A_r=1.20$ and the effective wavelengths
from \citep{SF2010} for the SDSS bands, and 1.25 \mic, 1.65 \mic, and 2.17 \mic\ for the
2MASS $JHK$ bands, respectively.  Both lines are presented for a comparison with the
adopted CCM extinction curve.}
\end{table}



\clearpage

\clearpage

\clearpage

\clearpage

\clearpage

\clearpage

\clearpage

\clearpage

\clearpage

\clearpage

\clearpage

\clearpage



\clearpage


\clearpage

\clearpage

\clearpage

\clearpage

\clearpage

\clearpage

\clearpage


\clearpage

\clearpage

\clearpage

\clearpage

\clearpage

\clearpage

\clearpage

\clearpage

\end{document}